\documentclass[apjl]{emulateapj}
\usepackage{comment}
\usepackage{ifthen}

\newcommand{\forloop}[5][1]%
{%
\setcounter{#2}{#3}%
\ifthenelse{#4}%
	{%
	#5%
	\addtocounter{#2}{#1}%
	\forloop[#1]{#2}{\value{#2}}{#4}{#5}%
	}%
	{%
	}%
}%

\newcommand{\ctbd}[1]{}

\newcommand{\lc}{light curve}
\newcommand{\lcs}{light curves}
\newcommand{\Lc}{Light curve}

\newcommand{\band}[1]{\ensuremath{#1}-band}

\newcommand{\chisq}{\ensuremath{\chi^2}}

\newcommand{\kms}{\ensuremath{\rm km\,s^{-1}}}
\newcommand{\ms}{\ensuremath{\rm m\,s^{-1}}}

\newcommand{\msd}{\ensuremath{\rm m\,s^{-1}\,d^{-1}}}
\newcommand{\gcmc}{\ensuremath{\rm g\,cm^{-3}}}
\newcommand{\ergscmsq}{\ensuremath{\rm erg\,s^{-1}\,cm^{-2}}}

\newcommand{\vsini}{\ensuremath{v \sin{i}}}
\newcommand{\feh}{\ensuremath{\rm [Fe/H]}}

\newcommand{\vmac}{\ensuremath{v_{\rm mac}}}
\newcommand{\vmic}{\ensuremath{v_{\rm mic}}}

\newcommand{\vic}{\ensuremath{V\!-\!I_C}}

\newcommand{\rsun}{\ensuremath{R_\sun}}
\newcommand{\msun}{\ensuremath{M_\sun}}
\newcommand{\lsun}{\ensuremath{L_\sun}}

\newcommand{\rstar}{\ensuremath{R_\star}}
\newcommand{\mstar}{\ensuremath{M_\star}}
\newcommand{\lstar}{\ensuremath{L_\star}}

\newcommand{\teffstar}{\ensuremath{T_{\rm eff\star}}}
\newcommand{\rhostar}{\ensuremath{\rho_\star}}
\newcommand{\loggstar}{\ensuremath{\log{g_{\star}}}}

\newcommand{\mearth}{\ensuremath{M_\earth}}

\newcommand{\rpl}{\ensuremath{R_{p}}}
\newcommand{\mpl}{\ensuremath{M_{p}}}

\newcommand{\rhopl}{\ensuremath{\rho_{p}}}

\newcommand{\arstar}{\ensuremath{a/\rstar}}
\newcommand{\zrstar}{\ensuremath{\zeta/\rstar}}

\newcommand{\rjup}{\ensuremath{R_{\rm J}}}
\newcommand{\mjup}{\ensuremath{M_{\rm J}}}

\newcommand{\refsec}[1]{\mbox{\S\ \ref{sec:#1}}}

\newcommand{\reffigl}[1]{Figure~\ref{fig:#1}}
\newcommand{\refsecl}[1]{\mbox{Section \ref{sec:#1}}}

\newcommand{\reftabl}[1]{Table~\ref{tab:#1}}

\newcommand{\flwof}{\mbox{FLWO 1.2\,m}}

\newcommand{\loopand}{\ifnum\value{planetcounter}=2 and \else\fi}
\newcommand{\loopcomma}{\ifnum\value{planetcounter}<2 ,\else. \fi}
\newcommand{\loopcommanoperiod}{\ifnum\value{planetcounter}<2 ,\else \space\fi}
\newcommand{\loopcommanospace}{\ifnum\value{planetcounter}<2 ,\else \fi}

\newcommand{\hatcurhtrxxxA}{HTR239-004}                                    
\newcommand{\hatcurfieldxxxA}{239}                                         
\newcommand{\hatcurCCraxxxA}{\ensuremath{17^{\mathrm h}05^{\mathrm m}23.28{\mathrm s}}}                                  
\newcommand{\hatcurCCdecxxxA}{\ensuremath{+33{\arcdeg}00{\arcmin}45.0{\arcsec}}}                                 
\newcommand{\hatcurCCmagxxxA}{12.759}                                      
\newcommand{\hatcurCCtwomassxxxA}{2MASS~17052315+3300450}                  
\newcommand{\hatcurCCgscxxxA}{GSC~2594-00646}                              
\newcommand{\hatcurCCtassmvxxxA}{12.759}                                   
\newcommand{\hatcurCCtwomassJmagxxxA}{\ensuremath{10.822\pm0.020}}         
\newcommand{\hatcurCCtwomassHmagxxxA}{\ensuremath{10.340\pm0.019}}         
\newcommand{\hatcurCCtwomassKmagxxxA}{\ensuremath{10.234\pm0.017}}         
\newcommand{\hatcurCCcitJmagxxxA}{\ensuremath{10.827\pm0.021}}             
\newcommand{\hatcurCCcitHmagxxxA}{\ensuremath{10.334\pm0.020}}             
\newcommand{\hatcurCCcitKmagxxxA}{\ensuremath{10.258\pm0.017}}             
\newcommand{\hatcurCCbbJmagxxxA}{\ensuremath{10.894\pm0.023}}              
\newcommand{\hatcurCCbbHmagxxxA}{\ensuremath{10.356\pm0.020}}              
\newcommand{\hatcurCCbbKmagxxxA}{\ensuremath{10.278\pm0.017}}              
\newcommand{\hatcurCCesoJmagxxxA}{\ensuremath{10.899\pm0.026}}             
\newcommand{\hatcurCCesoHmagxxxA}{\ensuremath{10.351\pm0.025}}             
\newcommand{\hatcurCCesoKmagxxxA}{\ensuremath{10.276\pm0.019}}             
\newcommand{\hatcurCCesoJHmagxxxA}{\ensuremath{0.547\pm0.033}}             
\newcommand{\hatcurCCesoJKmagxxxA}{\ensuremath{0.623\pm0.036}}             
\newcommand{\hatcurCCesoHKmagxxxA}{\ensuremath{0.075\pm0.030}}             
\newcommand{\hatcurLCdipxxxA}{\ensuremath{18.5}}                           
\newcommand{\hatcurLCrprstarxxxA}{\ensuremath{0.1365\pm0.0015}}            
\newcommand{\hatcurLCbsqxxxA}{\ensuremath{0.105_{-0.040}^{+0.040}}}        
\newcommand{\hatcurLCimpxxxA}{\ensuremath{0.324_{-0.078}^{+0.055}}}        
\newcommand{\hatcurLCzetaxxxA}{\ensuremath{20.36\pm0.08}}                  
\newcommand{\hatcurLCdurxxxA}{\ensuremath{0.1131\pm0.0009}}                
\newcommand{\hatcurLCdurshortxxxA}{\ensuremath{0.11310}}                    
\newcommand{\hatcurLCdurhrxxxA}{\ensuremath{2.716\pm0.021}}                
\newcommand{\hatcurLCdurhrshortxxxA}{\ensuremath{2.716}}                   
\newcommand{\hatcurLCqxxxA}{\ensuremath{0.0205\pm0.0002}}                  
\newcommand{\hatcurLCqshortxxxA}{\ensuremath{0.021}}                       
\newcommand{\hatcurLCingdurxxxA}{\ensuremath{0.0150\pm0.0008}}             
\newcommand{\hatcurLCPxxxA}{\ensuremath{5.508023\pm0.000006}}              
\newcommand{\hatcurLCPprecxxxA}{\ensuremath{5.5080228}}                    
\newcommand{\hatcurLCPshortxxxA}{\ensuremath{5.5080}}                      
\newcommand{\hatcurLCTxxxA}{\ensuremath{2454715.02174\pm0.00020}}          
\newcommand{\hatcurLCTAxxxA}{\ensuremath{2454191.75958\pm0.00057}}         
\newcommand{\hatcurLCTBxxxA}{\ensuremath{2454962.88277\pm0.00040}}         
\newcommand{\hatcurLCiblendxxxA}{\ensuremath{0.88\pm0.05}}                 
\newcommand{\hatcurLChatnetmxxxA}{\ensuremath{11.5534\pm0.0001}}           
\newcommand{\hatcurSMEiteffxxxA}{\ensuremath{4850\pm75}}                   
\newcommand{\hatcurSMEizfehxxxA}{\ensuremath{+0.12\pm0.05}}                 
\newcommand{\hatcurSMEizfehshortxxxA}{\ensuremath{+0.12}}                   
\newcommand{\hatcurSMEiloggxxxA}{\ensuremath{4.7\pm0.1}}                  
\newcommand{\hatcurSMEivsinxxxA}{\ensuremath{1.2\pm0.5}}                   
\newcommand{\hatcurSMEivmacxxxA}{\ensuremath{0.0}}                         
\newcommand{\hatcurSMEivmicxxxA}{\ensuremath{0.0}}                         
\newcommand{\hatcurSMEiiteffxxxA}{\ensuremath{4803\pm80}}                  
\newcommand{\hatcurSMEiizfehxxxA}{\ensuremath{+0.10\pm0.08}}               
\newcommand{\hatcurSMEiizfehshortxxxA}{\ensuremath{+0.10}}                 
\newcommand{\hatcurSMEiiloggxxxA}{\ensuremath{4.56\pm0.06}}                
\newcommand{\hatcurSMEiivsinxxxA}{\ensuremath{0.5\pm0.5}}                  
\newcommand{\hatcurSMEiivmacxxxA}{\ensuremath{2.53}}                       
\newcommand{\hatcurSMEiivmicxxxA}{\ensuremath{0.85}}                       
\newcommand{\hatcurDSteffxxxA}{\ensuremath{4750\pm100}}                    
\newcommand{\hatcurDSzfehxxxA}{\ensuremath{0.0\pm0.0}}                     
\newcommand{\hatcurDSloggxxxA}{\ensuremath{4.50\pm0.25}}                    
\newcommand{\hatcurDSvsinixxxA}{\ensuremath{1.5\pm1.0}}                    
\newcommand{\hatcurDSgammaxxxA}{\ensuremath{-11.92\pm0.28}}                
\newcommand{\hatcurDSnumspecxxxA}{\ensuremath{4}}                          
\newcommand{\hatcurDSspanxxxA}{\ensuremath{183}}                           
\newcommand{\hatcurDSrvrmsxxxA}{\ensuremath{0.28}}                         
\newcommand{\hatcurTRESteffxxxA}{\ensuremath{}}                            
\newcommand{\hatcurTRESzfehxxxA}{\ensuremath{}}                            
\newcommand{\hatcurTRESloggxxxA}{\ensuremath{}}                            
\newcommand{\hatcurTRESvsinixxxA}{\ensuremath{}}                           
\newcommand{\hatcurTRESgammaxxxA}{\ensuremath{}}                           
\newcommand{\hatcurTRESnumspecxxxA}{\ensuremath{}}                 
\newcommand{\hatcurTRESspanxxxA}{\ensuremath{}}                    
\newcommand{\hatcurTRESrvrmsxxxA}{\ensuremath{}}                   
\newcommand{\hatcurFIESteffxxxA}{\ensuremath{}}                            
\newcommand{\hatcurFIESzfehxxxA}{\ensuremath{}}                            
\newcommand{\hatcurFIESloggxxxA}{\ensuremath{}}                            
\newcommand{\hatcurFIESvsinixxxA}{\ensuremath{}}                           
\newcommand{\hatcurFIESgammaxxxA}{\ensuremath{}}                           
\newcommand{\hatcurFIESnumspecxxxA}{\ensuremath{}}                 
\newcommand{\hatcurFIESspanxxxA}{\ensuremath{}}                    
\newcommand{\hatcurFIESrvrmsxxxA}{\ensuremath{}}                   
\newcommand{\hatcurLBizxxxA}{\ensuremath{0.3441}}                          
\newcommand{\hatcurLBiizxxxA}{\ensuremath{0.2546}}                         
\newcommand{\hatcurLBiixxxA}{\ensuremath{0.4372}}                          
\newcommand{\hatcurLBiiixxxA}{\ensuremath{0.2276}}                         
\newcommand{\hatcurLBiIxxxA}{\ensuremath{0.4053}}                          
\newcommand{\hatcurLBiiIxxxA}{\ensuremath{0.2371}}                         
\newcommand{\hatcurLBigxxxA}{\ensuremath{0.8477}}                          
\newcommand{\hatcurLBiigxxxA}{\ensuremath{-0.0060}}                        
\newcommand{\hatcurLBikepxxxA}{\ensuremath{}}                      
\newcommand{\hatcurLBiikepxxxA}{\ensuremath{}}                     
\newcommand{\hatcurISOmxxxA}{\ensuremath{0.77\pm0.03}}                     
\newcommand{\hatcurISOmshortxxxA}{\ensuremath{0.77}}                       
\newcommand{\hatcurISOmlongxxxA}{\ensuremath{0.770\pm0.031}}               
\newcommand{\hatcurISOrxxxA}{\ensuremath{0.75\pm0.04}}                     
\newcommand{\hatcurISOrshortxxxA}{\ensuremath{0.75}}                       
\newcommand{\hatcurISOrlongxxxA}{\ensuremath{0.749\pm0.037}}               
\newcommand{\hatcurISOrhoxxxA}{\ensuremath{2.57\pm0.36}}                   
\newcommand{\hatcurISOloggxxxA}{\ensuremath{4.57\pm0.04}}                  
\newcommand{\hatcurISOlumxxxA}{\ensuremath{0.27\pm0.04}}                   
\newcommand{\hatcurISOlumshortxxxA}{\ensuremath{0.27}}                     
\newcommand{\hatcurISOmvxxxA}{\ensuremath{6.50\pm0.19}}                    
\newcommand{\hatcurISOvixxxA}{\ensuremath{0.977\pm0.037}}                  
\newcommand{\hatcurISOagexxxA}{\ensuremath{12.4_{-6.4}^{+4.4}}}            
\newcommand{\hatcurISOsigmaxxxA}{\ensuremath{0.00040\pm0.00009}}           
\newcommand{\hatcurISOMJxxxA}{\ensuremath{4.78\pm0.13}}                    
\newcommand{\hatcurISOMHxxxA}{\ensuremath{4.26\pm0.12}}                    
\newcommand{\hatcurISOMKxxxA}{\ensuremath{4.17\pm0.12}}                    
\newcommand{\hatcurISOJKxxxA}{\ensuremath{0.61\pm0.02}}                    
\newcommand{\hatcurISOspecxxxA}{K2}                                        
\newcommand{\hatcurRVKxxxA}{\ensuremath{27.1\pm1.6}}                       
\newcommand{\hatcurRVkxxxA}{\ensuremath{-0.035\pm0.038}}                   
\newcommand{\hatcurRVhxxxA}{\ensuremath{0.063\pm0.062}}                    
\newcommand{\hatcurRVgammaxxxA}{\ensuremath{-6.1\pm1.2}}                   
\newcommand{\hatcurRVjitterxxxA}{\ensuremath{5.0}}                         
\newcommand{\hatcurRVeccenxxxA}{\ensuremath{0.084\pm0.048}}                
\newcommand{\hatcurRVomegaxxxA}{\ensuremath{120\pm56}}                     
\newcommand{\hatcurPPixxxA}{\ensuremath{88.8\pm0.3}}                       
\newcommand{\hatcurPPgxxxA}{\ensuremath{4.9\pm0.6}}                        
\newcommand{\hatcurPPloggxxxA}{\ensuremath{2.69\pm0.05}}                   
\newcommand{\hatcurPParxxxA}{\ensuremath{16.04\pm0.75}}                    
\newcommand{\hatcurPParelxxxA}{\ensuremath{0.0559\pm0.0007}}               
\newcommand{\hatcurPPrhoxxxA}{\ensuremath{0.25\pm0.04}}                    
\newcommand{\hatcurPPmxxxA}{\ensuremath{0.20\pm0.01}}                      
\newcommand{\hatcurPPmshortxxxA}{\ensuremath{0.20}}                        
\newcommand{\hatcurPPmlongxxxA}{\ensuremath{0.197\pm0.013}}                
\newcommand{\hatcurPPmexxxA}{\ensuremath{62.5\pm4.0}}                      
\newcommand{\hatcurPPmeshortxxxA}{\ensuremath{62.5}}                       
\newcommand{\hatcurPPmelongxxxA}{\ensuremath{62.5\pm4.0}}                
\newcommand{\hatcurPPrxxxA}{\ensuremath{1.00\pm0.05}}                      
\newcommand{\hatcurPPrshortxxxA}{\ensuremath{1.00}}                        
\newcommand{\hatcurPPrlongxxxA}{\ensuremath{0.995\pm0.052}}                
\newcommand{\hatcurPPrexxxA}{\ensuremath{11.2\pm0.6}}                      
\newcommand{\hatcurPPreshortxxxA}{\ensuremath{11.2}}                       
\newcommand{\hatcurPPrelongxxxA}{\ensuremath{11.15\pm0.58}}                
\newcommand{\hatcurPPmrcorrxxxA}{\ensuremath{0.19}}                        
\newcommand{\hatcurPPteffxxxA}{\ensuremath{852\pm28}}                      
\newcommand{\hatcurPPthetaxxxA}{\ensuremath{0.029\pm0.002}}                
\newcommand{\hatcurPPfluxperixxxA}{\ensuremath{1.43\pm0.29}}              
\newcommand{\hatcurPPfluxperidimxxxA}{\ensuremath{8}}                      
\newcommand{\hatcurPPfluxapxxxA}{\ensuremath{1.0\pm0.1}}                   
\newcommand{\hatcurPPfluxapdimxxxA}{\ensuremath{8}}                        
\newcommand{\hatcurPPfluxavgxxxA}{\ensuremath{1.19\pm0.16}}               
\newcommand{\hatcurPPfluxavgdimxxxA}{\ensuremath{8}}                       
\newcommand{\hatcurXsecphasexxxA}{\ensuremath{0.478\pm0.024}}            
\newcommand{\hatcurXsecondaryxxxA}{\ensuremath{2454717.65\pm0.13}}       
\newcommand{\hatcurXsecdurxxxA}{\ensuremath{0.127\pm0.014}}              
\newcommand{\hatcurXsecingdurxxxA}{\ensuremath{0.0173\pm0.0029}}           
\newcommand{\hatcurPPphiconjxxxA}{\ensuremath{-0.06\pm0.15}}           
\newcommand{\hatcurPPperixxxA}{\ensuremath{2454715.36\pm0.81}}             
\newcommand{\hatcurPPaequivxxxA}{\ensuremath{0.1073\pm0.0071}}             
\newcommand{\hatcurXdistxxxA}{\ensuremath{166\pm9}}                        
\newcommand{\hatcurCCpmraxxxA}{\ensuremath{-21.2\pm5.4}}                   
\newcommand{\hatcurCCpmdecxxxA}{\ensuremath{-44.2\pm5.1}}                  
\newcommand{\hatcurCCpmxxxA}{\ensuremath{49.0\pm7.4}}               

\newcommand{\hatcurhtrxxxB}{HTR163-001}                                    
\newcommand{\hatcurfieldxxxB}{163}                                         
\newcommand{\hatcurCCraxxxB}{\ensuremath{00^{\mathrm h}38^{\mathrm m}04.02{\mathrm s}}}                                  
\newcommand{\hatcurCCdecxxxB}{\ensuremath{+34{\arcdeg}42{\arcmin}41.7{\arcsec}}}                                 
\newcommand{\hatcurCCmagxxxB}{12.901}                                      
\newcommand{\hatcurCCtwomassxxxB}{2MASS~00380401+3442416}                  
\newcommand{\hatcurCCgscxxxB}{GSC~2283-00589}                              
\newcommand{\hatcurCCtassmvxxxB}{12.901}                                   
\newcommand{\hatcurCCtwomassJmagxxxB}{\ensuremath{11.095\pm0.020}}         
\newcommand{\hatcurCCtwomassHmagxxxB}{\ensuremath{10.644\pm0.022}}         
\newcommand{\hatcurCCtwomassKmagxxxB}{\ensuremath{10.546\pm0.019}}         
\newcommand{\hatcurCCcitJmagxxxB}{\ensuremath{11.102\pm0.021}}             
\newcommand{\hatcurCCcitHmagxxxB}{\ensuremath{10.638\pm0.023}}             
\newcommand{\hatcurCCcitKmagxxxB}{\ensuremath{10.570\pm0.019}}             
\newcommand{\hatcurCCbbJmagxxxB}{\ensuremath{11.166\pm0.023}}              
\newcommand{\hatcurCCbbHmagxxxB}{\ensuremath{10.661\pm0.023}}              
\newcommand{\hatcurCCbbKmagxxxB}{\ensuremath{10.590\pm0.019}}              
\newcommand{\hatcurCCesoJmagxxxB}{\ensuremath{11.171\pm0.025}}             
\newcommand{\hatcurCCesoHmagxxxB}{\ensuremath{10.656\pm0.028}}             
\newcommand{\hatcurCCesoKmagxxxB}{\ensuremath{10.588\pm0.020}}             
\newcommand{\hatcurCCesoJHmagxxxB}{\ensuremath{0.515\pm0.034}}             
\newcommand{\hatcurCCesoJKmagxxxB}{\ensuremath{0.583\pm0.031}}             
\newcommand{\hatcurCCesoHKmagxxxB}{\ensuremath{0.068\pm0.034}}             
\newcommand{\hatcurLCdipxxxB}{\ensuremath{22.0}}                           
\newcommand{\hatcurLCrprstarxxxB}{\ensuremath{0.1418\pm0.0020}}            
\newcommand{\hatcurLCbsqxxxB}{\ensuremath{0.163_{-0.057}^{+0.055}}}        
\newcommand{\hatcurLCimpxxxB}{\ensuremath{0.404_{-0.088}^{+0.061}}}        
\newcommand{\hatcurLCzetaxxxB}{\ensuremath{19.76\pm0.12}}                  
\newcommand{\hatcurLCdurxxxB}{\ensuremath{0.1182\pm0.0014}}                
\newcommand{\hatcurLCdurshortxxxB}{\ensuremath{0.1182}}                    
\newcommand{\hatcurLCdurhrxxxB}{\ensuremath{2.837\pm0.034}}                
\newcommand{\hatcurLCdurhrshortxxxB}{\ensuremath{2.837}}                   
\newcommand{\hatcurLCqxxxB}{\ensuremath{0.0295\pm0.0004}}                  
\newcommand{\hatcurLCqshortxxxB}{\ensuremath{0.029}}                       
\newcommand{\hatcurLCingdurxxxB}{\ensuremath{0.0172\pm0.0014}}             
\newcommand{\hatcurLCPxxxB}{\ensuremath{4.008778\pm0.000006}}              
\newcommand{\hatcurLCPprecxxxB}{\ensuremath{4.0087782}}                    
\newcommand{\hatcurLCPshortxxxB}{\ensuremath{4.0088}}                      
\newcommand{\hatcurLCTxxxB}{\ensuremath{2455091.53417\pm0.00034}}          
\newcommand{\hatcurLCTAxxxB}{\ensuremath{2454349.9102\pm0.0012}}         
\newcommand{\hatcurLCTBxxxB}{\ensuremath{2455167.70095\pm0.00036}}         
\newcommand{\hatcurLChatnetmAxxxB}{\ensuremath{12.2539\pm0.0002}}          
\newcommand{\hatcurLCiblendAxxxB}{\ensuremath{0.96\pm0.03}}                
\newcommand{\hatcurLChatnetmBxxxB}{\ensuremath{12.2536\pm0.0002}}          
\newcommand{\hatcurLCiblendBxxxB}{\ensuremath{0.97\pm0.03}}                
\newcommand{\hatcurSMEiteffxxxB}{\ensuremath{5037\pm44}}                
\newcommand{\hatcurSMEizfehxxxB}{\ensuremath{+0.24\pm0.03}}                 
\newcommand{\hatcurSMEizfehshortxxxB}{\ensuremath{+0.24}}                   
\newcommand{\hatcurSMEiloggxxxB}{\ensuremath{4.7\pm0.1}}                 
\newcommand{\hatcurSMEivsinxxxB}{\ensuremath{2.4\pm0.5}}                  
\newcommand{\hatcurSMEivmacxxxB}{\ensuremath{2.85}}                        
\newcommand{\hatcurSMEivmicxxxB}{\ensuremath{0.85}}                        
\newcommand{\hatcurSMEiiteffxxxB}{\ensuremath{4990\pm130}}                 
\newcommand{\hatcurSMEiizfehxxxB}{\ensuremath{+0.23\pm0.08}}                
\newcommand{\hatcurSMEiizfehshortxxxB}{\ensuremath{+0.23}}                  
\newcommand{\hatcurSMEiiloggxxxB}{\ensuremath{4.53\pm0.06}}                
\newcommand{\hatcurSMEiivsinxxxB}{\ensuremath{0.7\pm0.5}}                  
\newcommand{\hatcurSMEiivmacxxxB}{\ensuremath{2.81}}                       
\newcommand{\hatcurSMEiivmicxxxB}{\ensuremath{0.85}}                       
\newcommand{\hatcurDSteffxxxB}{\ensuremath{4750\pm100}}                    
\newcommand{\hatcurDSzfehxxxB}{\ensuremath{0.0\pm0.0}}                     
\newcommand{\hatcurDSloggxxxB}{\ensuremath{4.0\pm0.25}}                    
\newcommand{\hatcurDSvsinixxxB}{\ensuremath{6.7\pm1.0}}                    
\newcommand{\hatcurDSgammaxxxB}{\ensuremath{-21.11\pm0.40}}                
\newcommand{\hatcurDSnumspecxxxB}{\ensuremath{3}}                          
\newcommand{\hatcurDSspanxxxB}{\ensuremath{37}}                            
\newcommand{\hatcurDSrvrmsxxxB}{\ensuremath{0.40}}                         
\newcommand{\hatcurTRESteffxxxB}{\ensuremath{5000\pm100}}                  
\newcommand{\hatcurTRESzfehxxxB}{\ensuremath{0.0\pm0.0}}                   
\newcommand{\hatcurTRESloggxxxB}{\ensuremath{4.5\pm0.25}}                  
\newcommand{\hatcurTRESvsinixxxB}{\ensuremath{2.0\pm1.0}}                  
\newcommand{\hatcurTRESgammaxxxB}{\ensuremath{-19.54\pm0.51}}              
\newcommand{\hatcurTRESnumspecxxxB}{\ensuremath{1}}                        
\newcommand{\hatcurTRESspanxxxB}{\ensuremath{1}}                           
\newcommand{\hatcurTRESrvrmsxxxB}{\ensuremath{0.00}}                       
\newcommand{\hatcurFIESteffxxxB}{\ensuremath{4750\pm100}}                  
\newcommand{\hatcurFIESzfehxxxB}{\ensuremath{0.0\pm0.0}}                   
\newcommand{\hatcurFIESloggxxxB}{\ensuremath{4.0\pm0.25}}                  
\newcommand{\hatcurFIESvsinixxxB}{\ensuremath{4.0\pm1.0}}                  
\newcommand{\hatcurFIESgammaxxxB}{\ensuremath{-19.64\pm0.04}}             
\newcommand{\hatcurFIESnumspecxxxB}{\ensuremath{4}}                        
\newcommand{\hatcurFIESspanxxxB}{\ensuremath{199}}                         
\newcommand{\hatcurFIESrvrmsxxxB}{\ensuremath{0.04}}                       
\newcommand{\hatcurLBizxxxB}{\ensuremath{0.3227}}                          
\newcommand{\hatcurLBiizxxxB}{\ensuremath{0.2713}}                         
\newcommand{\hatcurLBiixxxB}{\ensuremath{0.4135}}                          
\newcommand{\hatcurLBiiixxxB}{\ensuremath{0.2459}}                         
\newcommand{\hatcurLBiIxxxB}{\ensuremath{0.3831}}                          
\newcommand{\hatcurLBiiIxxxB}{\ensuremath{0.2542}}                         
\newcommand{\hatcurLBigxxxB}{\ensuremath{0.8016}}                          
\newcommand{\hatcurLBiigxxxB}{\ensuremath{0.0368}}                         
\newcommand{\hatcurLBikepxxxB}{\ensuremath{0.1000}}                        
\newcommand{\hatcurLBiikepxxxB}{\ensuremath{0.1000}}                       
\newcommand{\hatcurISOmxxxB}{\ensuremath{0.84\pm0.04}}                     
\newcommand{\hatcurISOmshortxxxB}{\ensuremath{0.84}}                       
\newcommand{\hatcurISOmlongxxxB}{\ensuremath{0.842\pm0.042}}               
\newcommand{\hatcurISOrxxxB}{\ensuremath{0.82\pm0.05}}                     
\newcommand{\hatcurISOrshortxxxB}{\ensuremath{0.82}}                       
\newcommand{\hatcurISOrlongxxxB}{\ensuremath{0.820\pm0.048}}               
\newcommand{\hatcurISOrhoxxxB}{\ensuremath{2.16\pm0.35}}                   
\newcommand{\hatcurISOloggxxxB}{\ensuremath{4.54\pm0.05}}                  
\newcommand{\hatcurISOlumxxxB}{\ensuremath{0.37_{-0.06}^{+0.08}}}          
\newcommand{\hatcurISOlumshortxxxB}{\ensuremath{0.37}}                     
\newcommand{\hatcurISOmvxxxB}{\ensuremath{6.08\pm0.24}}                    
\newcommand{\hatcurISOvixxxB}{\ensuremath{0.922\pm0.040}}                  
\newcommand{\hatcurISOagexxxB}{\ensuremath{8.8\pm5.2}}                     
\newcommand{\hatcurISOsigmaxxxB}{\ensuremath{0.00060\pm0.00010}}           
\newcommand{\hatcurISOMJxxxB}{\ensuremath{4.48\pm0.17}}                    
\newcommand{\hatcurISOMHxxxB}{\ensuremath{4.00\pm0.15}}                    
\newcommand{\hatcurISOMKxxxB}{\ensuremath{3.92\pm0.15}}                    
\newcommand{\hatcurISOJKxxxB}{\ensuremath{0.57\pm0.03}}                    
\newcommand{\hatcurISOspecxxxB}{K1}                                        
\newcommand{\hatcurRVKxxxB}{\ensuremath{42.0\pm2.1}}                       
\newcommand{\hatcurRVkxxxB}{\ensuremath{-0.009\pm0.029}}                   
\newcommand{\hatcurRVhxxxB}{\ensuremath{-0.058\pm0.054}}                   
\newcommand{\hatcurRVtronexxxB}{\ensuremath{0.439\pm0.048}}                
\newcommand{\hatcurRVtrtwoxxxB}{\ensuremath{0.0000\pm0.0000}}                
\newcommand{\hatcurRVgammaAxxxB}{\ensuremath{-0.4\pm2.2}}            
\newcommand{\hatcurRVgammaBxxxB}{\ensuremath{42.0\pm5.9}}            
\newcommand{\hatcurRVjitterxxxB}{\ensuremath{6.7}}                         
\newcommand{\hatcurRVeccenxxxB}{\ensuremath{0.067\pm0.042}}                
\newcommand{\hatcurRVomegaxxxB}{\ensuremath{256\pm77}}                     
\newcommand{\hatcurPPixxxB}{\ensuremath{88.2\pm0.4}}                       
\newcommand{\hatcurPPgxxxB}{\ensuremath{5.7\pm0.7}}                        
\newcommand{\hatcurPPloggxxxB}{\ensuremath{2.75\pm0.05}}                   
\newcommand{\hatcurPParxxxB}{\ensuremath{12.24\pm0.67}}                    
\newcommand{\hatcurPParelxxxB}{\ensuremath{0.0466\pm0.0008}}               
\newcommand{\hatcurPPrhoxxxB}{\ensuremath{0.25\pm0.04}}                    
\newcommand{\hatcurPPmxxxB}{\ensuremath{0.29\pm0.02}}                      
\newcommand{\hatcurPPmshortxxxB}{\ensuremath{0.29}}                        
\newcommand{\hatcurPPmlongxxxB}{\ensuremath{0.292\pm0.018}}                
\newcommand{\hatcurPPmexxxB}{\ensuremath{92.7\pm5.6}}                      
\newcommand{\hatcurPPmeshortxxxB}{\ensuremath{92.7}}                       
\newcommand{\hatcurPPmelongxxxB}{\ensuremath{92.7\pm5.6}}                
\newcommand{\hatcurPPrxxxB}{\ensuremath{1.13\pm0.07}}                      
\newcommand{\hatcurPPrshortxxxB}{\ensuremath{1.13}}                        
\newcommand{\hatcurPPrlongxxxB}{\ensuremath{1.132\pm0.072}}                
\newcommand{\hatcurPPrexxxB}{\ensuremath{12.7\pm0.8}}                      
\newcommand{\hatcurPPreshortxxxB}{\ensuremath{12.7}}                       
\newcommand{\hatcurPPrelongxxxB}{\ensuremath{12.69\pm0.81}}                
\newcommand{\hatcurPPmrcorrxxxB}{\ensuremath{0.35}}                        
\newcommand{\hatcurPPteffxxxB}{\ensuremath{1010\pm42}}                     
\newcommand{\hatcurPPthetaxxxB}{\ensuremath{0.028\pm0.002}}                
\newcommand{\hatcurPPfluxperixxxB}{\ensuremath{2.72\pm0.40}}              
\newcommand{\hatcurPPfluxperidimxxxB}{\ensuremath{8}}                      
\newcommand{\hatcurPPfluxapxxxB}{\ensuremath{2.06\pm0.46}}                
\newcommand{\hatcurPPfluxapdimxxxB}{\ensuremath{8}}                        
\newcommand{\hatcurPPfluxavgxxxB}{\ensuremath{2.35\pm0.41}}               
\newcommand{\hatcurPPfluxavgdimxxxB}{\ensuremath{8}}                       
\newcommand{\hatcurXsecphasexxxB}{\ensuremath{0.494\pm0.019}}            
\newcommand{\hatcurXsecondaryxxxB}{\ensuremath{2455093.515\pm0.074}}       
\newcommand{\hatcurXsecdurxxxB}{\ensuremath{0.107\pm0.010}}              
\newcommand{\hatcurXsecingdurxxxB}{\ensuremath{0.0149\pm0.0020}}           
\newcommand{\hatcurPPphiconjxxxB}{\ensuremath{-0.371_{-0.080}^{+0.644}}} 
\newcommand{\hatcurPPperixxxB}{\ensuremath{2455093.0\pm1.8}}             
\newcommand{\hatcurPPaequivxxxB}{\ensuremath{0.0763\pm0.0065}}             
\newcommand{\hatcurXdistxxxB}{\ensuremath{215\pm15}}                       
\newcommand{\hatcurCCpmraxxxB}{\ensuremath{-29.7\pm3.8}}                   
\newcommand{\hatcurCCpmdecxxxB}{\ensuremath{-36.4\pm4.1}}                  
\newcommand{\hatcurCCpmxxxB}{\ensuremath{47.0\pm5.6}}               

\newcommand{\hatcurCCbbHmag}[1]{\ifnum#1=18 %
\hatcurCCbbHmagxxxA
\else
\ifnum#1=19 %
\hatcurCCbbHmagxxxB
\else
??????\fi
\fi
}
\newcommand{\hatcurCCbbJmag}[1]{\ifnum#1=18 %
\hatcurCCbbJmagxxxA
\else
\ifnum#1=19 %
\hatcurCCbbJmagxxxB
\else
??????\fi
\fi
}
\newcommand{\hatcurCCbbKmag}[1]{\ifnum#1=18 %
\hatcurCCbbKmagxxxA
\else
\ifnum#1=19 %
\hatcurCCbbKmagxxxB
\else
??????\fi
\fi
}
\newcommand{\hatcurCCcitHmag}[1]{\ifnum#1=18 %
\hatcurCCcitHmagxxxA
\else
\ifnum#1=19 %
\hatcurCCcitHmagxxxB
\else
??????\fi
\fi
}
\newcommand{\hatcurCCcitJmag}[1]{\ifnum#1=18 %
\hatcurCCcitJmagxxxA
\else
\ifnum#1=19 %
\hatcurCCcitJmagxxxB
\else
??????\fi
\fi
}
\newcommand{\hatcurCCcitKmag}[1]{\ifnum#1=18 %
\hatcurCCcitKmagxxxA
\else
\ifnum#1=19 %
\hatcurCCcitKmagxxxB
\else
??????\fi
\fi
}
\newcommand{\hatcurCCdec}[1]{\ifnum#1=18 %
\hatcurCCdecxxxA
\else
\ifnum#1=19 %
\hatcurCCdecxxxB
\else
??????\fi
\fi
}
\newcommand{\hatcurCCesoHKmag}[1]{\ifnum#1=18 %
\hatcurCCesoHKmagxxxA
\else
\ifnum#1=19 %
\hatcurCCesoHKmagxxxB
\else
??????\fi
\fi
}
\newcommand{\hatcurCCesoHmag}[1]{\ifnum#1=18 %
\hatcurCCesoHmagxxxA
\else
\ifnum#1=19 %
\hatcurCCesoHmagxxxB
\else
??????\fi
\fi
}
\newcommand{\hatcurCCesoJHmag}[1]{\ifnum#1=18 %
\hatcurCCesoJHmagxxxA
\else
\ifnum#1=19 %
\hatcurCCesoJHmagxxxB
\else
??????\fi
\fi
}
\newcommand{\hatcurCCesoJKmag}[1]{\ifnum#1=18 %
\hatcurCCesoJKmagxxxA
\else
\ifnum#1=19 %
\hatcurCCesoJKmagxxxB
\else
??????\fi
\fi
}
\newcommand{\hatcurCCesoJmag}[1]{\ifnum#1=18 %
\hatcurCCesoJmagxxxA
\else
\ifnum#1=19 %
\hatcurCCesoJmagxxxB
\else
??????\fi
\fi
}
\newcommand{\hatcurCCesoKmag}[1]{\ifnum#1=18 %
\hatcurCCesoKmagxxxA
\else
\ifnum#1=19 %
\hatcurCCesoKmagxxxB
\else
??????\fi
\fi
}
\newcommand{\hatcurCCgsc}[1]{\ifnum#1=18 %
\hatcurCCgscxxxA
\else
\ifnum#1=19 %
\hatcurCCgscxxxB
\else
??????\fi
\fi
}
\newcommand{\hatcurCCmag}[1]{\ifnum#1=18 %
\hatcurCCmagxxxA
\else
\ifnum#1=19 %
\hatcurCCmagxxxB
\else
??????\fi
\fi
}
\newcommand{\hatcurCCpm}[1]{\ifnum#1=18 %
\hatcurCCpmxxxA
\else
\ifnum#1=19 %
\hatcurCCpmxxxB
\else
??????\fi
\fi
}
\newcommand{\hatcurCCpmdec}[1]{\ifnum#1=18 %
\hatcurCCpmdecxxxA
\else
\ifnum#1=19 %
\hatcurCCpmdecxxxB
\else
??????\fi
\fi
}
\newcommand{\hatcurCCpmra}[1]{\ifnum#1=18 %
\hatcurCCpmraxxxA
\else
\ifnum#1=19 %
\hatcurCCpmraxxxB
\else
??????\fi
\fi
}
\newcommand{\hatcurCCra}[1]{\ifnum#1=18 %
\hatcurCCraxxxA
\else
\ifnum#1=19 %
\hatcurCCraxxxB
\else
??????\fi
\fi
}
\newcommand{\hatcurCCtassmv}[1]{\ifnum#1=18 %
\hatcurCCtassmvxxxA
\else
\ifnum#1=19 %
\hatcurCCtassmvxxxB
\else
??????\fi
\fi
}
\newcommand{\hatcurCCtwomass}[1]{\ifnum#1=18 %
\hatcurCCtwomassxxxA
\else
\ifnum#1=19 %
\hatcurCCtwomassxxxB
\else
??????\fi
\fi
}
\newcommand{\hatcurCCtwomassHmag}[1]{\ifnum#1=18 %
\hatcurCCtwomassHmagxxxA
\else
\ifnum#1=19 %
\hatcurCCtwomassHmagxxxB
\else
??????\fi
\fi
}
\newcommand{\hatcurCCtwomassJmag}[1]{\ifnum#1=18 %
\hatcurCCtwomassJmagxxxA
\else
\ifnum#1=19 %
\hatcurCCtwomassJmagxxxB
\else
??????\fi
\fi
}
\newcommand{\hatcurCCtwomassKmag}[1]{\ifnum#1=18 %
\hatcurCCtwomassKmagxxxA
\else
\ifnum#1=19 %
\hatcurCCtwomassKmagxxxB
\else
??????\fi
\fi
}
\newcommand{\hatcurDSgamma}[1]{\ifnum#1=18 %
\hatcurDSgammaxxxA
\else
\ifnum#1=19 %
\hatcurDSgammaxxxB
\else
??????\fi
\fi
}
\newcommand{\hatcurDSlogg}[1]{\ifnum#1=18 %
\hatcurDSloggxxxA
\else
\ifnum#1=19 %
\hatcurDSloggxxxB
\else
??????\fi
\fi
}
\newcommand{\hatcurDSnumspec}[1]{\ifnum#1=18 %
\hatcurDSnumspecxxxA
\else
\ifnum#1=19 %
\hatcurDSnumspecxxxB
\else
??????\fi
\fi
}
\newcommand{\hatcurDSrvrms}[1]{\ifnum#1=18 %
\hatcurDSrvrmsxxxA
\else
\ifnum#1=19 %
\hatcurDSrvrmsxxxB
\else
??????\fi
\fi
}
\newcommand{\hatcurDSspan}[1]{\ifnum#1=18 %
\hatcurDSspanxxxA
\else
\ifnum#1=19 %
\hatcurDSspanxxxB
\else
??????\fi
\fi
}
\newcommand{\hatcurDSteff}[1]{\ifnum#1=18 %
\hatcurDSteffxxxA
\else
\ifnum#1=19 %
\hatcurDSteffxxxB
\else
??????\fi
\fi
}
\newcommand{\hatcurDSvsini}[1]{\ifnum#1=18 %
\hatcurDSvsinixxxA
\else
\ifnum#1=19 %
\hatcurDSvsinixxxB
\else
??????\fi
\fi
}
\newcommand{\hatcurDSzfeh}[1]{\ifnum#1=18 %
\hatcurDSzfehxxxA
\else
\ifnum#1=19 %
\hatcurDSzfehxxxB
\else
??????\fi
\fi
}
\newcommand{\hatcurfield}[1]{\ifnum#1=18 %
\hatcurfieldxxxA
\else
\ifnum#1=19 %
\hatcurfieldxxxB
\else
??????\fi
\fi
}
\newcommand{\hatcurFIESgamma}[1]{\ifnum#1=18 %
\hatcurFIESgammaxxxA
\else
\ifnum#1=19 %
\hatcurFIESgammaxxxB
\else
??????\fi
\fi
}
\newcommand{\hatcurFIESlogg}[1]{\ifnum#1=18 %
\hatcurFIESloggxxxA
\else
\ifnum#1=19 %
\hatcurFIESloggxxxB
\else
??????\fi
\fi
}
\newcommand{\hatcurFIESnumspec}[1]{\ifnum#1=18 %
\hatcurFIESnumspecxxxA
\else
\ifnum#1=19 %
\hatcurFIESnumspecxxxB
\else
??????\fi
\fi
}
\newcommand{\hatcurFIESrvrms}[1]{\ifnum#1=18 %
\hatcurFIESrvrmsxxxA
\else
\ifnum#1=19 %
\hatcurFIESrvrmsxxxB
\else
??????\fi
\fi
}
\newcommand{\hatcurFIESspan}[1]{\ifnum#1=18 %
\hatcurFIESspanxxxA
\else
\ifnum#1=19 %
\hatcurFIESspanxxxB
\else
??????\fi
\fi
}
\newcommand{\hatcurFIESteff}[1]{\ifnum#1=18 %
\hatcurFIESteffxxxA
\else
\ifnum#1=19 %
\hatcurFIESteffxxxB
\else
??????\fi
\fi
}
\newcommand{\hatcurFIESvsini}[1]{\ifnum#1=18 %
\hatcurFIESvsinixxxA
\else
\ifnum#1=19 %
\hatcurFIESvsinixxxB
\else
??????\fi
\fi
}
\newcommand{\hatcurFIESzfeh}[1]{\ifnum#1=18 %
\hatcurFIESzfehxxxA
\else
\ifnum#1=19 %
\hatcurFIESzfehxxxB
\else
??????\fi
\fi
}
\newcommand{\hatcurhtr}[1]{\ifnum#1=18 %
\hatcurhtrxxxA
\else
\ifnum#1=19 %
\hatcurhtrxxxB
\else
??????\fi
\fi
}
\newcommand{\hatcurISOage}[1]{\ifnum#1=18 %
\hatcurISOagexxxA
\else
\ifnum#1=19 %
\hatcurISOagexxxB
\else
??????\fi
\fi
}
\newcommand{\hatcurISOJK}[1]{\ifnum#1=18 %
\hatcurISOJKxxxA
\else
\ifnum#1=19 %
\hatcurISOJKxxxB
\else
??????\fi
\fi
}
\newcommand{\hatcurISOlogg}[1]{\ifnum#1=18 %
\hatcurISOloggxxxA
\else
\ifnum#1=19 %
\hatcurISOloggxxxB
\else
??????\fi
\fi
}
\newcommand{\hatcurISOlum}[1]{\ifnum#1=18 %
\hatcurISOlumxxxA
\else
\ifnum#1=19 %
\hatcurISOlumxxxB
\else
??????\fi
\fi
}
\newcommand{\hatcurISOlumshort}[1]{\ifnum#1=18 %
\hatcurISOlumshortxxxA
\else
\ifnum#1=19 %
\hatcurISOlumshortxxxB
\else
??????\fi
\fi
}
\newcommand{\hatcurISOm}[1]{\ifnum#1=18 %
\hatcurISOmxxxA
\else
\ifnum#1=19 %
\hatcurISOmxxxB
\else
??????\fi
\fi
}
\newcommand{\hatcurISOMH}[1]{\ifnum#1=18 %
\hatcurISOMHxxxA
\else
\ifnum#1=19 %
\hatcurISOMHxxxB
\else
??????\fi
\fi
}
\newcommand{\hatcurISOMJ}[1]{\ifnum#1=18 %
\hatcurISOMJxxxA
\else
\ifnum#1=19 %
\hatcurISOMJxxxB
\else
??????\fi
\fi
}
\newcommand{\hatcurISOMK}[1]{\ifnum#1=18 %
\hatcurISOMKxxxA
\else
\ifnum#1=19 %
\hatcurISOMKxxxB
\else
??????\fi
\fi
}
\newcommand{\hatcurISOmlong}[1]{\ifnum#1=18 %
\hatcurISOmlongxxxA
\else
\ifnum#1=19 %
\hatcurISOmlongxxxB
\else
??????\fi
\fi
}
\newcommand{\hatcurISOmshort}[1]{\ifnum#1=18 %
\hatcurISOmshortxxxA
\else
\ifnum#1=19 %
\hatcurISOmshortxxxB
\else
??????\fi
\fi
}
\newcommand{\hatcurISOmv}[1]{\ifnum#1=18 %
\hatcurISOmvxxxA
\else
\ifnum#1=19 %
\hatcurISOmvxxxB
\else
??????\fi
\fi
}
\newcommand{\hatcurISOr}[1]{\ifnum#1=18 %
\hatcurISOrxxxA
\else
\ifnum#1=19 %
\hatcurISOrxxxB
\else
??????\fi
\fi
}
\newcommand{\hatcurISOrho}[1]{\ifnum#1=18 %
\hatcurISOrhoxxxA
\else
\ifnum#1=19 %
\hatcurISOrhoxxxB
\else
??????\fi
\fi
}
\newcommand{\hatcurISOrlong}[1]{\ifnum#1=18 %
\hatcurISOrlongxxxA
\else
\ifnum#1=19 %
\hatcurISOrlongxxxB
\else
??????\fi
\fi
}
\newcommand{\hatcurISOrshort}[1]{\ifnum#1=18 %
\hatcurISOrshortxxxA
\else
\ifnum#1=19 %
\hatcurISOrshortxxxB
\else
??????\fi
\fi
}
\newcommand{\hatcurISOsigma}[1]{\ifnum#1=18 %
\hatcurISOsigmaxxxA
\else
\ifnum#1=19 %
\hatcurISOsigmaxxxB
\else
??????\fi
\fi
}
\newcommand{\hatcurISOspec}[1]{\ifnum#1=18 %
\hatcurISOspecxxxA
\else
\ifnum#1=19 %
\hatcurISOspecxxxB
\else
??????\fi
\fi
}
\newcommand{\hatcurISOvi}[1]{\ifnum#1=18 %
\hatcurISOvixxxA
\else
\ifnum#1=19 %
\hatcurISOvixxxB
\else
??????\fi
\fi
}
\newcommand{\hatcurLBig}[1]{\ifnum#1=18 %
\hatcurLBigxxxA
\else
\ifnum#1=19 %
\hatcurLBigxxxB
\else
??????\fi
\fi
}
\newcommand{\hatcurLBii}[1]{\ifnum#1=18 %
\hatcurLBiixxxA
\else
\ifnum#1=19 %
\hatcurLBiixxxB
\else
??????\fi
\fi
}
\newcommand{\hatcurLBiI}[1]{\ifnum#1=18 %
\hatcurLBiIxxxA
\else
\ifnum#1=19 %
\hatcurLBiIxxxB
\else
??????\fi
\fi
}
\newcommand{\hatcurLBiig}[1]{\ifnum#1=18 %
\hatcurLBiigxxxA
\else
\ifnum#1=19 %
\hatcurLBiigxxxB
\else
??????\fi
\fi
}
\newcommand{\hatcurLBiii}[1]{\ifnum#1=18 %
\hatcurLBiiixxxA
\else
\ifnum#1=19 %
\hatcurLBiiixxxB
\else
??????\fi
\fi
}
\newcommand{\hatcurLBiiI}[1]{\ifnum#1=18 %
\hatcurLBiiIxxxA
\else
\ifnum#1=19 %
\hatcurLBiiIxxxB
\else
??????\fi
\fi
}
\newcommand{\hatcurLBiikep}[1]{\ifnum#1=18 %
\hatcurLBiikepxxxA
\else
\ifnum#1=19 %
\hatcurLBiikepxxxB
\else
??????\fi
\fi
}
\newcommand{\hatcurLBiiz}[1]{\ifnum#1=18 %
\hatcurLBiizxxxA
\else
\ifnum#1=19 %
\hatcurLBiizxxxB
\else
??????\fi
\fi
}
\newcommand{\hatcurLBikep}[1]{\ifnum#1=18 %
\hatcurLBikepxxxA
\else
\ifnum#1=19 %
\hatcurLBikepxxxB
\else
??????\fi
\fi
}
\newcommand{\hatcurLBiz}[1]{\ifnum#1=18 %
\hatcurLBizxxxA
\else
\ifnum#1=19 %
\hatcurLBizxxxB
\else
??????\fi
\fi
}
\newcommand{\hatcurLCbsq}[1]{\ifnum#1=18 %
\hatcurLCbsqxxxA
\else
\ifnum#1=19 %
\hatcurLCbsqxxxB
\else
??????\fi
\fi
}
\newcommand{\hatcurLCdip}[1]{\ifnum#1=18 %
\hatcurLCdipxxxA
\else
\ifnum#1=19 %
\hatcurLCdipxxxB
\else
??????\fi
\fi
}
\newcommand{\hatcurLCdur}[1]{\ifnum#1=18 %
\hatcurLCdurxxxA
\else
\ifnum#1=19 %
\hatcurLCdurxxxB
\else
??????\fi
\fi
}
\newcommand{\hatcurLCdurhr}[1]{\ifnum#1=18 %
\hatcurLCdurhrxxxA
\else
\ifnum#1=19 %
\hatcurLCdurhrxxxB
\else
??????\fi
\fi
}
\newcommand{\hatcurLCdurhrshort}[1]{\ifnum#1=18 %
\hatcurLCdurhrshortxxxA
\else
\ifnum#1=19 %
\hatcurLCdurhrshortxxxB
\else
??????\fi
\fi
}
\newcommand{\hatcurLCdurshort}[1]{\ifnum#1=18 %
\hatcurLCdurshortxxxA
\else
\ifnum#1=19 %
\hatcurLCdurshortxxxB
\else
??????\fi
\fi
}
\newcommand{\hatcurLChatnetm}[1]{\ifnum#1=18 %
\hatcurLChatnetmxxxA
\else
??????\fi
}
\newcommand{\hatcurLChatnetmA}[1]{\ifnum#1=19 %
\hatcurLChatnetmAxxxB
\else
??????\fi
}
\newcommand{\hatcurLChatnetmB}[1]{\ifnum#1=19 %
\hatcurLChatnetmBxxxB
\else
??????\fi
}
\newcommand{\hatcurLCiblend}[1]{\ifnum#1=18 %
\hatcurLCiblendxxxA
\else
??????\fi
}
\newcommand{\hatcurLCiblendA}[1]{\ifnum#1=19 %
\hatcurLCiblendAxxxB
\else
??????\fi
}
\newcommand{\hatcurLCiblendB}[1]{\ifnum#1=19 %
\hatcurLCiblendBxxxB
\else
??????\fi
}
\newcommand{\hatcurLCimp}[1]{\ifnum#1=18 %
\hatcurLCimpxxxA
\else
\ifnum#1=19 %
\hatcurLCimpxxxB
\else
??????\fi
\fi
}
\newcommand{\hatcurLCingdur}[1]{\ifnum#1=18 %
\hatcurLCingdurxxxA
\else
\ifnum#1=19 %
\hatcurLCingdurxxxB
\else
??????\fi
\fi
}
\newcommand{\hatcurLCP}[1]{\ifnum#1=18 %
\hatcurLCPxxxA
\else
\ifnum#1=19 %
\hatcurLCPxxxB
\else
??????\fi
\fi
}
\newcommand{\hatcurLCPprec}[1]{\ifnum#1=18 %
\hatcurLCPprecxxxA
\else
\ifnum#1=19 %
\hatcurLCPprecxxxB
\else
??????\fi
\fi
}
\newcommand{\hatcurLCPshort}[1]{\ifnum#1=18 %
\hatcurLCPshortxxxA
\else
\ifnum#1=19 %
\hatcurLCPshortxxxB
\else
??????\fi
\fi
}
\newcommand{\hatcurLCq}[1]{\ifnum#1=18 %
\hatcurLCqxxxA
\else
\ifnum#1=19 %
\hatcurLCqxxxB
\else
??????\fi
\fi
}
\newcommand{\hatcurLCqshort}[1]{\ifnum#1=18 %
\hatcurLCqshortxxxA
\else
\ifnum#1=19 %
\hatcurLCqshortxxxB
\else
??????\fi
\fi
}
\newcommand{\hatcurLCrprstar}[1]{\ifnum#1=18 %
\hatcurLCrprstarxxxA
\else
\ifnum#1=19 %
\hatcurLCrprstarxxxB
\else
??????\fi
\fi
}
\newcommand{\hatcurLCT}[1]{\ifnum#1=18 %
\hatcurLCTxxxA
\else
\ifnum#1=19 %
\hatcurLCTxxxB
\else
??????\fi
\fi
}
\newcommand{\hatcurLCTA}[1]{\ifnum#1=18 %
\hatcurLCTAxxxA
\else
\ifnum#1=19 %
\hatcurLCTAxxxB
\else
??????\fi
\fi
}
\newcommand{\hatcurLCTB}[1]{\ifnum#1=18 %
\hatcurLCTBxxxA
\else
\ifnum#1=19 %
\hatcurLCTBxxxB
\else
??????\fi
\fi
}
\newcommand{\hatcurLCzeta}[1]{\ifnum#1=18 %
\hatcurLCzetaxxxA
\else
\ifnum#1=19 %
\hatcurLCzetaxxxB
\else
??????\fi
\fi
}
\newcommand{\hatcurPPaequiv}[1]{\ifnum#1=18 %
\hatcurPPaequivxxxA
\else
\ifnum#1=19 %
\hatcurPPaequivxxxB
\else
??????\fi
\fi
}
\newcommand{\hatcurPPar}[1]{\ifnum#1=18 %
\hatcurPParxxxA
\else
\ifnum#1=19 %
\hatcurPParxxxB
\else
??????\fi
\fi
}
\newcommand{\hatcurPParel}[1]{\ifnum#1=18 %
\hatcurPParelxxxA
\else
\ifnum#1=19 %
\hatcurPParelxxxB
\else
??????\fi
\fi
}
\newcommand{\hatcurPPfluxap}[1]{\ifnum#1=18 %
\hatcurPPfluxapxxxA
\else
\ifnum#1=19 %
\hatcurPPfluxapxxxB
\else
??????\fi
\fi
}
\newcommand{\hatcurPPfluxapdim}[1]{\ifnum#1=18 %
\hatcurPPfluxapdimxxxA
\else
\ifnum#1=19 %
\hatcurPPfluxapdimxxxB
\else
??????\fi
\fi
}
\newcommand{\hatcurPPfluxavg}[1]{\ifnum#1=18 %
\hatcurPPfluxavgxxxA
\else
\ifnum#1=19 %
\hatcurPPfluxavgxxxB
\else
??????\fi
\fi
}
\newcommand{\hatcurPPfluxavgdim}[1]{\ifnum#1=18 %
\hatcurPPfluxavgdimxxxA
\else
\ifnum#1=19 %
\hatcurPPfluxavgdimxxxB
\else
??????\fi
\fi
}
\newcommand{\hatcurPPfluxperi}[1]{\ifnum#1=18 %
\hatcurPPfluxperixxxA
\else
\ifnum#1=19 %
\hatcurPPfluxperixxxB
\else
??????\fi
\fi
}
\newcommand{\hatcurPPfluxperidim}[1]{\ifnum#1=18 %
\hatcurPPfluxperidimxxxA
\else
\ifnum#1=19 %
\hatcurPPfluxperidimxxxB
\else
??????\fi
\fi
}
\newcommand{\hatcurPPg}[1]{\ifnum#1=18 %
\hatcurPPgxxxA
\else
\ifnum#1=19 %
\hatcurPPgxxxB
\else
??????\fi
\fi
}
\newcommand{\hatcurPPi}[1]{\ifnum#1=18 %
\hatcurPPixxxA
\else
\ifnum#1=19 %
\hatcurPPixxxB
\else
??????\fi
\fi
}
\newcommand{\hatcurPPlogg}[1]{\ifnum#1=18 %
\hatcurPPloggxxxA
\else
\ifnum#1=19 %
\hatcurPPloggxxxB
\else
??????\fi
\fi
}
\newcommand{\hatcurPPm}[1]{\ifnum#1=18 %
\hatcurPPmxxxA
\else
\ifnum#1=19 %
\hatcurPPmxxxB
\else
??????\fi
\fi
}
\newcommand{\hatcurPPme}[1]{\ifnum#1=18 %
\hatcurPPmexxxA
\else
\ifnum#1=19 %
\hatcurPPmexxxB
\else
??????\fi
\fi
}
\newcommand{\hatcurPPmelong}[1]{\ifnum#1=18 %
\hatcurPPmelongxxxA
\else
\ifnum#1=19 %
\hatcurPPmelongxxxB
\else
??????\fi
\fi
}
\newcommand{\hatcurPPmeshort}[1]{\ifnum#1=18 %
\hatcurPPmeshortxxxA
\else
\ifnum#1=19 %
\hatcurPPmeshortxxxB
\else
??????\fi
\fi
}
\newcommand{\hatcurPPmlong}[1]{\ifnum#1=18 %
\hatcurPPmlongxxxA
\else
\ifnum#1=19 %
\hatcurPPmlongxxxB
\else
??????\fi
\fi
}
\newcommand{\hatcurPPmrcorr}[1]{\ifnum#1=18 %
\hatcurPPmrcorrxxxA
\else
\ifnum#1=19 %
\hatcurPPmrcorrxxxB
\else
??????\fi
\fi
}
\newcommand{\hatcurPPmshort}[1]{\ifnum#1=18 %
\hatcurPPmshortxxxA
\else
\ifnum#1=19 %
\hatcurPPmshortxxxB
\else
??????\fi
\fi
}
\newcommand{\hatcurPPperi}[1]{\ifnum#1=18 %
\hatcurPPperixxxA
\else
\ifnum#1=19 %
\hatcurPPperixxxB
\else
??????\fi
\fi
}
\newcommand{\hatcurPPphiconj}[1]{\ifnum#1=18 %
\hatcurPPphiconjxxxA
\else
\ifnum#1=19 %
\hatcurPPphiconjxxxB
\else
??????\fi
\fi
}
\newcommand{\hatcurPPr}[1]{\ifnum#1=18 %
\hatcurPPrxxxA
\else
\ifnum#1=19 %
\hatcurPPrxxxB
\else
??????\fi
\fi
}
\newcommand{\hatcurPPre}[1]{\ifnum#1=18 %
\hatcurPPrexxxA
\else
\ifnum#1=19 %
\hatcurPPrexxxB
\else
??????\fi
\fi
}
\newcommand{\hatcurPPrelong}[1]{\ifnum#1=18 %
\hatcurPPrelongxxxA
\else
\ifnum#1=19 %
\hatcurPPrelongxxxB
\else
??????\fi
\fi
}
\newcommand{\hatcurPPreshort}[1]{\ifnum#1=18 %
\hatcurPPreshortxxxA
\else
\ifnum#1=19 %
\hatcurPPreshortxxxB
\else
??????\fi
\fi
}
\newcommand{\hatcurPPrho}[1]{\ifnum#1=18 %
\hatcurPPrhoxxxA
\else
\ifnum#1=19 %
\hatcurPPrhoxxxB
\else
??????\fi
\fi
}
\newcommand{\hatcurPPrlong}[1]{\ifnum#1=18 %
\hatcurPPrlongxxxA
\else
\ifnum#1=19 %
\hatcurPPrlongxxxB
\else
??????\fi
\fi
}
\newcommand{\hatcurPPrshort}[1]{\ifnum#1=18 %
\hatcurPPrshortxxxA
\else
\ifnum#1=19 %
\hatcurPPrshortxxxB
\else
??????\fi
\fi
}
\newcommand{\hatcurPPteff}[1]{\ifnum#1=18 %
\hatcurPPteffxxxA
\else
\ifnum#1=19 %
\hatcurPPteffxxxB
\else
??????\fi
\fi
}
\newcommand{\hatcurPPtheta}[1]{\ifnum#1=18 %
\hatcurPPthetaxxxA
\else
\ifnum#1=19 %
\hatcurPPthetaxxxB
\else
??????\fi
\fi
}
\newcommand{\hatcurRVeccen}[1]{\ifnum#1=18 %
\hatcurRVeccenxxxA
\else
\ifnum#1=19 %
\hatcurRVeccenxxxB
\else
??????\fi
\fi
}
\newcommand{\hatcurRVgamma}[1]{\ifnum#1=18 %
\hatcurRVgammaxxxA
\else
??????\fi
}
\newcommand{\hatcurRVgammaA}[1]{\ifnum#1=19 %
\hatcurRVgammaAxxxB
\else
??????\fi
}
\newcommand{\hatcurRVgammaB}[1]{\ifnum#1=19 %
\hatcurRVgammaBxxxB
\else
??????\fi
}
\newcommand{\hatcurRVh}[1]{\ifnum#1=18 %
\hatcurRVhxxxA
\else
\ifnum#1=19 %
\hatcurRVhxxxB
\else
??????\fi
\fi
}
\newcommand{\hatcurRVjitter}[1]{\ifnum#1=18 %
\hatcurRVjitterxxxA
\else
\ifnum#1=19 %
\hatcurRVjitterxxxB
\else
??????\fi
\fi
}
\newcommand{\hatcurRVk}[1]{\ifnum#1=18 %
\hatcurRVkxxxA
\else
\ifnum#1=19 %
\hatcurRVkxxxB
\else
??????\fi
\fi
}
\newcommand{\hatcurRVK}[1]{\ifnum#1=18 %
\hatcurRVKxxxA
\else
\ifnum#1=19 %
\hatcurRVKxxxB
\else
??????\fi
\fi
}
\newcommand{\hatcurRVomega}[1]{\ifnum#1=18 %
\hatcurRVomegaxxxA
\else
\ifnum#1=19 %
\hatcurRVomegaxxxB
\else
??????\fi
\fi
}
\newcommand{\hatcurRVtrone}[1]{\ifnum#1=19 %
\hatcurRVtronexxxB
\else
??????\fi
}
\newcommand{\hatcurRVtrtwo}[1]{\ifnum#1=19 %
\hatcurRVtrtwoxxxB
\else
??????\fi
}
\newcommand{\hatcurSMEiilogg}[1]{\ifnum#1=18 %
\hatcurSMEiiloggxxxA
\else
\ifnum#1=19 %
\hatcurSMEiiloggxxxB
\else
??????\fi
\fi
}
\newcommand{\hatcurSMEiiteff}[1]{\ifnum#1=18 %
\hatcurSMEiiteffxxxA
\else
\ifnum#1=19 %
\hatcurSMEiiteffxxxB
\else
??????\fi
\fi
}
\newcommand{\hatcurSMEiivmac}[1]{\ifnum#1=18 %
\hatcurSMEiivmacxxxA
\else
\ifnum#1=19 %
\hatcurSMEiivmacxxxB
\else
??????\fi
\fi
}
\newcommand{\hatcurSMEiivmic}[1]{\ifnum#1=18 %
\hatcurSMEiivmicxxxA
\else
\ifnum#1=19 %
\hatcurSMEiivmicxxxB
\else
??????\fi
\fi
}
\newcommand{\hatcurSMEiivsin}[1]{\ifnum#1=18 %
\hatcurSMEiivsinxxxA
\else
\ifnum#1=19 %
\hatcurSMEiivsinxxxB
\else
??????\fi
\fi
}
\newcommand{\hatcurSMEiizfeh}[1]{\ifnum#1=18 %
\hatcurSMEiizfehxxxA
\else
\ifnum#1=19 %
\hatcurSMEiizfehxxxB
\else
??????\fi
\fi
}
\newcommand{\hatcurSMEiizfehshort}[1]{\ifnum#1=18 %
\hatcurSMEiizfehshortxxxA
\else
\ifnum#1=19 %
\hatcurSMEiizfehshortxxxB
\else
??????\fi
\fi
}
\newcommand{\hatcurSMEilogg}[1]{\ifnum#1=18 %
\hatcurSMEiloggxxxA
\else
\ifnum#1=19 %
\hatcurSMEiloggxxxB
\else
??????\fi
\fi
}
\newcommand{\hatcurSMEiteff}[1]{\ifnum#1=18 %
\hatcurSMEiteffxxxA
\else
\ifnum#1=19 %
\hatcurSMEiteffxxxB
\else
??????\fi
\fi
}
\newcommand{\hatcurSMEivmac}[1]{\ifnum#1=18 %
\hatcurSMEivmacxxxA
\else
\ifnum#1=19 %
\hatcurSMEivmacxxxB
\else
??????\fi
\fi
}
\newcommand{\hatcurSMEivmic}[1]{\ifnum#1=18 %
\hatcurSMEivmicxxxA
\else
\ifnum#1=19 %
\hatcurSMEivmicxxxB
\else
??????\fi
\fi
}
\newcommand{\hatcurSMEivsin}[1]{\ifnum#1=18 %
\hatcurSMEivsinxxxA
\else
\ifnum#1=19 %
\hatcurSMEivsinxxxB
\else
??????\fi
\fi
}
\newcommand{\hatcurSMEizfeh}[1]{\ifnum#1=18 %
\hatcurSMEizfehxxxA
\else
\ifnum#1=19 %
\hatcurSMEizfehxxxB
\else
??????\fi
\fi
}
\newcommand{\hatcurSMEizfehshort}[1]{\ifnum#1=18 %
\hatcurSMEizfehshortxxxA
\else
\ifnum#1=19 %
\hatcurSMEizfehshortxxxB
\else
??????\fi
\fi
}
\newcommand{\hatcurTRESgamma}[1]{\ifnum#1=18 %
\hatcurTRESgammaxxxA
\else
\ifnum#1=19 %
\hatcurTRESgammaxxxB
\else
??????\fi
\fi
}
\newcommand{\hatcurTRESlogg}[1]{\ifnum#1=18 %
\hatcurTRESloggxxxA
\else
\ifnum#1=19 %
\hatcurTRESloggxxxB
\else
??????\fi
\fi
}
\newcommand{\hatcurTRESnumspec}[1]{\ifnum#1=18 %
\hatcurTRESnumspecxxxA
\else
\ifnum#1=19 %
\hatcurTRESnumspecxxxB
\else
??????\fi
\fi
}
\newcommand{\hatcurTRESrvrms}[1]{\ifnum#1=18 %
\hatcurTRESrvrmsxxxA
\else
\ifnum#1=19 %
\hatcurTRESrvrmsxxxB
\else
??????\fi
\fi
}
\newcommand{\hatcurTRESspan}[1]{\ifnum#1=18 %
\hatcurTRESspanxxxA
\else
\ifnum#1=19 %
\hatcurTRESspanxxxB
\else
??????\fi
\fi
}
\newcommand{\hatcurTRESteff}[1]{\ifnum#1=18 %
\hatcurTRESteffxxxA
\else
\ifnum#1=19 %
\hatcurTRESteffxxxB
\else
??????\fi
\fi
}
\newcommand{\hatcurTRESvsini}[1]{\ifnum#1=18 %
\hatcurTRESvsinixxxA
\else
\ifnum#1=19 %
\hatcurTRESvsinixxxB
\else
??????\fi
\fi
}
\newcommand{\hatcurTRESzfeh}[1]{\ifnum#1=18 %
\hatcurTRESzfehxxxA
\else
\ifnum#1=19 %
\hatcurTRESzfehxxxB
\else
??????\fi
\fi
}
\newcommand{\hatcurXdist}[1]{\ifnum#1=18 %
\hatcurXdistxxxA
\else
\ifnum#1=19 %
\hatcurXdistxxxB
\else
??????\fi
\fi
}
\newcommand{\hatcurXsecdur}[1]{\ifnum#1=18 %
\hatcurXsecdurxxxA
\else
\ifnum#1=19 %
\hatcurXsecdurxxxB
\else
??????\fi
\fi
}
\newcommand{\hatcurXsecingdur}[1]{\ifnum#1=18 %
\hatcurXsecingdurxxxA
\else
\ifnum#1=19 %
\hatcurXsecingdurxxxB
\else
??????\fi
\fi
}
\newcommand{\hatcurXsecondary}[1]{\ifnum#1=18 %
\hatcurXsecondaryxxxA
\else
\ifnum#1=19 %
\hatcurXsecondaryxxxB
\else
??????\fi
\fi
}
\newcommand{\hatcurXsecphase}[1]{\ifnum#1=18 %
\hatcurXsecphasexxxA
\else
\ifnum#1=19 %
\hatcurXsecphasexxxB
\else
??????\fi
\fi
}
\newcommand{\hatcurxxxA}{HAT-P-18}
\newcommand{\hatcurbxxxA}{HAT-P-18b}
\newcommand{\hatcurcxxxA}{HAT-P-18c}

\newcommand{\hatcurplanetnumxxxA}{18}

\newcommand{\hatcurRVgammaabsxxxA}{\hatcurDSgamma{\hatcurplanetnumxxxA}}                           

\newcommand{\hatcurRVgammarelxxxA}{\hatcurRVgamma{\hatcurplanetnumxxxA}}                           

\newcommand{\hatcurCCtassvixxxA}{\ensuremath{1.18\pm0.13}}                  

\newcommand{\hatcurSMEversionxxxA}{ii}                                       

\newcommand{\hatcurisoshortxxxA}{YY}
\newcommand{\hatcurisofullxxxA}{Yonsei-Yale \citep[YY;][]{yi:2001}}
\newcommand{\hatcurisocitexxxA}{yi:2001}

\newcommand{\hatcurlumindxxxA}{\arstar}

\newcommand{\hatcurjhkfilsetxxxA}{ESO}

\newcommand{\hatcurSMEteffxxxA}{\ifthenelse{\equal{\hatcurSMEversionxxxA}{i}}{\hatcurSMEiteff{\hatcurplanetnumxxxA}}{\hatcurSMEiiteff{\hatcurplanetnumxxxA}}}
\newcommand{\hatcurSMEzfehxxxA}{\ifthenelse{\equal{\hatcurSMEversionxxxA}{i}}{\hatcurSMEizfeh{\hatcurplanetnumxxxA}}{\hatcurSMEiizfeh{\hatcurplanetnumxxxA}}}
\newcommand{\hatcurSMEzfehshortxxxA}{\ifthenelse{\equal{\hatcurSMEversionxxxA}{i}}{\hatcurSMEizfehshort{\hatcurplanetnumxxxA}}{\hatcurSMEiizfehshort{\hatcurplanetnumxxxA}}}
\newcommand{\hatcurSMEloggxxxA}{\ifthenelse{\equal{\hatcurSMEversionxxxA}{i}}{\hatcurSMEilogg{\hatcurplanetnumxxxA}}{\hatcurSMEiilogg{\hatcurplanetnumxxxA}}}
\newcommand{\hatcurSMEvsinxxxA}{\ifthenelse{\equal{\hatcurSMEversionxxxA}{i}}{\hatcurSMEivsin{\hatcurplanetnumxxxA}}{\hatcurSMEiivsin{\hatcurplanetnumxxxA}}}
\newcommand{\hatcurSMEvmacxxxA}{\ifthenelse{\equal{\hatcurSMEversionxxxA}{i}}{\hatcurSMEivmac{\hatcurplanetnumxxxA}}{\hatcurSMEiivmac{\hatcurplanetnumxxxA}}}
\newcommand{\hatcurSMEvmicxxxA}{\ifthenelse{\equal{\hatcurSMEversionxxxA}{i}}{\hatcurSMEivmic{\hatcurplanetnumxxxA}}{\hatcurSMEiivmic{\hatcurplanetnumxxxA}}}

\newcommand{\hatcurxxxB}{HAT-P-19}
\newcommand{\hatcurbxxxB}{HAT-P-19b}
\newcommand{\hatcurcxxxB}{HAT-P-19c}

\newcommand{\hatcurplanetnumxxxB}{19}

\newcommand{\hatcurRVgammaabsxxxB}{\hatcurDSgamma{\hatcurplanetnumxxxB}}                           

\newcommand{\hatcurRVgammarelxxxB}{\hatcurRVgamma{\hatcurplanetnumxxxB}}                           

\newcommand{\hatcurCCtassvixxxB}{\ensuremath{1.04\pm0.13}}                  

\newcommand{\hatcurSMEversionxxxB}{ii}                                       

\newcommand{\hatcurisoshortxxxB}{YY}
\newcommand{\hatcurisofullxxxB}{Yonsei-Yale (YY)}
\newcommand{\hatcurisocitexxxB}{yi:2001}

\newcommand{\hatcurlumindxxxB}{\arstar}

\newcommand{\hatcurjhkfilsetxxxB}{ESO}

\newcommand{\hatcurSMEteffxxxB}{\ifthenelse{\equal{\hatcurSMEversionxxxB}{i}}{\hatcurSMEiteff{\hatcurplanetnumxxxB}}{\hatcurSMEiiteff{\hatcurplanetnumxxxB}}}
\newcommand{\hatcurSMEzfehxxxB}{\ifthenelse{\equal{\hatcurSMEversionxxxB}{i}}{\hatcurSMEizfeh{\hatcurplanetnumxxxB}}{\hatcurSMEiizfeh{\hatcurplanetnumxxxB}}}
\newcommand{\hatcurSMEzfehshortxxxB}{\ifthenelse{\equal{\hatcurSMEversionxxxB}{i}}{\hatcurSMEizfehshort{\hatcurplanetnumxxxB}}{\hatcurSMEiizfehshort{\hatcurplanetnumxxxB}}}
\newcommand{\hatcurSMEloggxxxB}{\ifthenelse{\equal{\hatcurSMEversionxxxB}{i}}{\hatcurSMEilogg{\hatcurplanetnumxxxB}}{\hatcurSMEiilogg{\hatcurplanetnumxxxB}}}
\newcommand{\hatcurSMEvsinxxxB}{\ifthenelse{\equal{\hatcurSMEversionxxxB}{i}}{\hatcurSMEivsin{\hatcurplanetnumxxxB}}{\hatcurSMEiivsin{\hatcurplanetnumxxxB}}}
\newcommand{\hatcurSMEvmacxxxB}{\ifthenelse{\equal{\hatcurSMEversionxxxB}{i}}{\hatcurSMEivmac{\hatcurplanetnumxxxB}}{\hatcurSMEiivmac{\hatcurplanetnumxxxB}}}
\newcommand{\hatcurSMEvmicxxxB}{\ifthenelse{\equal{\hatcurSMEversionxxxB}{i}}{\hatcurSMEivmic{\hatcurplanetnumxxxB}}{\hatcurSMEiivmic{\hatcurplanetnumxxxB}}}

\newcommand{\hatcur}[1]{\ifnum#1=18 %
\hatcurxxxA
\else
\ifnum#1=19 %
\hatcurxxxB
\else
??????\fi
\fi
}
\newcommand{\hatcurb}[1]{\ifnum#1=18 %
\hatcurbxxxA
\else
\ifnum#1=19 %
\hatcurbxxxB
\else
??????\fi
\fi
}
\newcommand{\hatcurc}[1]{\ifnum#1=18 %
\hatcurcxxxA
\else
\ifnum#1=19 %
\hatcurcxxxB
\else
??????\fi
\fi
}
\newcommand{\hatcurCCtassvi}[1]{\ifnum#1=18 %
\hatcurCCtassvixxxA
\else
\ifnum#1=19 %
\hatcurCCtassvixxxB
\else
??????\fi
\fi
}
\newcommand{\hatcurisocite}[1]{\ifnum#1=18 %
\hatcurisocitexxxA
\else
\ifnum#1=19 %
\hatcurisocitexxxB
\else
??????\fi
\fi
}
\newcommand{\hatcurisofull}[1]{\ifnum#1=18 %
\hatcurisofullxxxA
\else
\ifnum#1=19 %
\hatcurisofullxxxB
\else
??????\fi
\fi
}
\newcommand{\hatcurisoshort}[1]{\ifnum#1=18 %
\hatcurisoshortxxxA
\else
\ifnum#1=19 %
\hatcurisoshortxxxB
\else
??????\fi
\fi
}
\newcommand{\hatcurjhkfilset}[1]{\ifnum#1=18 %
\hatcurjhkfilsetxxxA
\else
\ifnum#1=19 %
\hatcurjhkfilsetxxxB
\else
??????\fi
\fi
}
\newcommand{\hatcurlumind}[1]{\ifnum#1=18 %
\hatcurlumindxxxA
\else
\ifnum#1=19 %
\hatcurlumindxxxB
\else
??????\fi
\fi
}
\newcommand{\hatcurplanetnum}[1]{\ifnum#1=18 %
\hatcurplanetnumxxxA
\else
\ifnum#1=19 %
\hatcurplanetnumxxxB
\else
??????\fi
\fi
}
\newcommand{\hatcurRVgammaabs}[1]{\ifnum#1=18 %
\hatcurRVgammaabsxxxA
\else
\ifnum#1=19 %
\hatcurRVgammaabsxxxB
\else
??????\fi
\fi
}
\newcommand{\hatcurRVgammarel}[1]{\ifnum#1=18 %
\hatcurRVgammarelxxxA
\else
\ifnum#1=19 %
\hatcurRVgammarelxxxB
\else
??????\fi
\fi
}
\newcommand{\hatcurSMElogg}[1]{\ifnum#1=18 %
\hatcurSMEloggxxxA
\else
\ifnum#1=19 %
\hatcurSMEloggxxxB
\else
??????\fi
\fi
}
\newcommand{\hatcurSMEteff}[1]{\ifnum#1=18 %
\hatcurSMEteffxxxA
\else
\ifnum#1=19 %
\hatcurSMEteffxxxB
\else
??????\fi
\fi
}
\newcommand{\hatcurSMEversion}[1]{\ifnum#1=18 %
\hatcurSMEversionxxxA
\else
\ifnum#1=19 %
\hatcurSMEversionxxxB
\else
??????\fi
\fi
}
\newcommand{\hatcurSMEvmac}[1]{\ifnum#1=18 %
\hatcurSMEvmacxxxA
\else
\ifnum#1=19 %
\hatcurSMEvmacxxxB
\else
??????\fi
\fi
}
\newcommand{\hatcurSMEvmic}[1]{\ifnum#1=18 %
\hatcurSMEvmicxxxA
\else
\ifnum#1=19 %
\hatcurSMEvmicxxxB
\else
??????\fi
\fi
}
\newcommand{\hatcurSMEvsin}[1]{\ifnum#1=18 %
\hatcurSMEvsinxxxA
\else
\ifnum#1=19 %
\hatcurSMEvsinxxxB
\else
??????\fi
\fi
}
\newcommand{\hatcurSMEzfeh}[1]{\ifnum#1=18 %
\hatcurSMEzfehxxxA
\else
\ifnum#1=19 %
\hatcurSMEzfehxxxB
\else
??????\fi
\fi
}
\newcommand{\hatcurSMEzfehshort}[1]{\ifnum#1=18 %
\hatcurSMEzfehshortxxxA
\else
\ifnum#1=19 %
\hatcurSMEzfehshortxxxB
\else
??????\fi
\fi
}

\newcounter{planetcounter}

\newboolean{emulateapj}
\setboolean{emulateapj}{true}

\newboolean{rvtablelong}
\setboolean{rvtablelong}{true}

\newboolean{astroph}
\setboolean{astroph}{true}

\shortauthors{Hartman et al.}
\shorttitle{
\setcounter{planetcounter}{1}
\loopand\hatcur{18}\lowercase{b}\loopcommanospace
\setcounter{planetcounter}{2}
\loopand\hatcur{19}\lowercase{b}\loopcommanospace
}
\ifthenelse{\boolean{emulateapj}}{
    \newcommand{\titledag}{$\dagger$}
}{
    \newcommand{\titledag}{\dagger}
}

\begin{document}

\title{
\hatcur{18}\lowercase{b} and \hatcur{19}\lowercase{b}: Two Low-Density
Saturn-Mass Planets\\
Transiting Metal-Rich K Stars\altaffilmark{\titledag}
}

\author{
   J.~D.~Hartman\altaffilmark{1},
   G.~\'A.~Bakos\altaffilmark{1,2},
   B.~Sato\altaffilmark{3},
   G.~Torres\altaffilmark{1},
   R.~W.~Noyes\altaffilmark{1},
   D.~W.~Latham\altaffilmark{1},
   G.~Kov\'acs\altaffilmark{4},
   D.~A.~Fischer\altaffilmark{5},
   A.~W.~Howard\altaffilmark{6},
   J.~A.~Johnson\altaffilmark{7},
   G.~W.~Marcy\altaffilmark{6},
   L.~A.~Buchhave\altaffilmark{1,8},
   G.~F\"uresz\altaffilmark{1},
   G.~Perumpilly\altaffilmark{1,9},
   B.~B\'eky\altaffilmark{1},
   R.~P.~Stefanik\altaffilmark{1},
   D.~D.~Sasselov\altaffilmark{1},
   G.~A.~Esquerdo\altaffilmark{1},
   M.~Everett\altaffilmark{1},
   Z.~Csubry\altaffilmark{1},
   J.~L\'az\'ar\altaffilmark{10},
   I.~Papp\altaffilmark{10},
   P.~S\'ari\altaffilmark{10}
}
\altaffiltext{1}{Harvard-Smithsonian Center for Astrophysics,
    Cambridge, MA; email: gbakos@cfa.harvard.edu}

\altaffiltext{2}{NSF Fellow}

\altaffiltext{3}{Global Edge Institute, Tokyo Institute of Technology, Tokyo, Japan}

\altaffiltext{4}{Konkoly Observatory, Budapest, Hungary}

\altaffiltext{5}{Department of Physics and Astronomy, San Francisco
    State University, San Francisco, CA}

\altaffiltext{6}{Department of Astronomy, University of California,
    Berkeley, CA}

\altaffiltext{7}{Department of Astrophysics, and NASA Exoplanet Science Institute, California Institute of Technology, Pasadena, CA}

\altaffiltext{8}{Niels Bohr Institute, Copenhagen University, DK-2100 Copenhagen, Denmark}

\altaffiltext{9}{Department of Physics, University of South Dakota, Vermillion, South Dakota}

\altaffiltext{10}{Hungarian Astronomical Association, Budapest, 
    Hungary}

\altaffiltext{$\dagger$}{
  Based in part on observations obtained at the W.~M.~Keck
  Observatory, which is operated by the University of California and
  the California Institute of Technology. Keck time has been granted
  by NOAO (A146Hr, A201Hr, and A264Hr), NASA (N018Hr, N049Hr, N128Hr,
  and N167Hr), and by the NOAO Keck-Gemini time exchange program
  (G329Hr). Based in part on data collected at Subaru Telescope, which
  is operated by the National Astronomical Observatory of Japan. Based
  in part on observations made with the Nordic Optical Telescope,
  operated on the island of La Palma jointly by Denmark, Finland,
  Iceland, Norway, and Sweden, in the Spanish Observatorio del Roque
  de los Muchachos of the Instituto de Astrofisica de Canarias.
}


\begin{abstract}

\setcounter{footnote}{10}
We report the discovery of two new transiting extrasolar planets.
\hatcurb{18} orbits the
$V=$\hatcurCCtassmv{18}\ \hatcurISOspec{18}\ dwarf
star \hatcurCCgsc{18}, with a period
$P=\hatcurLCP{18}\,d$, transit epoch $T_c =
\hatcurLCT{18}$
(BJD), and transit duration
\hatcurLCdur{18}\,d. The host star has a mass of
\hatcurISOm{18}\,\msun, radius of
\hatcurISOr{18}\,\rsun, effective temperature
\hatcurSMEteff{18}\,K, and metallicity $\feh =
\hatcurSMEzfeh{18}$. The planetary companion has a mass of
\hatcurPPmlong{18}\,\mjup, and radius of
\hatcurPPrlong{18}\,\rjup\ yielding a mean density of
\hatcurPPrho{18}\,\gcmc.
\hatcurb{19} orbits the
$V=$\hatcurCCtassmv{19}\ \hatcurISOspec{19}\ dwarf star
\hatcurCCgsc{19}, with a period $P=\hatcurLCP{19}\,d$, transit epoch
$T_c = \hatcurLCT{19}$ (BJD), and transit duration
\hatcurLCdur{19}\,d. The host star has a mass of
\hatcurISOm{19}\,\msun, radius of \hatcurISOr{19}\,\rsun, effective
temperature \hatcurSMEteff{19}\,K, and metallicity $\feh =
\hatcurSMEzfeh{19}$. The planetary companion has a mass of
\hatcurPPmlong{19}\,\mjup, and radius of
\hatcurPPrlong{19}\,\rjup\ yielding a mean density of
\hatcurPPrho{19}\,\gcmc. The radial velocity residuals for \hatcur{19}
exhibit a linear trend in time, which indicates the presence of a
third body in the system. Comparing these observations with
theoretical models, we find that \hatcurb{18} and \hatcurb{19} are
each consistent with a hydrogen-helium dominated gas giant planet with
negligible core mass. \hatcurb{18} and \hatcurb{19} join HAT-P-12b and
WASP-21b in an emerging group of low-density Saturn-mass planets, with
negligible inferred core masses. However, unlike HAT-P-12b and
WASP-21b, both \hatcurb{18} and \hatcurb{19} orbit stars with
super-solar metallicity. This calls into question the heretofore
suggestive correlation between the inferred core mass and host star
metallicity for Saturn-mass planets.
\setcounter{footnote}{0}
\end{abstract}

\keywords{
    planetary systems ---
    stars: individual (
\setcounter{planetcounter}{1}
\hatcur{18},
\hatcurCCgsc{18}\loopcommanoperiod
\setcounter{planetcounter}{2}
\hatcur{19},
\hatcurCCgsc{19}\loopcommanoperiod
) 
    techniques: spectroscopic, photometric
}


\section{Introduction}
\label{sec:introduction}

Extrasolar planets which transit their host stars (Transiting
Extrasolar Planets, or TEPs) provide a unique opportunity to determine
the bulk physical properties (mass, radius and average density) of
planetary bodies outside the Solar System
\citep[e.g.][]{dc:2009}. From the more than 90 such planets that have
been announced to date\footnote{e.g. see http://exoplanet.eu}, it has
become apparent that gas giant planets more massive than $0.4\mjup$
exhibit a wide range of radii (from $0.885\rjup$ for CoRoT-13b,
\citealp{cabrera:2010}, to $1.79\rjup$ for WASP-12b,
\citealp{hebb:2009}). Below this mass fewer planets are known; however
the seven known TEPs with masses similar to Saturn ($0.15\mjup < M <
0.4\mjup$; the mass of Saturn is $0.299$\,\mjup,
\citealp{standish:1995}) also appear to have diverse bulk
properties. Two of these TEPs have densities much less than that of
Saturn (HAT-P-12b and WASP-21b both have $\rho \sim 0.3$\,\gcmc, while
Saturn has $\rho \sim 0.7$\,\gcmc; \citealp{hartman:2009};
\citealp{bouchy:2010}), three have densities that are somewhat less
than that of Saturn (Kepler-9b and Kepler-9c have densities of $\rho
\sim 0.5$\,\gcmc\ and $\rho \sim 0.4$\,\gcmc\ respectively,
\citealp{holman:2010}; and WASP-29b has $\rho \sim 0.65$\,\gcmc,
\citealp{hellier:2010}), and two have densities that are greater than
that of Saturn (HD~149026b has $\rho \sim 0.85$\,\gcmc,
\citealp{sato:2005,carter:2009}; and CoRoT-8b has $\rho \sim
1.6$\,\gcmc, \citealp{borde:2010}). The inferred core masses of these
planets also differ dramatically, with the two low-density planets
having negligible cores of $M_{C} \la 10\mearth$, the three
intermediate-density planets having cores that are perhaps several
tens of Earth masses, and the two high-density planets having cores
that represent a substantial fraction of their respective masses. The
inferred core masses of planets in this mass range appear to correlate
with the metallicity of the host star. The two low density planets
orbit stars with sub-solar metallicity ([Fe/H]$=-0.29$ for HAT-P-12,
and [Fe/H]$=-0.4$ for WASP-21), while the five higher density planets
orbit stars with super-solar metallicity ([Fe/H]$=0.3$ for CoRoT-8,
[Fe/H]$=0.36$ for HD~149026, [Fe/H]$=0.11$ for WASP-29, and
[Fe/H]$=0.12$ for Kepler-9). This has been taken as suggestive
evidence for the core-accretion scenario for planet formation
\citep{alibert:2005,guillot:2006,hartman:2009,bouchy:2010}.

In this work we present the discovery of two new low-density planets
with masses comparable to that of Saturn. The new planets \hatcurb{18}
and \hatcurb{19} have masses that are very similar to HAT-P-12b and
WASP-21b respectively, and have densities that are slightly less than
each of these planets. However, both new planets orbit stars with
super-solar metallicity, casting doubt on the correlation between
planetary core mass and stellar metallicity for Saturn-mass planets.

The planets presented in this paper were discovered by the
Hungarian-made Automated Telescope Network
\citep[HATNet;][]{bakos:2004} survey, which has been one of the main
contributors to the discovery of TEPs.  In operation since 2003, it
has now covered approximately 14\% of the sky, searching for TEPs
around bright stars ($8\lesssim I \lesssim 14$).  HATNet operates six
wide-field instruments: four at the Fred Lawrence Whipple Observatory
(FLWO) in Arizona, and two on the roof of the hangar servicing the
Smithsonian Astrophysical Observatory's Submillimeter Array, in
Hawaii.  Since 2006, HATNet has found seventeen TEPs.  In this work we
report our eighteenth and nineteenth discoveries, around the
relatively bright stars also known as
\setcounter{planetcounter}{1}
\loopand\hatcurCCgsc{18}\loopcomma
\setcounter{planetcounter}{2}
\loopand\hatcurCCgsc{19}\loopcomma

The layout of the paper is as follows. In \refsecl{obs} we report the
detections of the photometric signals and the follow-up spectroscopic
and photometric observations for each of the planets.  In
\refsecl{analysis} we describe the analysis of the data, beginning
with the determination of the stellar parameters, continuing with a
discussion of the methods used to rule out nonplanetary, false
positive scenarios which could mimic the photometric and spectroscopic
observations, and finishing with a description of our global modeling
of the photometry and radial velocities (RVs).  Our findings are discussed
in \refsecl{discussion}.


\section{Observations}
\label{sec:obs}

\subsection{Photometric detection}
\label{sec:detection}

\begin{figure}[!ht]
\plotone{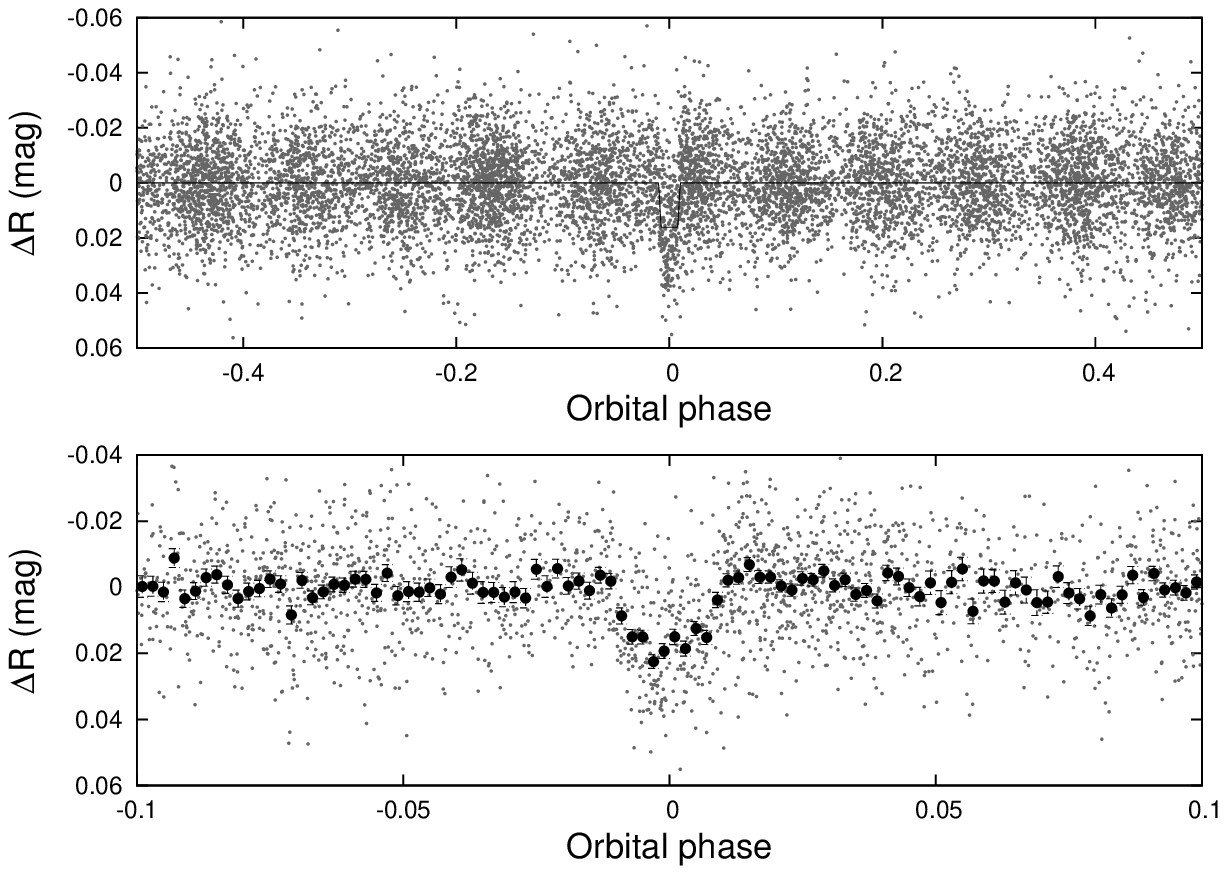}
\caption[]{
    Unbinned \lc{} of \hatcur{18} including all 10,000 instrumental
    \band{I} 5.5 minute cadence measurements obtained with the HAT-6
    and HAT-9 telescopes of HATNet (see \reftabl{photobs}), and folded
    with the period $P = \hatcurLCPprec{18}$\,days resulting from the
    global fit described in \refsecl{analysis}. The solid line shows
    a simplified transit model fit to the light curve
    (\refsecl{globmod}). The bold points in the lower panel show the
    \lc{} binned in phase with a bin-size of 0.002.
\label{fig:hatnet18}}
\end{figure}

\begin{figure}[!ht]
\plotone{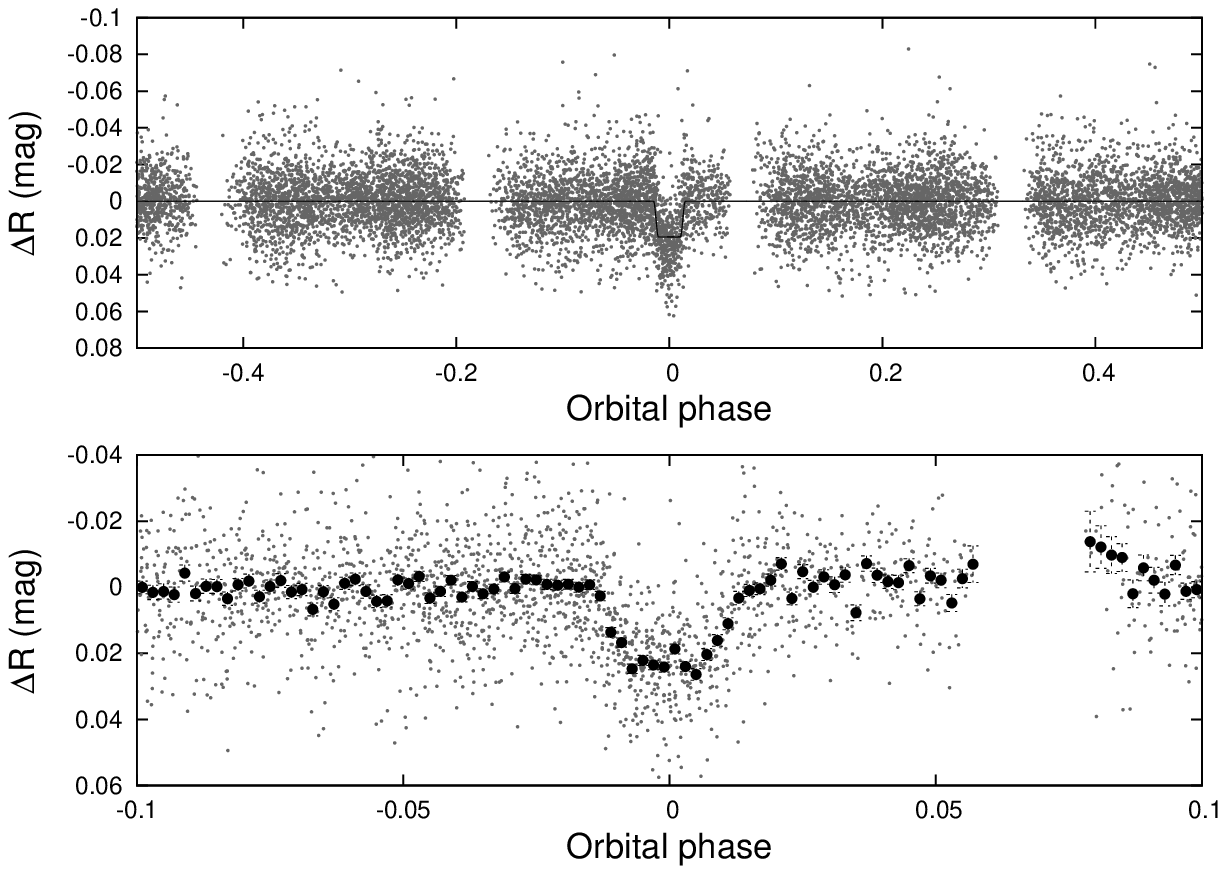}
\caption[]{
    Unbinned \lc{} of \hatcur{19} including all 10,000 instrumental
    $R$-band 5.5 minute cadence measurements obtained with the HAT-6,
    HAT-7, HAT-8 and HAT-9 telescopes of HATNet (see
    \reftabl{photobs}), and folded with the period $P =
    \hatcurLCPprec{19}$\,days resulting from the global fit described
    in \refsecl{analysis}. The solid line shows a simplified transit
    model fit to the light curve (\refsecl{globmod}). The bold points
    in the lower panel show the \lc{} binned in phase with a bin-size
    of 0.002.
\label{fig:hatnet19}}
\end{figure}

\reftabl{photobs} summarizes the HATNet discovery observations of each
new planetary system. The calibration of the HATNet frames was carried
out using standard photometric procedures.  The calibrated images were
then subjected to star detection and astrometric determination, as
described in \cite{pal:2006}.  Aperture photometry was performed on
each image at the stellar centroids derived from the Two Micron All
Sky Survey \citep[2MASS;][]{skrutskie:2006} catalog and the individual
astrometric solutions.  The resulting \lcs\ were decorrelated (cleaned
of trends) using the External Parameter Decorrelation \citep[EPD;
  see][]{bakos:2010} technique in ``constant'' mode and the Trend
Filtering Algorithm \citep[TFA; see][]{kovacs:2005}.  The \lcs{} were
searched for periodic box-shaped signals using the Box Least-Squares
\citep[BLS; see][]{kovacs:2002} method. We detected significant
signals in the \lcs\ of the stars summarized below:

\begin{itemize}
\item {\em \hatcur{18}} -- \hatcurCCgsc{18} (also known as
  \hatcurCCtwomass{18}; $\alpha = \hatcurCCra{18}$, $\delta =
  \hatcurCCdec{18}$; J2000; $V=\hatcurCCtassmv{18}$
  \citealp{droege:2006}). A signal was detected for this star with an
  apparent depth of $\sim$\hatcurLCdip{18}\,mmag, and a period of
  $P=$\hatcurLCPshort{18}\,days (see \reffigl{hatnet18}).  The drop in
  brightness had a first-to-last-contact duration, relative to the
  total period, of $q = $\hatcurLCq{18}, corresponding to a total
  duration of $Pq = $\hatcurLCdurhr{18}~hr.
\item {\em \hatcur{19}} -- \hatcurCCgsc{19} (also known as
  \hatcurCCtwomass{19}; $\alpha = \hatcurCCra{19}$, $\delta =
  \hatcurCCdec{19}$; J2000; $V=\hatcurCCtassmv{19}$
  \citealp{droege:2006}). A signal was detected for this star with an
  apparent depth of $\sim$\hatcurLCdip{19}\,mmag, and a period of
  $P=$\hatcurLCPshort{19}\,days (see \reffigl{hatnet19}).  The drop in
  brightness had a first-to-last-contact duration, relative to the
  total period, of $q = $\hatcurLCq{19}, corresponding to a total
  duration of $Pq = $\hatcurLCdurhr{19}~hr.
\end{itemize}

%
%
%

\ifthenelse{\boolean{emulateapj}}{
    \begin{deluxetable*}{llrrr}
}{
    \begin{deluxetable}{llrrr}
}
\tablewidth{0pc}
\tabletypesize{\scriptsize}
\tablecaption{
    Summary of photometric observations
    \label{tab:photobs}
}
\tablehead{
    \colhead{~~~~~~~~Instrument/Field~~~~~~~~}  &
    \colhead{Date(s)} &
    \colhead{Number of Images} &
    \colhead{Cadence (s)} &
    \colhead{Filter}
}
\startdata
\sidehead{\textbf{\hatcur{18}}}
~~~~HAT-6/G239 & 2007 Mar--2007 Jun & 4383 & 330 & $I$ \\
~~~~HAT-9/G239 & 2007 Mar--2007 Jun & 5719 & 330 & $I$ \\
~~~~KeplerCam  & 2008 Apr 25        &  215 &  73 & Sloan $i$ \\
~~~~KeplerCam  & 2009 May 10        &  185 & 133 & Sloan $g$ \\
\sidehead{\textbf{\hatcur{19}}}
~~~~HAT-7/G163 & 2007 Sep--2008 Jan & 2324 & 330 & $R$ \\
~~~~HAT-8/G163 & 2007 Sep--2008 Jan & 1617 & 330 & $R$ \\
~~~~HAT-6/G164 & 2007 Sep--2008 Feb & 3676 & 330 & $R$ \\
~~~~HAT-9/G164 & 2007 Sep--2008 Feb & 2711 & 330 & $R$ \\
~~~~KeplerCam  & 2009 Oct 06        &   37 &  84 & Sloan $i$ \\
~~~~KeplerCam  & 2009 Oct 30        &   97 & 150 & Sloan $i$ \\
~~~~KeplerCam  & 2009 Nov 27        &   76 &  84 & Sloan $i$ \\
~~~~KeplerCam  & 2009 Dec 01        &  194 &  89 & Sloan $i$ \\
[-1.5ex]
\enddata
\ifthenelse{\boolean{emulateapj}}{
    \end{deluxetable*}
}{
    \end{deluxetable}
}

\subsection{Reconnaissance Spectroscopy}
\label{sec:recspec}

As is routine in the HATNet project, all candidates are subjected to
careful scrutiny before investing valuable time on large telescopes.
This includes spectroscopic observations at relatively modest
facilities to establish whether the transit-like feature in the light
curve of a candidate might be due to astrophysical phenomena other
than a planet transiting a star.  Many of such false positives are
associated with large RV variations in the star (tens of \kms) that
are easily recognized.  We made use of three different facilities to
conduct these observations, including the Harvard-Smithsonian Center
for Astrophysics (CfA) Digital Speedometer \citep[DS;][]{latham:1992},
and the Tillinghast Reflector Echelle Spectrograph
\citep[TRES;][]{furesz:2008}, both on the 1.5\,m Tillinghast Reflector
at the Whipple Observatory on Mount Hopkins, Arizona, and the
FIbre-fed \'Echelle Spectrograph \citep[FIES;][]{frandsen:1999} on the
2.5\,m Nordic Optical Telescope \citep[NOT;][]{djupvik:2010} at La
Palma, Spain. We used these facilities to obtain high-resolution
spectra, with typically low signal-to-noise (S/N) ratios that are
nevertheless sufficient to derive RVs with moderate precisions of
0.5--1.0\,\kms\ for slowly rotating stars. We also use these spectra
to estimate the effective temperatures, surface gravities, and
projected rotational velocities of the stars. With these observations
we are able to reject many types of false positives, such as F dwarfs
orbited by M dwarfs, grazing eclipsing binaries, or triple or
quadruple star systems. Additional tests are performed with other
spectroscopic material described in the next section. The observations
and results for both stars are summarized in \reftabl{reconspecobs}.
Below we provide a brief description of each of the instruments used,
the data reduction, and the analysis procedure.

We used the DS to conduct observations of both \hatcur{18} and
\hatcur{19}. This instrument delivers high-resolution spectra
($\lambda/\Delta\lambda \approx 35,\!000$) over a single order
centered on the \ion{Mg}{1}\,b triplet ($\sim$5187\,\AA).  We measure
the RV and stellar atmospheric parameters from the spectra following
the method described by \cite{torres:2002}.

We used FIES to conduct observations of \hatcur{19}. We used the
medium and the high-resolution fibers with resolving powers of
$\lambda/\Delta\lambda \approx 46,\!000$ and $67,\!000$ respectively,
giving a wavelength coverage of $\sim$\,3600-7400\,\AA. The spectra
were extracted and analyzed to measure the RV and stellar atmospheric
parameters following the procedures described by
\cite{buchhave:2010}. The velocities were corrected to the same system
as the DS observations (heliocentric velocities with the gravitational
redshift of the Sun subtracted) using 10 observations of the velocity
standard HD~182488 obtained on the same nights as observations of
\hatcur{19}.

A single observation of \hatcur{19} was obtained with TRES. We used
the medium-resolution fiber to obtain a spectrum with a resolution of
$\lambda/\Delta\lambda \approx 44,000$ and a wavelength coverage of
$\sim$\,3900-8900\,\AA. The spectrum was extracted and analyzed in a
similar manner to the FIES observations. The velocity was corrected to
the same system as the DS observations using a single TRES measurement
of HD~182488 obtained on the same night. The velocity uncertainty
reported in this case is our estimate of the systematic error based on
the rms of multiple observations of HD~182488 obtained on other nights
close in time.

Based on the reconnaissance spectroscopy observations we find that
both systems have rms residuals consistent with no detectable RV
variation within the precision of the measurements. All spectra were
single-lined, i.e., there is no evidence that either target star has a
stellar companion. Additionally, both stars have surface gravity
measurements which indicate that they are dwarfs. We note that for
\hatcur{19} all three instruments yielded similar results for the RV
and stellar parameters. There is a $\sim 1$\,\kms\ difference between
the DS and the TRES/FIES observations of \hatcur{19}. The last DS
observation was obtained only four nights before the first FIES
observation, and the DS and FIES data-sets each span significantly
more than four nights, but do not show internal variations at the
$\sim 1$\,\kms\ level. We conclude that the velocity difference
between the instruments does not indicate a physical variation in the
velocity of \hatcur{19}. The DS observations are all weak, with only a
few counts per pixel, and the sky velocity is slightly more negative
than the system velocity in all cases. This velocity difference may be
due to a systematic error in the DS velocities due to sky
contamination.

\ifthenelse{\boolean{emulateapj}}{
    \begin{deluxetable*}{llrrrrr}
}{
    \begin{deluxetable}{llrrrrr}
}
\tablewidth{0pc}
\tabletypesize{\scriptsize}
\tablecaption{
    Summary of reconnaissance spectroscopy observations
    \label{tab:reconspecobs}
}
\tablehead{
    \multicolumn{1}{c}{Instrument}          &
    \multicolumn{1}{c}{Date(s)}             &
    \multicolumn{1}{c}{Number of Spectra}   &
    \multicolumn{1}{c}{$\teffstar$}         &
    \multicolumn{1}{c}{$\loggstar$}         &
    \multicolumn{1}{c}{$\vsini$}            &
    \multicolumn{1}{c}{$\gamma_{\rm RV}$\tablenotemark{a}} \\
    &
    &
    &
    \multicolumn{1}{c}{(K)}                 &
    \multicolumn{1}{c}{(cgs)}               &
    \multicolumn{1}{c}{(\kms)}              &
    \multicolumn{1}{c}{(\kms)}
}
\startdata
\sidehead{\textbf{\hatcur{18}}}
~~~~DS   & 2007 Sep--2008 Mar & 4 & \hatcurDSteff{18}  & \hatcurDSlogg{18} & \hatcurDSvsini{18} & \hatcurDSgamma{18} \\
\sidehead{\textbf{\hatcur{19}}}
~~~~DS   & 2008 Dec--2009 Jan & 3 & $4875 \pm 125$    & $4.25 \pm 0.25$ & $5 \pm 5$ & $-21.2 \pm 0.5$   \\
~~~~TRES & 2009 Sep 04        & 1 & $5000 \pm 125$    & $4.5 \pm 0.25$ & $2 \pm 2$ & $-20.20 \pm 0.05$ \\
~~~~FIES & 2009 Jan--2009 Oct & 7 & $4875 \pm 133$    & $4.25 \pm 0.27$ & $2.1 \pm 2$ & $-20.22 \pm 0.02$ \\
[-1.5ex]
\enddata 
\tablenotetext{a}{
    The mean heliocentric RV of the target.
}
\ifthenelse{\boolean{emulateapj}}{
    \end{deluxetable*}
}{
    \end{deluxetable}
}

\subsection{High resolution, high S/N spectroscopy}
\label{sec:hispec}

\begin{figure} [ht]
\ifthenelse{\boolean{emulateapj}}{
\epsscale{1.0}
}{
\epsscale{0.65}
}
\plotone{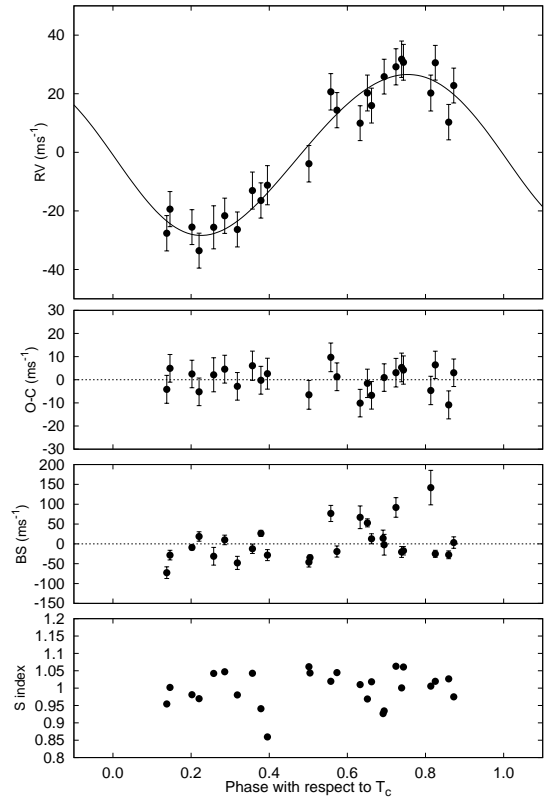}
\ifthenelse{\value{planetcounter}=1}{
\caption{
    {\em Top panel:} Keck/HIRES RV measurements for
        \hbox{\hatcur{18}{}} shown as a function of orbital
        phase, along with our best-fit model (see
        \reftabl{planetparam}).  Zero phase corresponds to the
        time of mid-transit.  The center-of-mass velocity has been
        subtracted.
    {\em Second panel:} Velocity $O\!-\!C$ residuals from the best
        fit. The error bars include a component from velocity
        jitter (\hatcurRVjitter{18}\,\ms) added in quadrature to
        the formal errors (see \refsecl{globmod}).
    {\em Third panel:} Bisector spans (BS), with the mean value
    subtracted. The measurement from the template spectrum is included
    (see \refsecl{bisec}). These measurements have not been corrected
    for contamination from moonlight; the corrected BS are shown in
    \reffigl{scf}.
    {\em Bottom panel:} Relative chromospheric activity index $S$
        measured from the Keck spectra. The formal errors on $S$ based on
    photon statistics are comparable to the size of the displayed
    symbols. The scatter, however, is likely dominated by systematic errors
    in the measurements.
    Note the different vertical scales of the panels.
}}{
\caption{
    Keck/HIRES observations of \hatcur{18}. The panels are as in
    \reffigl{rvbis18}.  The parameters used in the
    best-fit model are given in \reftabl{planetparam}.
}}
\label{fig:rvbis18}
\end{figure}

\begin{figure} [ht]
\ifthenelse{\boolean{emulateapj}}{
\epsscale{1.0}
}{
\epsscale{0.35}
}
\plotone{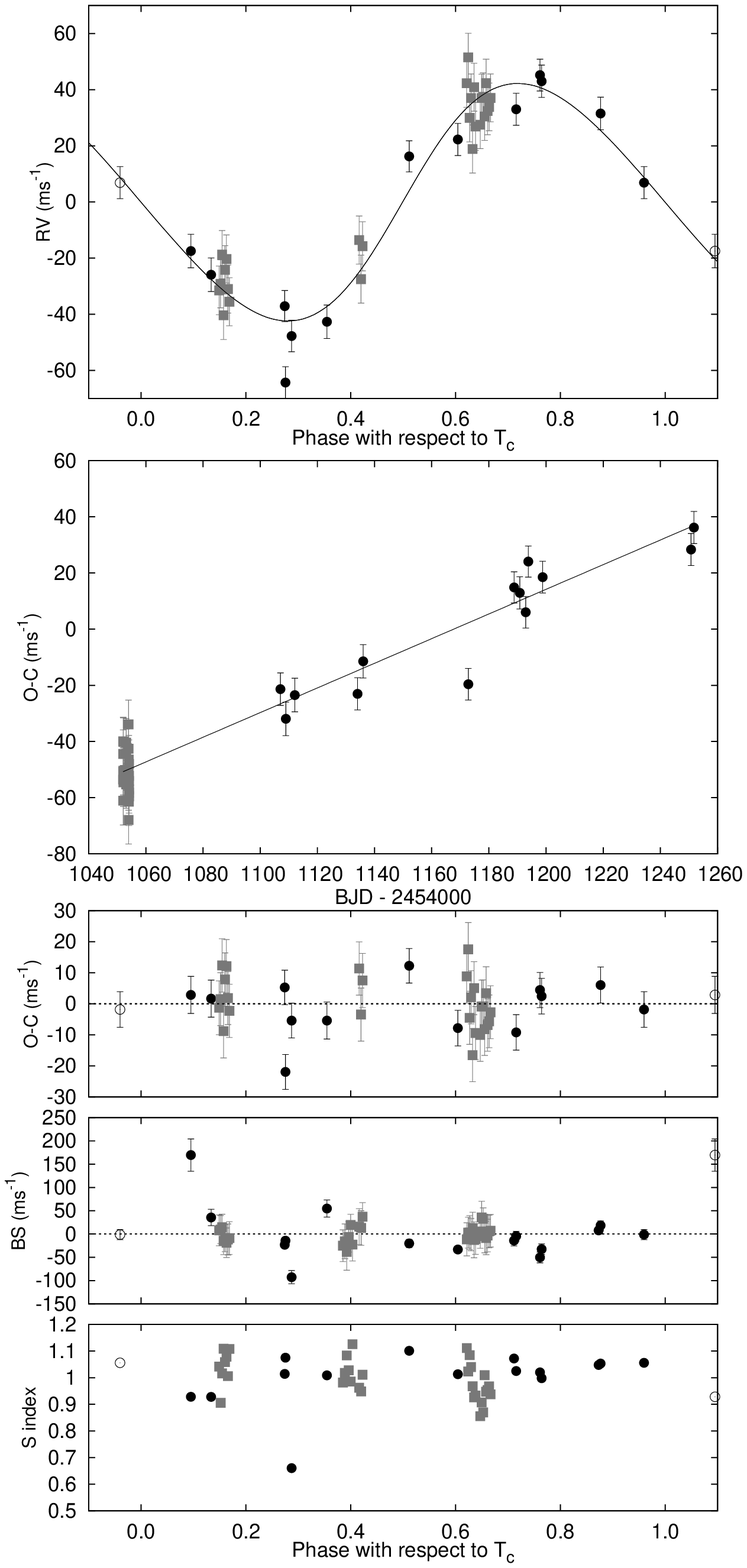}
\caption{
    {\em Top panel:} High-precision RV measurements for
    \hbox{\hatcur{19}{}} from Keck/HIRES (dark filled circles) and
    Subaru/HDS (light filled squares) shown as a function of orbital
    phase, along with our best-fit model (see
    \reftabl{planetparam}). Zero phase corresponds to the time of
    mid-transit.  The center-of-mass velocity and a linear trend
    (second panel) have been subtracted.
    {\em Second panel:} Velocity $O\!-\!C$ residuals from the best fit
    single Keplerian orbit model as a function of time. The residuals
    show a linear trend, indicative of a third body in the
    system. Note that the velocity zero points of the Subaru and Keck
    observations are independently free parameters. The linear trend
    is thus not constrained by the Subaru observations which span only
    3 days.
    {\em Third panel:} Velocity $O\!-\!C$ residuals from the best fit
    including both the Keplerian orbit and linear trend, shown as a
    function of orbital phase. The error bars include a component from
    velocity jitter (\hatcurRVjitter{19}\,\ms) added in
    quadrature to the formal errors (see \refsecl{globmod}).
    {\em Fourth panel:} Bisector spans (BS), with the mean value
    subtracted. The measurement from the template spectrum is included
    (see \refsecl{bisec}). These measurements have not been corrected
    for contamination from moonlight; the corrected BS are shown in
    \reffigl{scf}.
    {\em Bottom panel:} Relative chromospheric activity index $S$
    measured from the Keck spectra. The formal errors on $S$ based on
    photon statistics are comparable to the size of the displayed
    symbols. The scatter, however, is likely dominated by systematic errors
    in the measurements.
    Note the different vertical scales of the panels. Observations
    shown twice are represented with open symbols.
}
\label{fig:rvbis19}
\end{figure}

We proceeded with the follow-up of each candidate by obtaining
high-resolution, high-S/N spectra to characterize the RV variations,
and to refine the determination of the stellar parameters. These
observations are summarized in \reftabl{highsnspecobs}. The RV
measurements and uncertainties
\setcounter{planetcounter}{1}
\loopand{}for \hatcur{18}\ifthenelse{\value{planetcounter}=1}{ are given in }{ in }\reftabl{rvs18}\loopcomma
\setcounter{planetcounter}{2}
\loopand{}for \hatcur{19}\ifthenelse{\value{planetcounter}=1}{ are given in }{ in }\reftabl{rvs19}\loopcomma
The period-folded data, along with our best fit described below in
\refsecl{analysis}, are displayed
\setcounter{planetcounter}{1}
\loopand{}in \reffigl{rvbis18} for \hatcur{18}\loopcomma
\setcounter{planetcounter}{2}
\loopand{}in \reffigl{rvbis19} for \hatcur{19}\loopcomma For
\hatcur{18}, we exclude five RV measurements that are significant
outliers from the best fit model. These points are all strongly
affected by contamination from scattered moonlight (see
\refsecl{bisec}). Below we briefly describe the instruments used, the
data reduction, and the analysis procedure.

\ifthenelse{\boolean{emulateapj}}{
    \begin{deluxetable}{llrr}
}{
    \begin{deluxetable}{llrr}
}
\tablewidth{0pc}
\tabletypesize{\scriptsize}
\tablecaption{
    Summary of high-resolution/high-SN spectroscopic observations
    \label{tab:highsnspecobs}
}
\tablehead{
    \multicolumn{1}{c}{Instrument}  &
    \multicolumn{1}{c}{Date(s)}     &
    \multicolumn{1}{c}{Number of}   \\
    &
    &
    \multicolumn{1}{c}{RV obs.}
}
\startdata
\sidehead{\textbf{\hatcur{18}}}
~~~~Keck/HIRES           & 2007 Oct--2010 Mar & 29\tablenotemark{a}  \\
\sidehead{\textbf{\hatcur{19}}}
~~~~Keck/HIRES           & 2009 Oct--2010 Feb & 13  \\
~~~~Subaru/HDS           & 2009 Aug 8--2009 Aug 10 & 26  \\
[-1.5ex]
\enddata 
\tablenotetext{a}{This number includes five outlier RV points which were excluded from the analysis for \hatcur{18}.}
\ifthenelse{\boolean{emulateapj}}{
    \end{deluxetable}
}{
    \end{deluxetable}
}

Observations were made of \hatcur{18} and \hatcur{19} with the HIRES
instrument \citep{vogt:1994} on the Keck~I telescope located on Mauna
Kea, Hawaii.  The width of the spectrometer slit was $0\farcs86$,
resulting in a resolving power of $\lambda/\Delta\lambda \approx
55,\!000$, with a wavelength coverage of
$\sim$3800--8000\,\AA\@. Exposures were obtained through an iodine gas
absorption cell, which was used to superimpose a dense forest of
$\mathrm{I}_2$ lines on the stellar spectrum and establish an accurate
wavelength fiducial \citep[see][]{marcy:1992}. For each target two
additional exposures were taken without the iodine cell; in both cases
we used the second, higher S/N, observation as the template in the
reductions. Relative RVs in the solar system barycentric frame were
derived as described by \cite{butler:1996}, incorporating full
modeling of the spatial and temporal variations of the instrumental
profile.

We also made use of the High-Dispersion Spectrograph
\citep[HDS;][]{noguchi:2002} on the Subaru telescope on Mauna Kea,
Hawaii to obtain high-S/N spectroscopic observations of \hatcur{19}
from which we derived high-precision RV measurements. Observations
were made over three consecutive nights using a slit width of
$0\farcs6$, yielding a resolving power of $\lambda/\Delta\lambda
\approx 60,\!000$. We used the I2b setup which provides a wavelength
coverage of $\sim3500-6200$\,\AA\@. To reduce the effect of changes in
the barycentric velocity correction during an exposure, we limited
exposure times to 15 minutes. As for Keck/HIRES, we made use of an
iodine gas absorption cell to establish an accurate wavelength
fiducial for each exposure. We also obtained six spectra without the
iodine cell, which were combined to form the template observation. The
spectra were extracted, and reduced to relative RVs in the solar
system barycentric frame following the methods described by
\citet{sato:2002,sato:2005}.

\setcounter{planetcounter}{1}
%
\setcounter{planetcounter}{2}
%

In each figure we show also the relative $S$ index, which is a measure
of the chromospheric activity of the star derived from the flux in the
cores of the \ion{Ca}{2} H and K lines.  This index was computed
following the prescription given by \citet{vaughan:1978}, after
matching each spectrum to a reference spectrum using a transformation
that includes a wavelength shift and a flux scaling that is a
polynomial as a function of wavelength.  The transformation was
determined on regions of the spectra that are not used in computing
this indicator.
Note that our relative $S$ index has not been calibrated to the scale
of \citet{vaughan:1978}. We do not detect any significant variation of
the index correlated with orbital phase; such a correlation might have
indicated that the RV variations could be due to stellar activity,
casting doubt on the planetary nature of the candidate.

\ifthenelse{\boolean{emulateapj}}{
    \begin{deluxetable*}{lrrrrrr}
}{
    \begin{deluxetable}{lrrrrrr}
}
\tablewidth{0pc}
\tablecaption{
    Relative radial velocities, bisector spans, and activity index
    measurements of \hatcur{18}.
    \label{tab:rvs18}
}
\tablehead{
    \colhead{BJD\tablenotemark{a}} &
    \colhead{RV\tablenotemark{b}} &
    \colhead{\ensuremath{\sigma_{\rm RV}}\tablenotemark{c}} &
    \colhead{BS} &
    \colhead{\ensuremath{\sigma_{\rm BS}}} &
    \colhead{S\tablenotemark{d}} &
    \colhead{\ensuremath{\sigma_{\rm S}}}\\
    \colhead{\hbox{(2,454,000$+$)}} &
    \colhead{(\ms)} &
    \colhead{(\ms)} &
    \colhead{(\ms)} &
    \colhead{(\ms)} &
    \colhead{} &
    \colhead{}
}
\startdata
$ 397.73420 $ \dotfill & $   -11.23 $ & $     3.79 $ & $   -28.29 $ & $    13.90 $ & $    0.8595 $ & $   0.0099 $\\
$ 548.08129 $ \dotfill & \nodata      & \nodata      & $    14.12 $ & $    20.54 $ & $    0.9267 $ & $   0.0051 $\\
$ 548.09633 $ \dotfill & $    25.83 $ & $     2.26 $ & $    -2.20 $ & $    26.16 $ & $    0.9344 $ & $   0.0057 $\\
$ 549.07743 $ \dotfill & $    22.80 $ & $     2.27 $ & $     3.36 $ & $    14.43 $ & $    0.9748 $ & $   0.0048 $\\
$ 602.83762 $ \dotfill & $     9.95 $ & $     2.25 $ & $    67.02 $ & $    28.63 $ & $    1.0102 $ & $   0.0048 $\\
$ 602.99835 $ \dotfill & $    15.96 $ & $     2.29 $ & $    12.79 $ & $    12.66 $ & $    1.0183 $ & $   0.0045 $\\
$ 603.83449 $ \dotfill & $    20.24 $ & $     2.72 $ & $   141.82 $ & $    43.52 $ & $    1.0057 $ & $   0.0061 $\\
$ 604.08627 $ \dotfill & $    10.30 $ & $     2.51 $ & $   -27.50 $ & $     9.88 $ & $    1.0266 $ & $   0.0049 $\\
$ 633.97708 $ \dotfill & $   -21.65 $ & $     2.45 $ & $     9.96 $ & $    12.03 $ & $    1.0472 $ & $   0.0146 $\\
$ 635.98889 $ \dotfill & $    20.26 $ & $     2.69 $ & $    52.95 $ & $    10.05 $ & $    0.9688 $ & $   0.0058 $\\
$ 639.02280 $ \dotfill & $   -25.53 $ & $     2.29 $ & $    -8.96 $ & $     6.28 $ & $    0.9811 $ & $   0.0058 $\\
$ 641.97905 $ \dotfill & $    31.77 $ & $     2.88 $ & $   -20.46 $ & $    13.72 $ & $    1.0007 $ & $   0.0075 $\\
$ 724.84068\tablenotemark{e} $ \dotfill & $    6.49 $ & $    3.39 $ & $ -198.81 $ & $   36.35 $ & $ 0.9206 $ & $ 0.0158 $ \\
$ 726.79620 $ \dotfill & $   -27.63 $ & $     2.46 $ & $   -72.66 $ & $    14.74 $ & $    0.9544 $ & $   0.0091 $\\
$ 727.79088 $ \dotfill & $   -26.34 $ & $     2.31 $ & $   -47.93 $ & $    16.52 $ & $    0.9805 $ & $   0.0078 $\\
$ 777.69605 $ \dotfill & $   -16.44 $ & $     2.46 $ & $    26.52 $ & $     7.61 $ & $    0.9408 $ & $   0.0076 $\\
$ 778.69476\tablenotemark{e} $ \dotfill & $   35.06 $ & $    3.21 $ & $   45.44 $ & $   11.89 $ & $ 0.8932 $ & $ 0.0127 $ \\
$ 779.70147\tablenotemark{e} $ \dotfill & $   56.60 $ & $    4.98 $ & $   60.37 $ & $   20.83 $ & $ 0.8978 $ & $ 0.0212 $ \\
$ 779.74051\tablenotemark{e} $ \dotfill & $   61.15 $ & $    5.71 $ & $  134.75 $ & $   51.14 $ & $ 0.8017 $ & $ 0.0458 $ \\
$ 865.15876 $ \dotfill & $   -25.60 $ & $     4.88 $ & $   -31.22 $ & $    22.33 $ & $    1.0424 $ & $   0.0166 $\\
$ 955.02444 $ \dotfill & $    14.38 $ & $     2.41 $ & $   -19.22 $ & $    14.02 $ & $    1.0448 $ & $   0.0056 $\\
$ 955.96262 $ \dotfill & $    30.73 $ & $     2.64 $ & $   -17.35 $ & $    10.67 $ & $    1.0609 $ & $   0.0050 $\\
$ 964.09646 $ \dotfill & $   -33.57 $ & $     2.32 $ & $    18.65 $ & $    11.88 $ & $    0.9697 $ & $   0.0047 $\\
$ 987.98665 $ \dotfill & $    20.65 $ & $     2.88 $ & $    76.71 $ & $    20.29 $ & $    1.0195 $ & $   0.0057 $\\
$ 988.90490 $ \dotfill & $    29.18 $ & $     2.82 $ & $    91.83 $ & $    24.70 $ & $    1.0630 $ & $   0.0058 $\\
$ 1109.76458\tablenotemark{e} $ \dotfill & $  -49.58 $ & $    3.25 $ & $ -135.43 $ & $   28.38 $ & $ 0.7973 $ & $ 0.0308 $ \\
$ 1016.99998 $ \dotfill & $    30.57 $ & $     2.20 $ & $   -25.04 $ & $     8.82 $ & $    1.0197 $ & $   0.0076 $\\
$ 1041.96078 $ \dotfill & $   -13.06 $ & $     3.12 $ & $   -12.39 $ & $    11.79 $ & $    1.0428 $ & $   0.0162 $\\
$ 1252.06235 $ \dotfill & $    -3.90 $ & $     2.92 $ & $   -45.97 $ & $    12.38 $ & $    1.0616 $ & $   0.0065 $\\
$ 1252.07745 $ \dotfill & \nodata      & \nodata      & $   -34.63 $ & $     6.47 $ & $    1.0437 $ & $   0.0055 $\\
$ 1261.11905 $ \dotfill & $   -19.39 $ & $     2.39 $ & $   -28.25 $ & $    12.16 $ & $    1.0019 $ & $   0.0053 $\\
[-1.5ex]
\enddata
\tablenotetext{a}{Barycentric
    Julian dates throughout the paper are calculated from Coordinated
    Universal Time (UTC)}
\tablenotetext{b}{
    The zero-point of these velocities is arbitrary. An overall offset
    $\gamma_{\rm rel}$ fitted to these velocities in \refsecl{globmod}
    has {\em not} been subtracted.
}
\tablenotetext{c}{
    Internal errors excluding the component of velocity jitter
    considered in \refsecl{globmod}.
}
\tablenotetext{d}{
    Relative chromospheric activity index, not calibrated to the scale
    of \citet{vaughan:1978}.
}
\tablenotetext{e}{
    Outlier RV measurements excluded from the analysis.
}
\ifthenelse{\boolean{rvtablelong}}{
    \tablecomments{
        Note that for the iodine-free template exposures we do not
        measure the RV but do measure the BS and S index.  Such
        template exposures can be distinguished by the missing RV
        value.
    }
}{
    \tablecomments{
        Note that for the iodine-free template exposures we do not
        measure the RV but do measure the BS and S index.  Such
        template exposures can be distinguished by the missing RV
        value.  This table is presented in its entirety in the
        electronic edition of the Astrophysical Journal.  A portion is
        shown here for guidance regarding its form and content.
    }
} 
\ifthenelse{\boolean{emulateapj}}{
    \end{deluxetable*}
}{
    \end{deluxetable}
}
%
\ifthenelse{\boolean{emulateapj}}{
    \begin{deluxetable*}{lrrrrrrr}
}{
    \begin{deluxetable}{lrrrrrrr}
}
\tablewidth{0pc}
\tablecaption{
    Relative radial velocities, bisector spans, and activity index
    measurements of \hatcur{19}.
    \label{tab:rvs19}
}
\tablehead{
    \colhead{BJD\tablenotemark{a}} &
    \colhead{RV\tablenotemark{b}} &
    \colhead{\ensuremath{\sigma_{\rm RV}}\tablenotemark{c}} &
    \colhead{BS} &
    \colhead{\ensuremath{\sigma_{\rm BS}}} &
    \colhead{S\tablenotemark{d}} &
    \colhead{\ensuremath{\sigma_{\rm S}}} &
    \colhead{Inst.}\\
    \colhead{\hbox{(2,454,000$+$)}} &
    \colhead{(\ms)} &
    \colhead{(\ms)} &
    \colhead{(\ms)} &
    \colhead{(\ms)} &
    \colhead{} &
    \colhead{} &
    \colhead{}
}
\startdata
$ 1052.04317 $ \dotfill & $   -83.78 $ & $     7.01 $ & $     7.91 $ & $    33.17 $ & $ 1.042 $ & $   0.324 $     &  Subaru \\
$ 1052.05438 $ \dotfill & $   -81.45 $ & $     6.85 $ & $     9.71 $ & $    30.12 $ & $ 0.905 $ & $   0.294 $    &  Subaru \\
$ 1052.06559 $ \dotfill & $   -71.03 $ & $     6.92 $ & $    14.66 $ & $    27.93 $ & $ 1.016 $ & $   0.315 $    &  Subaru \\
$ 1052.07680 $ \dotfill & $   -92.65 $ & $     6.96 $ & $   -14.95 $ & $    29.17 $ & $ 1.109 $ & $   0.328 $    &  Subaru \\
$ 1052.08801 $ \dotfill & $   -76.43 $ & $     6.91 $ & $    -8.57 $ & $    29.23 $ & $ 1.059 $ & $   0.330 $    &  Subaru \\
$ 1052.09923 $ \dotfill & $   -72.59 $ & $     6.90 $ & $   -18.67 $ & $    31.68 $ & $ 1.079 $ & $   0.329 $    &  Subaru \\
$ 1052.11044 $ \dotfill & $   -83.27 $ & $     6.87 $ & $   -11.72 $ & $    33.40 $ & $ 1.006 $ & $   0.304 $    &  Subaru \\
$ 1052.12165 $ \dotfill & $   -87.81 $ & $     6.88 $ & $    -8.79 $ & $    35.31 $ & $ 1.108 $ & $   0.340 $    &  Subaru \\
$ 1052.99045 $ \dotfill & \nodata      & \nodata      & $   -25.11 $ & $    34.01 $ & $ 0.982 $ & $   0.313 $    &  Subaru \\
$ 1053.00513 $ \dotfill & \nodata      & \nodata      & $   -15.75 $ & $    37.37 $ & $ 1.018 $ & $   0.318 $    &  Subaru \\
$ 1053.01981 $ \dotfill & \nodata      & \nodata      & $   -38.58 $ & $    39.13 $ & $ 1.083 $ & $   0.330 $    &  Subaru \\
$ 1053.03449 $ \dotfill & \nodata      & \nodata      & $    -6.01 $ & $    32.64 $ & $ 1.028 $ & $   0.332 $    &  Subaru \\
$ 1053.04918 $ \dotfill & \nodata      & \nodata      & $    19.45 $ & $    23.05 $ & $ 0.986 $ & $   0.315 $    &  Subaru \\
$ 1053.06386 $ \dotfill & \nodata      & \nodata      & $   -22.72 $ & $    35.09 $ & $ 1.126 $ & $   0.347 $    &  Subaru \\
$ 1053.11425 $ \dotfill & $   -65.38 $ & $     6.90 $ & $    15.97 $ & $    38.62 $ & $ 0.962 $ & $   0.309 $    &  Subaru \\
$ 1053.12894 $ \dotfill & $   -79.28 $ & $     6.85 $ & $    13.01 $ & $    37.16 $ & $ 0.948 $ & $   0.313 $    &  Subaru \\
$ 1053.14023 $ \dotfill & $   -67.57 $ & $     7.11 $ & $    37.01 $ & $    30.50 $ & $ 1.011 $ & $   0.331 $    &  Subaru \\
$ 1053.93729 $ \dotfill & $    -9.14 $ & $     6.91 $ & $   -11.41 $ & $    35.52 $ & $ 1.111 $ & $   0.338 $    &  Subaru \\
$ 1053.94850 $ \dotfill & $     0.07 $ & $     6.93 $ & $     2.85 $ & $    33.24 $ & $ 1.023 $ & $   0.320 $    &  Subaru \\
$ 1053.95970 $ \dotfill & $   -21.48 $ & $     6.88 $ & $     3.11 $ & $    36.49 $ & $ 1.085 $ & $   0.334 $    &  Subaru \\
$ 1053.97091 $ \dotfill & $   -14.40 $ & $     6.87 $ & $    -1.28 $ & $    34.68 $ & $ 1.040 $ & $   0.314 $    &  Subaru \\
$ 1053.98213 $ \dotfill & $   -32.56 $ & $     6.87 $ & $    12.91 $ & $    33.85 $ & $ 0.968 $ & $   0.301 $    &  Subaru \\
$ 1053.99335 $ \dotfill & $   -10.54 $ & $     6.84 $ & $   -12.62 $ & $    38.43 $ & $ 0.926 $ & $   0.304 $    &  Subaru \\
$ 1054.00457 $ \dotfill & $   -24.54 $ & $     6.83 $ & $   -10.77 $ & $    32.87 $ & $ 0.933 $ & $   0.306 $    &  Subaru \\
$ 1054.04057 $ \dotfill & $   -23.88 $ & $     6.80 $ & $     5.45 $ & $    35.97 $ & $ 0.855 $ & $   0.273 $    &  Subaru \\
$ 1054.05178 $ \dotfill & $   -14.33 $ & $     6.78 $ & $    35.76 $ & $    34.71 $ & $ 0.907 $ & $   0.295 $    &  Subaru \\
$ 1054.06300 $ \dotfill & $   -13.93 $ & $     6.82 $ & $    32.75 $ & $    23.55 $ & $ 0.870 $ & $   0.285 $    &  Subaru \\
$ 1054.07421 $ \dotfill & $   -20.91 $ & $     6.81 $ & $    -4.30 $ & $    35.97 $ & $ 1.009 $ & $   0.323 $    &  Subaru \\
$ 1054.08543 $ \dotfill & $    -9.03 $ & $     6.81 $ & $    -8.63 $ & $    35.77 $ & $ 0.947 $ & $   0.310 $    &  Subaru \\
$ 1054.09665 $ \dotfill & $   -18.96 $ & $     6.75 $ & $    -3.25 $ & $    35.11 $ & $ 0.953 $ & $   0.304 $    &  Subaru \\
$ 1054.10787 $ \dotfill & $   -17.49 $ & $     6.78 $ & $     5.49 $ & $    35.24 $ & $ 0.967 $ & $   0.308 $    &  Subaru \\
$ 1054.11908 $ \dotfill & $   -14.29 $ & $     6.82 $ & $     7.07 $ & $    34.89 $ & $ 0.937 $ & $   0.297 $    &  Subaru \\
$ 1107.06075 $ \dotfill & \nodata      & \nodata      & $     7.74 $ & $     7.81 $ & $ 1.048 $ & $   0.011 $    &  Keck \\
$ 1107.07559 $ \dotfill & $     4.11 $ & $     2.77 $ & $    18.29 $ & $     9.69 $ & $ 1.053 $ & $   0.044 $    &  Keck \\
$ 1108.99050 $ \dotfill & $   -69.19 $ & $     3.09 $ & $    54.83 $ & $    18.48 $ & $ 1.009 $ & $   0.014 $    &  Keck \\
$ 1112.11372 $ \dotfill & $   -51.05 $ & $     3.13 $ & $    35.61 $ & $    17.65 $ & $ 0.928 $ & $   0.020 $    &  Keck \\
$ 1134.04526 $ \dotfill & $     7.06 $ & $     2.60 $ & $   -33.21 $ & $     6.43 $ & $ 1.013 $ & $   0.024 $    &  Keck \\
$ 1136.01174 $ \dotfill & $   -31.79 $ & $     3.10 $ & $   169.45 $ & $    34.81 $ & $ 0.928 $ & $   0.034 $    &  Keck \\
$ 1172.81382 $ \dotfill & $   -61.98 $ & $     2.36 $ & $   -14.18 $ & $     3.85 $ & $ 1.075 $ & $   0.011 $    &  Keck \\
$ 1188.84358 $ \dotfill & $   -27.51 $ & $     2.19 $ & $   -22.96 $ & $     5.20 $ & $ 1.014 $ & $   0.010 $    &  Keck \\
$ 1190.80886 $ \dotfill & $    53.47 $ & $     2.62 $ & $   -32.46 $ & $    11.07 $ & $ 0.998 $ & $   0.011 $    &  Keck \\
$ 1192.90496 $ \dotfill & $   -36.33 $ & $     2.34 $ & $   -92.65 $ & $    14.20 $ & $ 0.660 $ & $   0.020 $    &  Keck \\
$ 1193.80419 $ \dotfill & $    28.07 $ & $     2.18 $ & $   -20.27 $ & $     8.65 $ & $ 1.101 $ & $   0.011 $    &  Keck \\
$ 1198.81373 $ \dotfill & $    59.29 $ & $     2.49 $ & $   -50.27 $ & $    11.66 $ & $ 1.020 $ & $   0.010 $    &  Keck \\
$ 1250.72949 $ \dotfill & \nodata      & \nodata      & $   -14.06 $ & $    11.36 $ & $ 1.072 $ & $   0.011 $    &  Keck \\
$ 1250.74616 $ \dotfill & $    70.58 $ & $     2.56 $ & $    -4.56 $ & $    10.41 $ & $ 1.025 $ & $   0.015 $    &  Keck \\
$ 1251.72400 $ \dotfill & $    44.90 $ & $     2.60 $ & $    -1.30 $ & $    10.91 $ & $ 1.056 $ & $   0.016 $    &  Keck \\
[-1.5ex]
\enddata
\tablenotetext{a}{Barycentric
    Julian dates throughout the paper are calculated from Coordinated
    Universal Time (UTC)}
\tablenotetext{b}{
    The zero-point of these velocities is arbitrary. An overall offset
    $\gamma_{\rm rel}$ fitted to these velocities in \refsecl{globmod}
    has {\em not} been subtracted.
}
\tablenotetext{c}{
    Internal errors excluding the component of velocity jitter
    considered in \refsecl{globmod}.
}
\tablenotetext{d}{
    Relative chromospheric activity index, not calibrated to the scale
    of \citet{vaughan:1978}. Note the values for the Keck and Subaru
    observations have independently been scaled to have a mean of 1.0.
}
\ifthenelse{\boolean{rvtablelong}}{
    \tablecomments{
        Note that for the iodine-free template exposures we do not
        measure the RV but do measure the BS and S index.  Such
        template exposures can be distinguished by the missing RV
        value.
    }
}{
    \tablecomments{
        Note that for the iodine-free template exposures we do not
        measure the RV but do measure the BS and S index.  Such
        template exposures can be distinguished by the missing RV
        value.  This table is presented in its entirety in the
        electronic edition of the Astrophysical Journal.  A portion is
        shown here for guidance regarding its form and content.
    }
} 
\ifthenelse{\boolean{emulateapj}}{
    \end{deluxetable*}
}{
    \end{deluxetable}
}

\subsection{Photometric follow-up observations}
\label{sec:phot}

%
\setcounter{planetcounter}{1}
%
\begin{figure}[!ht]
\ifthenelse{\boolean{emulateapj}}{
\epsscale{1.0}
}{
\epsscale{1.0}
}
\plotone{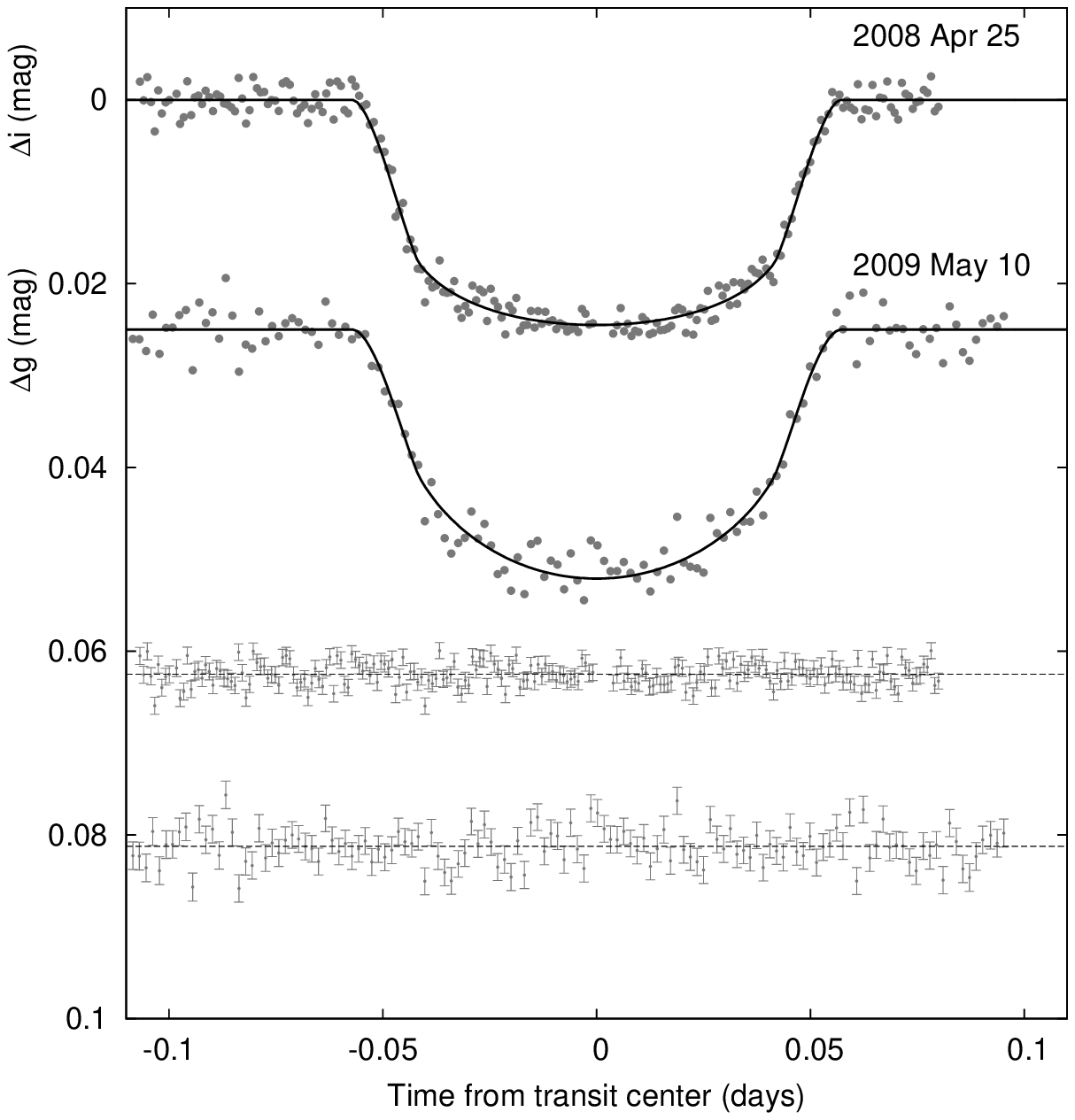}
\ifthenelse{\value{planetcounter}=1}{
\caption{
    Unbinned transit \lcs{} for \hatcur{18}, acquired with
    KeplerCam at the \flwof{} telescope.  The light curves have been
    EPD and TFA processed, as described in \refsec{globmod}.
    The dates of the events are indicated.  The second curve is
    displaced vertically for clarity.  Our best fit from the global
    modeling described in \refsecl{globmod} is shown by the solid
    lines.  Residuals from the fits are displayed at the bottom, in
    the same order as the top curves.  The error bars represent the
    photon and background shot noise, plus the readout noise.
}}{
\caption{
    Similar to \reffigl{lc18}; here we show the follow-up
    \lcs{} for \hatcur{18}.
}}
\label{fig:lc18}
\end{figure}
\setcounter{planetcounter}{2}
%
\begin{figure}[!ht]
\ifthenelse{\boolean{emulateapj}}{
\epsscale{1.0}
}{
\epsscale{0.75}
}
\plotone{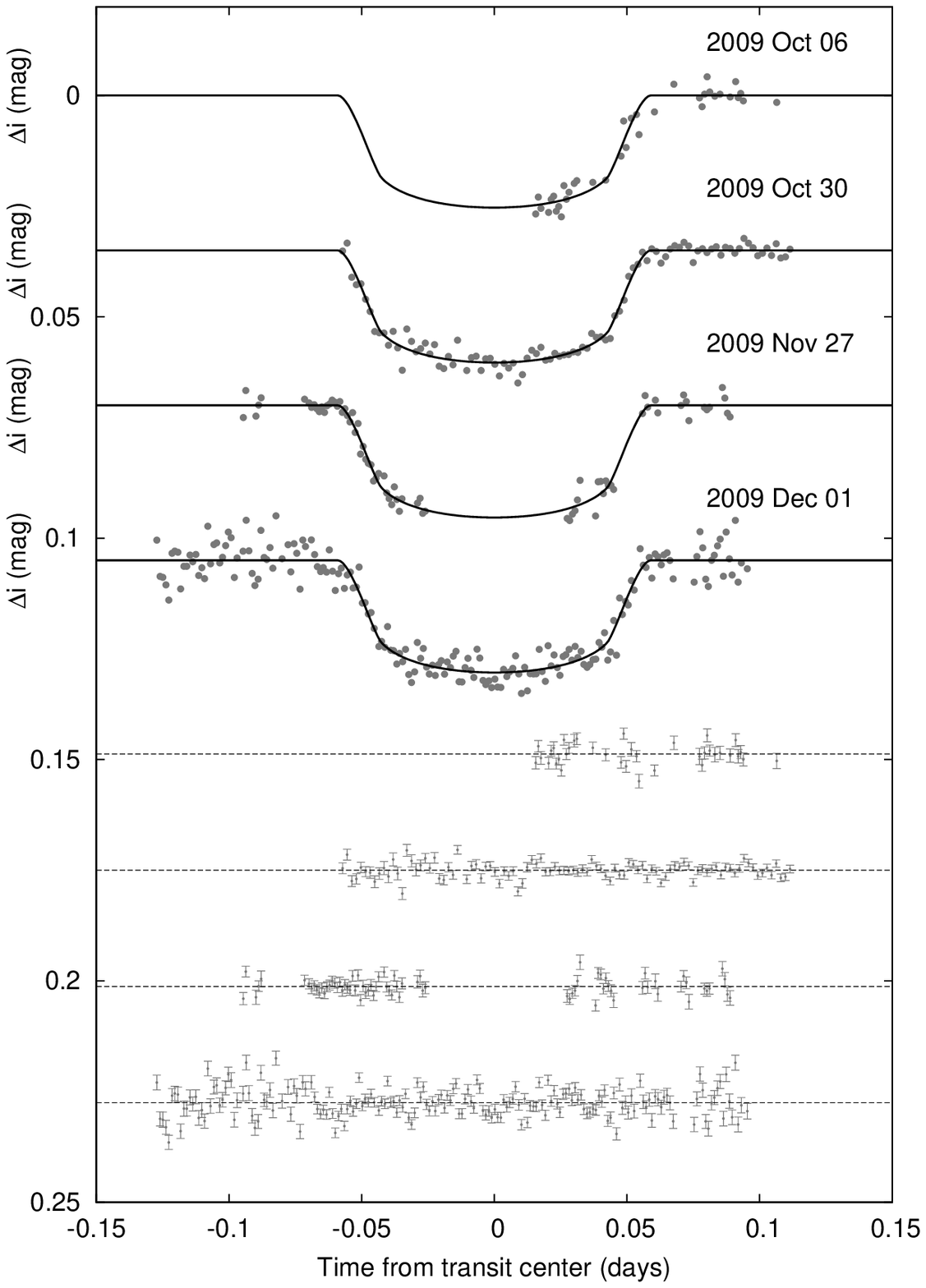}
\ifthenelse{\value{planetcounter}=1}{
\caption{
    Unbinned transit \lcs{} for \hatcur{19}, acquired with
    KeplerCam at the \flwof{} telescope.  The light curves have been
    EPD and TFA processed, as described in \refsec{globmod}.
    The dates of the events are indicated.  Curves after the first are
    displaced vertically for clarity.  Our best fit from the global
    modeling described in \refsecl{globmod} is shown by the solid
    lines.  Residuals from the fits are displayed at the bottom, in the
    same order as the top curves.  The error bars represent the photon
    and background shot noise, plus the readout noise.
}}{
\caption{
    Similar to \reffigl{lc18}; here we show the follow-up
    \lcs{} for \hatcur{19}.
}}
\label{fig:lc19}
\end{figure}

In order to permit a more accurate modeling of the light curves, we
conducted additional photometric observations with the KeplerCam CCD
camera on the \flwof{} telescope.  The observations for each target are
summarized in \reftabl{photobs}.

The reduction of these images, including basic calibration,
astrometry, and aperture photometry, was performed as described by
\citet{bakos:2010}.  We performed EPD and TFA to remove trends
simultaneously with the light curve modeling (for more details, see
\refsecl{analysis}, and \citealp{bakos:2010}).  The final time series,
together with our best-fit transit \lc{} model, are shown in the top
portion of Figures~\ref{fig:lc18}~and~\ref{fig:lc19} for \hatcur{18}
and \hatcur{19} respectively; the individual measurements are reported
in Tables~\ref{tab:phfu18}~and~\ref{tab:phfu19}.


%
%
\begin{deluxetable}{lrrrr}
\tablewidth{0pc}
\tablecaption{
    High-precision differential photometry of
    \hatcur{18}\label{tab:phfu18}.
}
\tablehead{
    \colhead{BJD} & 
    \colhead{Mag\tablenotemark{a}} & 
    \colhead{\ensuremath{\sigma_{\rm Mag}}} &
    \colhead{Mag(orig)\tablenotemark{b}} & 
    \colhead{Filter} \\
    \colhead{\hbox{~~~~(2,400,000$+$)~~~~}} & 
    \colhead{} & 
    \colhead{} &
    \colhead{} & 
    \colhead{}
}
\startdata
$ 54582.72249 $ & $  -0.00198 $ & $   0.00093 $ & $  10.79530 $ & $ i$\\
$ 54582.72334 $ & $   0.00006 $ & $   0.00093 $ & $  10.79800 $ & $ i$\\
$ 54582.72420 $ & $  -0.00247 $ & $   0.00093 $ & $  10.79660 $ & $ i$\\
$ 54582.72506 $ & $   0.00024 $ & $   0.00093 $ & $  10.79950 $ & $ i$\\
$ 54582.72592 $ & $   0.00344 $ & $   0.00093 $ & $  10.80160 $ & $ i$\\
$ 54582.72677 $ & $  -0.00104 $ & $   0.00093 $ & $  10.79740 $ & $ i$\\
$ 54582.72760 $ & $   0.00148 $ & $   0.00092 $ & $  10.79980 $ & $ i$\\
$ 54582.72848 $ & $   0.00027 $ & $   0.00092 $ & $  10.79820 $ & $ i$\\
$ 54582.72933 $ & $   0.00002 $ & $   0.00092 $ & $  10.79590 $ & $ i$\\
$ 54582.73102 $ & $  -0.00066 $ & $   0.00092 $ & $  10.79560 $ & $ i$\\
[-1.5ex]
\enddata
\tablenotetext{a}{
    The out-of-transit level has been subtracted. These magnitudes have
    been subjected to the EPD and TFA procedures, carried out
    simultaneously with the transit fit.
}
\tablenotetext{b}{
    Raw magnitude values without application of the EPD and TFA
    procedures.
}
\tablecomments{
    This table is available in a machine-readable form in the online
    journal.  A portion is shown here for guidance regarding its form
    and content.
}
\end{deluxetable}
%
\begin{deluxetable}{lrrrr}
\tablewidth{0pc}
\tablecaption{
    High-precision differential photometry of
    \hatcur{19}\label{tab:phfu19}.
}
\tablehead{
    \colhead{BJD} & 
    \colhead{Mag\tablenotemark{a}} & 
    \colhead{\ensuremath{\sigma_{\rm Mag}}} &
    \colhead{Mag(orig)\tablenotemark{b}} & 
    \colhead{Filter} \\
    \colhead{\hbox{~~~~(2,400,000$+$)~~~~}} & 
    \colhead{} & 
    \colhead{} &
    \colhead{} & 
    \colhead{}
}
\startdata
$ 55111.59368 $ & $   0.02677 $ & $   0.00139 $ & $  11.66270 $ & $ i$\\
$ 55111.59465 $ & $   0.02295 $ & $   0.00130 $ & $  11.65390 $ & $ i$\\
$ 55111.59562 $ & $   0.02551 $ & $   0.00142 $ & $  11.65450 $ & $ i$\\
$ 55111.59853 $ & $   0.02643 $ & $   0.00133 $ & $  11.66000 $ & $ i$\\
$ 55111.59949 $ & $   0.02348 $ & $   0.00120 $ & $  11.65220 $ & $ i$\\
$ 55111.60045 $ & $   0.02275 $ & $   0.00140 $ & $  11.65930 $ & $ i$\\
$ 55111.60140 $ & $   0.02615 $ & $   0.00126 $ & $  11.65320 $ & $ i$\\
$ 55111.60236 $ & $   0.02510 $ & $   0.00126 $ & $  11.65380 $ & $ i$\\
$ 55111.60333 $ & $   0.02741 $ & $   0.00130 $ & $  11.66230 $ & $ i$\\
$ 55111.60430 $ & $   0.02038 $ & $   0.00121 $ & $  11.64810 $ & $ i$\\
[-1.5ex]
\enddata
\tablenotetext{a}{
    The out-of-transit level has been subtracted. These magnitudes have
    been subjected to the EPD and TFA procedures, carried out
    simultaneously with the transit fit.
}
\tablenotetext{b}{
    Raw magnitude values without application of the EPD and TFA
    procedures.
}
\tablecomments{
    This table is available in a machine-readable form in the online
    journal.  A portion is shown here for guidance regarding its form
    and content.
}
\end{deluxetable}

\section{Analysis}
\label{sec:analysis}

\subsection{Properties of the parent star}
\label{sec:stelparam}

Fundamental parameters for each of the host stars, including the mass
(\mstar) and radius (\rstar), which are needed to infer the planetary
properties, depend strongly on other stellar quantities that can be
derived spectroscopically.  For this we have relied on our template
spectra obtained with the Keck/HIRES instrument, and the analysis
package known as Spectroscopy Made Easy \citep[SME;][]{valenti:1996},
along with the atomic line database of \cite{valenti:2005}.  For each
star, SME yielded the following {\em initial} values and uncertainties
(which we have conservatively increased by a factor of two to include
our estimates of the systematic errors):
\begin{itemize}
\item {\em \hatcur{18}} --
effective temperature $\teffstar=$\hatcurSMEiteff{18}\,K, 
stellar surface gravity $\loggstar=$\hatcurSMEilogg{18}\,(cgs),
metallicity $\feh=$\hatcurSMEizfeh{18}\,dex, and 
projected rotational velocity $\vsini=$\hatcurSMEivsin{18}\,\kms.
\item {\em \hatcur{19}} --
effective temperature $\teffstar=$\hatcurSMEiteff{19}\,K, 
stellar surface gravity $\loggstar=$\hatcurSMEilogg{19}\,(cgs),
metallicity $\feh=$\hatcurSMEizfeh{19}\,dex, and 
projected rotational velocity $\vsini=$\hatcurSMEivsin{19}\,\kms.
\end{itemize}
As discussed in \refsecl{bisec}, contamination from scattered
moonlight affects the bisector spans and radial velocities measured
for \hatcur{18} and the bisector spans measured for \hatcur{19}. For
\hatcur{18} the moon was below the horizon when the template used for
the SME analysis was obtained, so it is not affected by
contamination. For \hatcur{19} we estimate that scattered moonlight
may have contributed $\sim 0.1\%$ of the total flux to the template
spectrum used for SME analysis. The error in the parameters that
results from this contamination is likely dwarfed by other systematic
errors in the parameter determination.

In principle the effective temperature and metallicity, along with the
surface gravity taken as a luminosity indicator, could be used as
constraints to infer the stellar mass and radius by comparison with
stellar evolution models.  However, the effect of \loggstar\ on the
spectral line shapes is rather subtle, and as a result it is typically
difficult to determine accurately, so that it is a rather poor
luminosity indicator in practice.  For planetary transits a stronger
constraint is often provided by the \arstar\ normalized semimajor
axis, which is closely related to \rhostar, the mean stellar density.
The quantity \arstar\ can be derived directly from the transit
\lcs\ \citep[see][and also \refsecl{globmod}]{sozzetti:2007}.  This,
in turn, allows us to improve on the determination of the
spectroscopic parameters by supplying an indirect constraint on the
weakly determined spectroscopic value of \loggstar, which removes
degeneracies.  We take this approach here, as described below.  The
validity of our assumption, namely that the best physical model
describing our data is a planetary transit (as opposed to a blend), is
shown later in \refsecl{bisec}.

For each system, our initial values of \teffstar, \loggstar, and
\feh\ were used to determine auxiliary quantities needed in the global
modeling of the follow-up photometry and radial velocities
(specifically, the limb-darkening coefficients).  This modeling, the
details of which are described in \refsecl{globmod}, uses a Monte
Carlo approach to deliver the numerical probability distribution of
\arstar\ and other fitted variables.  For further details we refer the
reader to \cite{pal:2009b}.  When combining \arstar\ (used as a proxy
for luminosity) with assumed Gaussian distributions for \teffstar\ and
\feh\ based on the SME determinations, a comparison with stellar
evolution models allows the probability distributions of other stellar
properties to be inferred, including \loggstar.  Here we use the
stellar evolution calculations from \hatcurisofull{18} for both stars.
The comparison against the model isochrones was carried out for each
of 10,000 Monte Carlo trial sets for \hatcur{18}, and 20,000 Monte
Carlo trial sets for \hatcur{19} (see \refsecl{globmod}).  Parameter
combinations corresponding to unphysical locations in the \hbox{H-R}
diagram (41\% of the trials for \hatcur{18} and 31\% of the trials for
\hatcur{19}) were ignored, and replaced with another randomly drawn
parameter set.  For each system we carried out a second SME iteration
in which we adopted the value of \loggstar\ so determined and held it
fixed in a new SME analysis (coupled with a new global modeling of the
RV and \lcs), adjusting only \teffstar, \feh, and \vsini.  This gave
\begin{itemize}
\item {\em \hatcur{18}} --
$\teffstar=$\hatcurSMEiiteff{18}\,K, 
$\loggstar=$\hatcurSMEiilogg{18},
$\feh=$\hatcurSMEiizfeh{18}, and
$\vsini=$\hatcurSMEiivsin{18}\,\kms.
\item {\em \hatcur{19}} --
$\teffstar=$\hatcurSMEiiteff{19}\,K, 
$\loggstar=$\hatcurSMEiilogg{19} (fixed),
$\feh=$\hatcurSMEiizfeh{19}, and
$\vsini=$\hatcurSMEiivsin{19}\,\kms.
\end{itemize}
In each case the conservative uncertainties for $\teffstar$ and $\feh$
have been increased by a factor of two over their formal values, as
before.
For each system, a further iteration did not change
\loggstar\ significantly, so we adopted the values stated above,
together with the new \loggstar\ values resulting from the global
modeling, as the final atmospheric properties of the stars.  They are
collected in Table~\ref{tab:stellar} for both stars.

\begin{figure*}[!ht]
\ifthenelse{\boolean{emulateapj}}{
\epsscale{1.0}
}{
\epsscale{1.0}
}
\plottwo{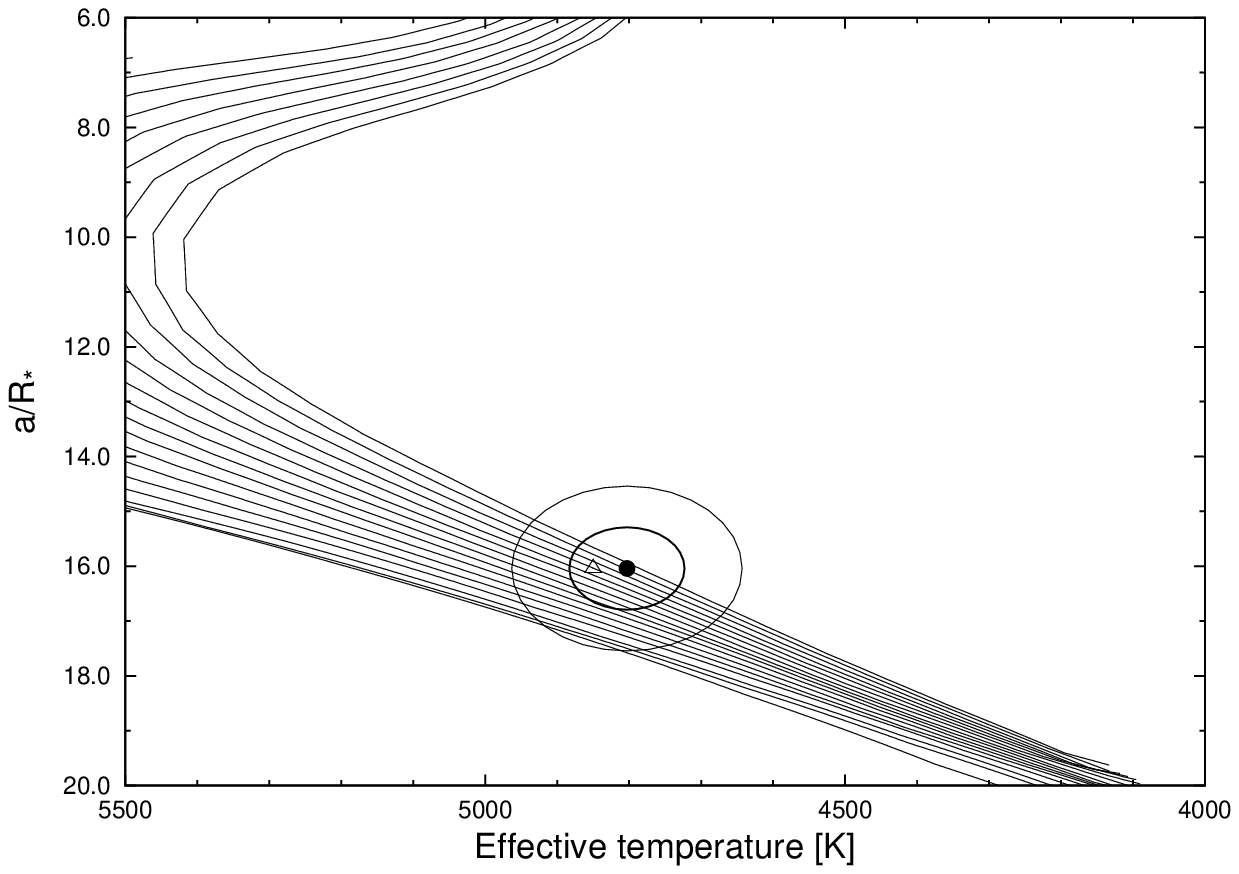}{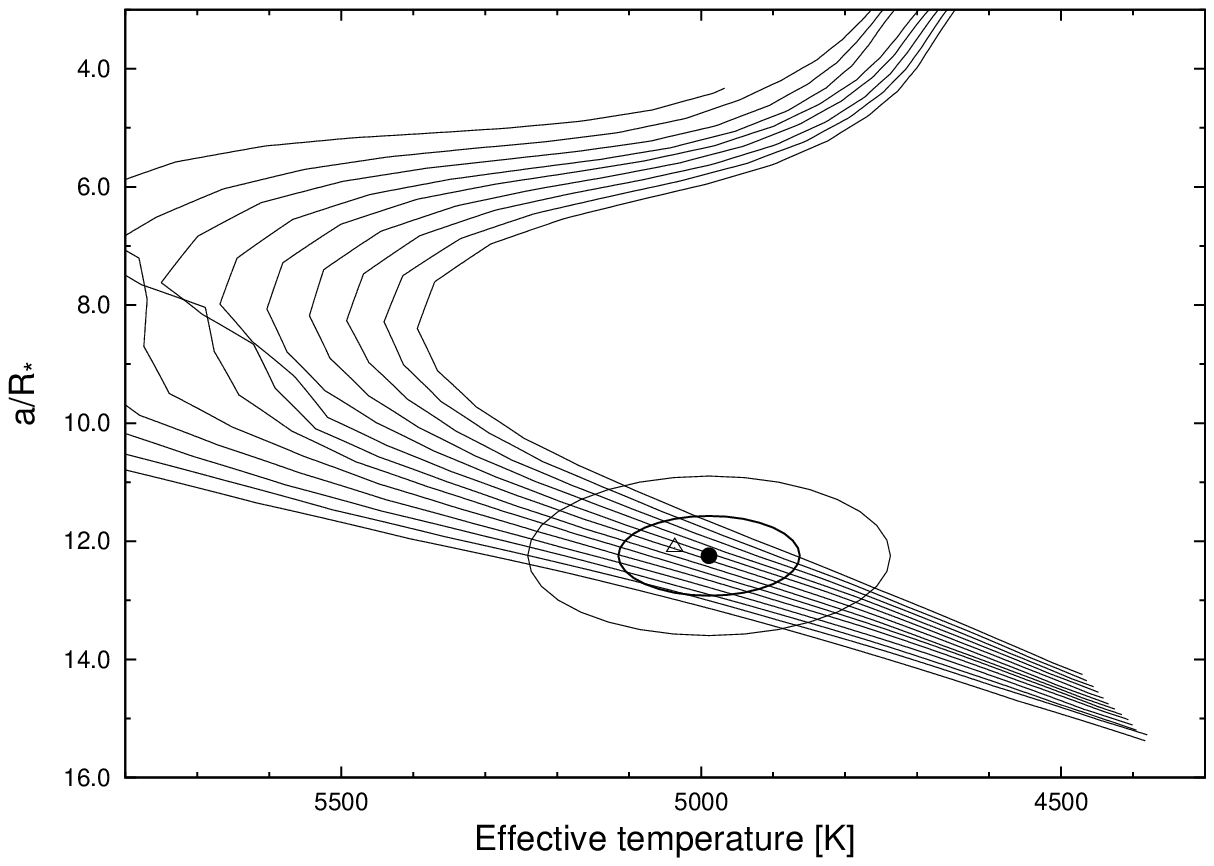}
\caption{
    Left: model isochrones from \cite{\hatcurisocite{18}} for the
    measured metallicity of \hatcur{18}, \feh =
    \hatcurSMEiizfehshort{18}, and ages of 0.2, 0.5, and 1 to 14\,Gyr,
    in 1\,Gyr increments (left to right). The adopted values of
    $\teffstar$ and \arstar\ are shown together with their 1$\sigma$
    and 2$\sigma$ confidence ellipsoids.  The initial values of
    \teffstar\ and \arstar\ from the first SME and \lc\ analyses are
    represented with a triangle. Right: same as left, here we show the
    results for \hatcur{19}, with \feh = \hatcurSMEiizfehshort{19},
    and ages of 1 to 13\,Gyr in steps of 1\,Gyr (left to right).
}
\label{fig:iso}
\end{figure*}

With the adopted spectroscopic parameters the model isochrones yield
the stellar mass and radius, and other properties.  These are listed
for each of the systems in Table~\ref{tab:stellar}.  According to
these models
\setcounter{planetcounter}{1}
\loopand\hatcur{18} is a dwarf star with an
estimated age of
\hatcurISOage{18}\,Gyr\loopcomma
\setcounter{planetcounter}{2}
\loopand\hatcur{19} is a dwarf star with an estimated age of
\hatcurISOage{19}\,Gyr\loopcomma 
The inferred location of each star in a diagram of \arstar\ versus
\teffstar, analogous to the classical H-R diagram, is shown in
\reffigl{iso}.  The stellar properties and their 1$\sigma$ and
2$\sigma$ confidence ellipsoids are displayed against the backdrop of
model isochrones for a range of ages, and the appropriate stellar
metallicity.  For comparison, the locations implied by the initial SME
results are also shown with triangles.

The stellar evolution modeling provides color indices that may be
compared against the measured values as a sanity check.  For each
star, the best available measurements are the near-infrared magnitudes
from the 2MASS Catalogue \citep{skrutskie:2006}, which are given in
Table~\ref{tab:stellar}.  These are converted to the photometric
system of the models (ESO system) using the transformations by
\citet{carpenter:2001}.  The resulting color index is
\setcounter{planetcounter}{1}
\loopand$J-K = \hatcurCCesoJKmag{18}$\loopcommanoperiod
\setcounter{planetcounter}{2}
\loopand$J-K = \hatcurCCesoJKmag{19}$\loopcommanoperiod
for
\setcounter{planetcounter}{1}
\loopand\hatcur{18}\loopcommanoperiod
\setcounter{planetcounter}{2}
\loopand\hatcur{19}\loopcommanoperiod respectively. These are both
within $1\sigma$ of the predicted values from the isochrones of $J-K =
\hatcurISOJK{18}$ and $J-K = \hatcurISOJK{19}$. The distance to each
object may be computed from the absolute $K$ magnitude from the models
and the 2MASS $K_s$ magnitudes, which has the advantage of being less
affected by extinction than optical magnitudes.  The results are given
in Table~\ref{tab:stellar}, where in each case the uncertainty
excludes possible systematics in the model isochrones that are
difficult to quantify.

%
%
\ifthenelse{\boolean{emulateapj}}{
    \begin{deluxetable*}{lccl}
}{
    \begin{deluxetable}{lccl}
}
\tablewidth{0pc}
\tabletypesize{\scriptsize}
\tablecaption{
    Stellar parameters for \hatcur{18} and \hatcur{19}
    \label{tab:stellar}
}
\tablehead{
    \colhead{~~~~~~~~Parameter~~~~~~~~} &
    \colhead{\hatcur{18}}               &
    \colhead{\hatcur{19}}               &
    \colhead{Source}
}
\startdata
\noalign{\vskip -3pt}
\sidehead{Spectroscopic properties}
~~~~$\teffstar$ (K)\dotfill         &  \hatcurSMEteff{18}   &  \hatcurSMEteff{19}   & SME\tablenotemark{a}\\
~~~~$\feh$\dotfill                  &  \hatcurSMEzfeh{18}   &  \hatcurSMEzfeh{19}   & SME                 \\
~~~~$\vsini$ (\kms)\dotfill         &  \hatcurSMEvsin{18}   &  \hatcurSMEvsin{19}   & SME                 \\
~~~~$\vmac$ (\kms)\dotfill          &  \hatcurSMEvmac{18}   &  \hatcurSMEvmac{19}   & SME                 \\
~~~~$\vmic$ (\kms)\dotfill          &  \hatcurSMEvmic{18}   &  \hatcurSMEvmic{19}   & SME                 \\
~~~~$\gamma_{\rm RV}$ (\kms)\dotfill&  \hatcurDSgamma{18}   &  $-20.22 \pm 0.02$   & DS/FIES\tablenotemark{b}                  \\
\sidehead{Photometric properties}
~~~~$V$ (mag)\dotfill               &  \hatcurCCtassmv{18}  &  \hatcurCCtassmv{19}  & TASS                \\
~~~~$\vic$ (mag)\dotfill            &  \hatcurCCtassvi{18}  &  \hatcurCCtassvi{19}  & TASS                \\
~~~~$J$ (mag)\dotfill               &  \hatcurCCtwomassJmag{18} &  \hatcurCCtwomassJmag{19} & 2MASS           \\
~~~~$H$ (mag)\dotfill               &  \hatcurCCtwomassHmag{18} &  \hatcurCCtwomassHmag{19} & 2MASS           \\
~~~~$K_s$ (mag)\dotfill             &  \hatcurCCtwomassKmag{18} &  \hatcurCCtwomassKmag{19} & 2MASS           \\
\sidehead{Derived properties}
~~~~$\mstar$ ($\msun$)\dotfill      &  \hatcurISOmlong{18}   &  \hatcurISOmlong{19}   & \hatcurisoshort{18}+\hatcurlumind{18}+SME \tablenotemark{c}\\
~~~~$\rstar$ ($\rsun$)\dotfill      &  \hatcurISOrlong{18}   &  \hatcurISOrlong{19}   & \hatcurisoshort{18}+\hatcurlumind{18}+SME         \\
~~~~$\loggstar$ (cgs)\dotfill       &  \hatcurISOlogg{18}    &  \hatcurISOlogg{19}    & \hatcurisoshort{18}+\hatcurlumind{18}+SME         \\
~~~~$\lstar$ ($\lsun$)\dotfill      &  \hatcurISOlum{18}     &  \hatcurISOlum{19}     & \hatcurisoshort{18}+\hatcurlumind{18}+SME         \\
~~~~$M_V$ (mag)\dotfill             &  \hatcurISOmv{18}      &  \hatcurISOmv{19}      & \hatcurisoshort{18}+\hatcurlumind{18}+SME         \\
~~~~$M_K$ (mag,\hatcurjhkfilset{18})\dotfill &  \hatcurISOMK{18} &  \hatcurISOMK{19} & \hatcurisoshort{18}+\hatcurlumind{18}+SME         \\
~~~~Age (Gyr)\dotfill               &  \hatcurISOage{18}     &  \hatcurISOage{19}     & \hatcurisoshort{18}+\hatcurlumind{18}+SME         \\
~~~~Distance (pc)\dotfill           &  \hatcurXdist{18}  &  \hatcurXdist{19}  & \hatcurisoshort{18}+\hatcurlumind{18}+SME\\ [-1.5ex]
\enddata
\tablenotetext{a}{
    SME = ``Spectroscopy Made Easy'' package for the analysis of
    high-resolution spectra \citep{valenti:1996}.  These parameters
    rely primarily on SME, but have a small dependence also on the
    iterative analysis incorporating the isochrone search and global
    modeling of the data, as described in the text.
}
\tablenotetext{b}{
    Based on DS observations for \hatcur{18} and FIES observations for \hatcur{19}.
}
\tablenotetext{c}{
    \hatcurisoshort{18}+\hatcurlumind{18}+SME = Based on the \hatcurisoshort{18}
    isochrones \citep{\hatcurisocite{18}}, \hatcurlumind{18} as a luminosity
    indicator, and the SME results.
}
\ifthenelse{\boolean{emulateapj}}{
    \end{deluxetable*}
}{
    \end{deluxetable}
}

\subsection{Rejecting Blend Scenarios}
\label{sec:blend}

Our initial spectroscopic analyses discussed in \refsecl{recspec} and
\refsecl{hispec} rule out the most obvious astrophysical false
positive scenarios.  However, more subtle phenomena such as blends
(contamination by an unresolved eclipsing binary, whether in the
background or associated with the target) can still mimic both the
photometric and spectroscopic signatures we see. In the following
sections we consider and rule out the possibility that such scenarios
may have caused the observed photometric and spectroscopic features.

\subsubsection{Spectral line-bisector analysis}
\label{sec:bisec}

Following \cite{torres:2007}, we explored the possibility that the
measured radial velocities are not real, but are instead caused by
distortions in the spectral line profiles due to contamination from a
nearby unresolved eclipsing binary.  A bisector analysis for each
system based on the Keck spectra (and the Subaru spectra for
\hatcur{19}) was done as described in \S 5 of \cite{bakos:2007a}.

Each system shows excess scatter in the bisector spans (BS), above
what is expected from the measurement errors (see \reffigl{rvbis18},
third panel, and \reffigl{rvbis19}, fourth panel). For \hatcur{18}
there may be a slight correlation between the RV and the BS, while for
\hatcur{19} no correlation is apparent. Such a correlation could
indicate that the photometric and spectroscopic signatures are due to
a blend scenario rather than a single planet transiting a single
star. We note that for \hatcur{19}, the Keck spectra show excess BS
variation, while the Subaru spectra do not.

\begin{figure*}[!ht]
\plotone{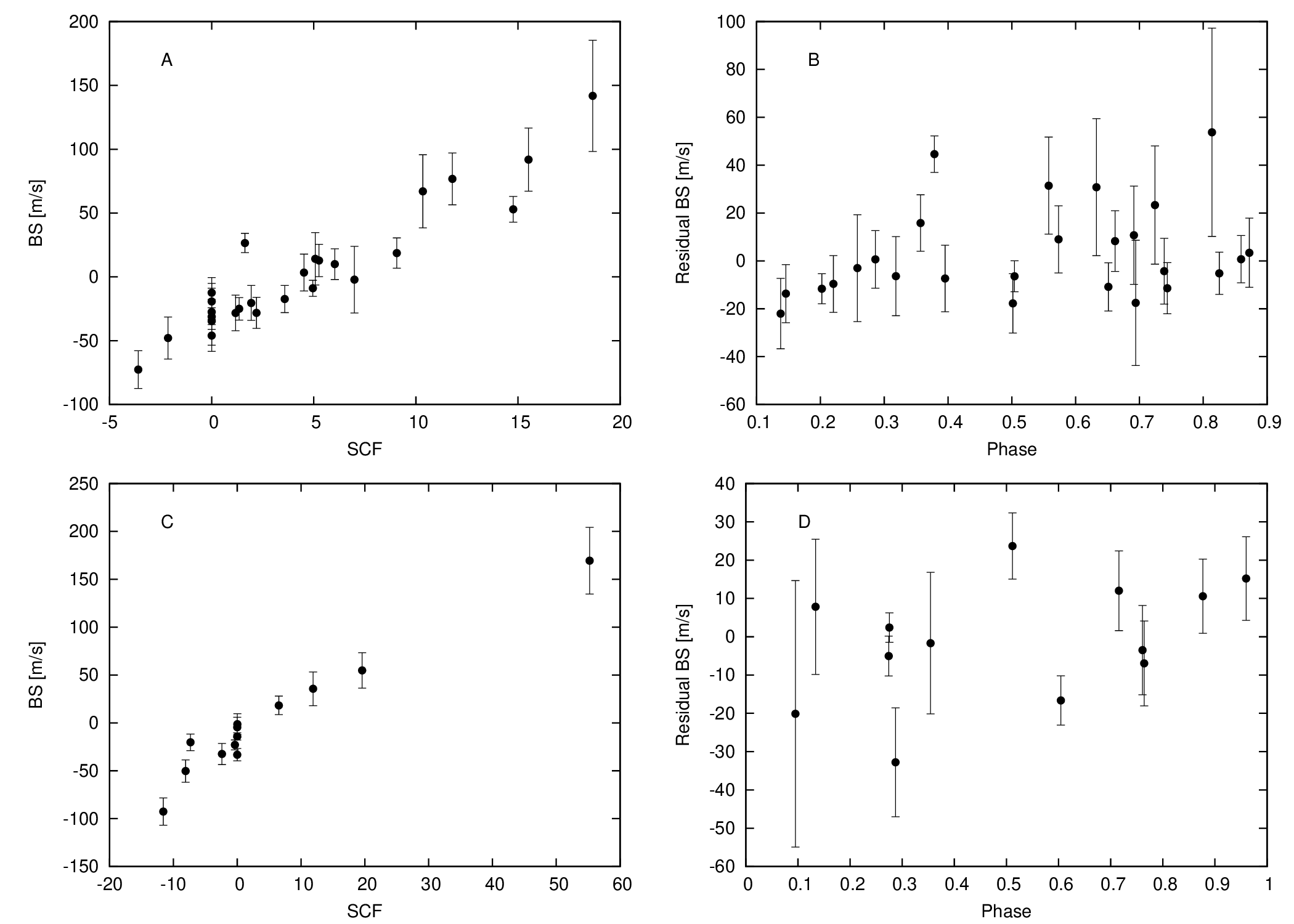}
\caption[]{
    A: BS vs SCF for Keck spectra of \hatcur{18}. The strong
    correlation between these quantities indicates that much of the BS
    variation for this object can be accounted for by changes in the
    sky contamination of the spectra. B: BS vs orbital phase for
    \hatcur{18} after fitting and subtracting a linear relation
    between BS and SCF. The residual BS is uncorrelated with the
    orbital phase, justifying our conclusion that this system is not a
    blend. Bottom: BS vs SCF for Keck spectra of \hatcur{19}. Again
    the BS variation for this target can be accounted for by the
    changing sky contamination. The Subaru spectra of \hatcur{19} do
    not show significant BS variations; these spectra were taken
    within a span of 3 days, under similar sky conditions. D: Same as
    panel B, here shown for \hatcur{19}.
\label{fig:scf}}
\end{figure*}

Following our earlier work \citep{kovacs:2010,hartman:2009} we
investigated the effect of contamination from moonlight on the
measured BS values.  As in \citet{kovacs:2010}, we estimate the
expected BS value for each spectrum by modeling the spectrum
cross-correlation function (CCF) as the sum of two Lorentzian
functions, shifted by the known velocity difference between the star
and the moon, and scaled by their expected flux ratio (estimated
following equation 3 of \citealp{hartman:2009}). We refer to the
simulated BS value as the sky contamination factor (SCF). We find a
strong correlation between the SCF and BS for both systems (see
\reffigl{scf}).  After correcting for this correlation, we find that
the BS show no significant variations, and the correlation between the
RV and BS variations is insignificant for both systems. Therefore, we
conclude that the the velocity variations are real for both stars, and
that both stars are orbited by close-in giant planets. An independent
method for arriving at this same conclusion is also presented in the
following section.

\begin{figure}[!ht]
\ifthenelse{\boolean{emulateapj}}{
\epsscale{1.0}
}{
\epsscale{0.75}
}
\plotone{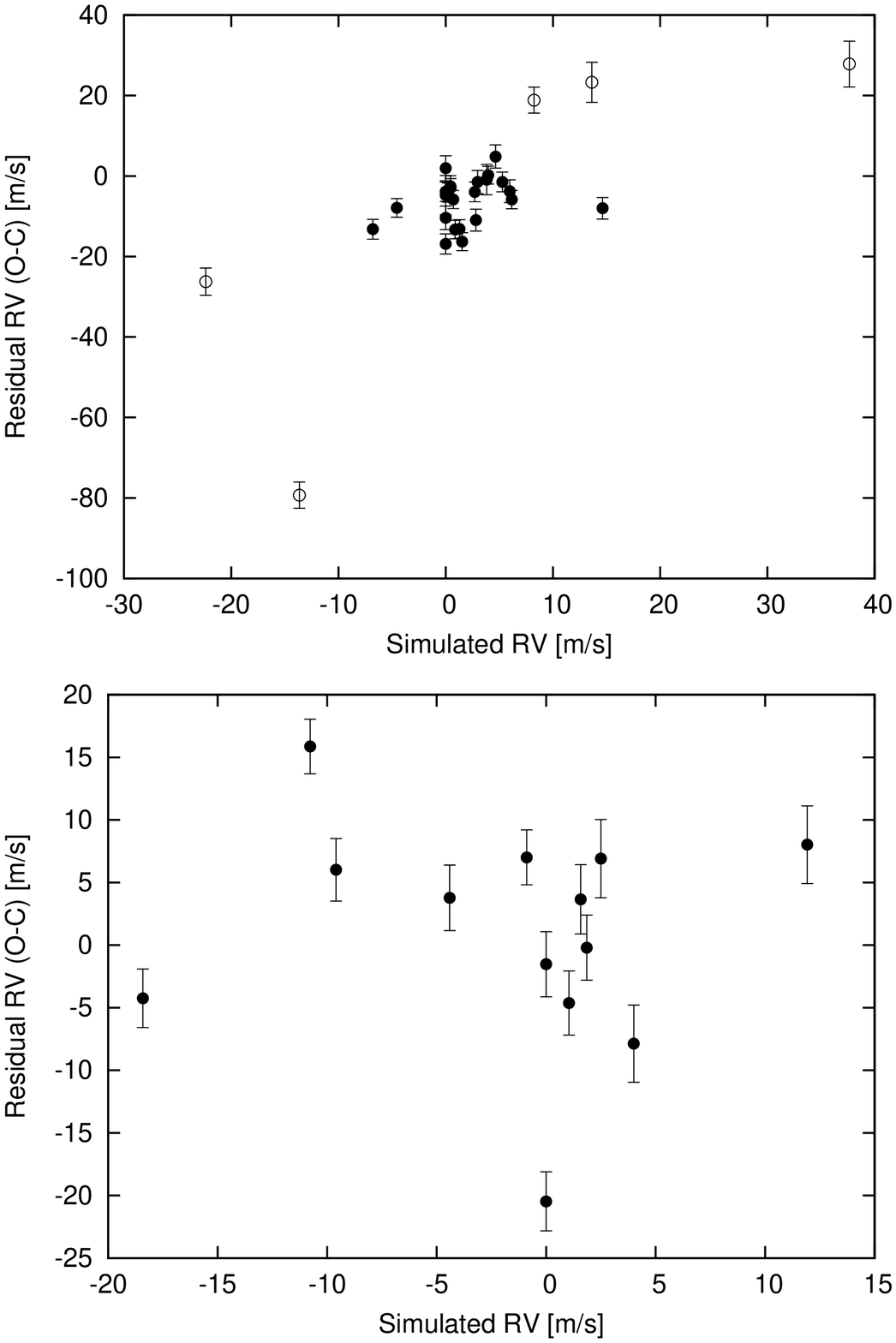}
\caption[]{
    Top: Residual RV from best-fit model vs estimated RV due to sky
    contamination for Keck spectra of \hatcur{18}. There is a rough
    correlation between the values. Open circles show RV outliers
    which were excluded when modeling the orbit, filled circles show
    all other measurements. The excluded measurements come from
    observations which are expected to be strongly contaminated by the
    sky. Bottom: Same as above, here we show the results for
    \hatcur{19}. In this case there is no apparent correlation between
    the observed RV residuals and the expected RVs due to sky
    contamination.
\label{fig:scfrv}}
\end{figure}

We have also investigated the effect of sky contamination on the
measured RVs. The expected RV due to sky contamination for a given
spectrum is estimated by finding the peak of the simulated CCF. Note
that the real RV measurements are obtained by directly modeling the
spectra and not by performing cross-correlation. We therefore only
expect a crude agreement between the estimated RVs due to sky
contamination, and the real RVs. \reffigl{scfrv} compares the expected
RVs to the measured RV residuals from the best-fit model for
\hatcur{18} and \hatcur{19}. For \hatcur{18} we find a rough
correlation between the estimated and measured RV residuals. Five of
the RV measurements which are significant outliers from the best-fit
model are rejected. These spectra are also among the most strongly
affected by sky contamination. For \hatcur{19} the values do not
appear to be correlated.

\subsubsection{Blend Modeling of the Photometry}
\label{sec:blendmod}

As an independent test on the possibility that the observations for
either \hatcur{18} or \hatcur{19} could be caused by a blend scenario,
we follow \citet{torres:2005}, \citet{hartman:2009}, and
\citet{bakos:2010} in attempting to model the photometric observations
for each object as either a hierarchical triple system, or a blend
between a bright foreground star and a background eclipsing binary. We
will show that for both \hatcur{18} and \hatcur{19} blend scenarios
that do not include a transiting planet may be rejected from the
photometric observations alone. We consider 5 possibilities:
\begin{enumerate}
\item One star orbited by a planet,
\item Hierarchical system, 3 stars, 2 fainter stars are eclipsing,
\item Hierarchical system, 2 stars, 1 planet, planet orbits the fainter star,
\item Hierarchical system, 2 stars, 1 planet, planet orbits the brighter star,
\item Chance alignment, 3 stars, 2 background stars are eclipsing.
\end{enumerate}
Here case 1 is the fiducial model to which we compare the various
blend models. We model the observed follow-up and HATNet light curves
(including only points that are within one transit duration of the
primary transit or secondary eclipse assuming zero eccentricity)
together with the 2MASS and TASS photometry. In all cases we vary the
distance to the brightest star in the system, parameters allowing for
dilution in the HATNet light curves, and we include simultaneous EPD
and TFA in fitting the light curves (see \refsecl{globmod}). We draw
the stellar radii and magnitudes from the Padova isochrones
\citep{girardi:2000}, extended below 0.15\,\msun\ with the
\citet{baraffe:1998} isochrones. We use these rather than the YY
isochrones for this analysis because of the need to allow for stars
with $M < 0.4$\,\msun, which is the lower limit available for the YY
models. We use the JKTEBOP program
\citep{southworth:2004a,southworth:2004b} which is based on the
Eclipsing Binary Orbit Program
\citep[EBOP;][]{popper:1981,etzel:1981,nelson:1972} to generate the
model light curves. We optimize the free parameters using the Downhill
Simplex Algorithm together with the classical linear least squares
algorithm for the EPD and TFA parameters. We rescale the errors for
each light curve such that $\chi^2$ per degree of freedom is 1.0 for
the out of transit portion of the light curve. Note that this is done
prior to applying EPD/TFA corrections for systematic errors. As a
result, the $\chi^2$ per degree of freedom is less than 1.0 for many
of the best-fit models discussed below. If the rescaling is not
performed, the difference in $\chi^2$ between the best-fit models is
even more significant than what is given below, and the blend models
may be rejected with even higher confidence. For \hatcur{18} we fix
the mass, age, and [Fe/H] metallicity of the brightest star in the
system to $0.76$\,\msun, $12.4$\,Gyr, and $+0.10$ respectively to
reproduce the effective temperature, metallicity, and surface gravity
of the bright star as determined from the SME analysis when using the
Padova isochrones. For \hatcur{19} we fix the mass, age, and
metallicity to $0.83$\,\msun, $8.8$\,Gyr, and $+0.23$ respectively.

{\em Case 1: 1 star, 1 planet}: In addition to the parameters
mentioned above, in this case we vary the radius of the planet and the
impact parameter of the transit. For \hatcur{18} the best-fit model
has $\chi^2_{\rm 18,Case 1} = 1542.4$ for 1712 degrees of freedom. For
\hatcur{19} the best-fit model has $\chi^2_{\rm 19,Case 1} = 2710.0$ for
2746 degrees of freedom. The parameters that we obtain for both
objects are comparable to those obtained from the global modelling
described in \refsecl{globmod}.

{\em Case 2: Hierarchical system, 3 stars}: For case 2 we vary the
masses of the eclipsing components, and the impact parameter of the
eclipse. We take the radii and magnitudes of all three stars from the
same isochrone. For \hatcur{18} we find $\chi^2_{\rm 18,Case 2} =
1566.9$ for 1711 degrees of freedom, while for \hatcur{19} we find
$\chi^2_{\rm 19,Case 2} = 2753.6$ for 2745 degrees of freedom. The
best-fit model for \hatcur{18} consists of a $0.76$\,\msun\ star that
is blended with a eclipsing binary with components of mass
$0.74$\,\msun\ and $0.12$\,\msun. For \hatcur{19} the best-fit model
consists of two equal $0.83$\,\msun\ stars with a $0.13$\,\msun\ M
dwarf eclipsing one of the two K stars. For both \hatcur{18} and
\hatcur{19} the best-fit case 2 model has higher $\chi^2$ with fewer
degrees of freedom than the best-fit case 1 model, so for both objects
the case 1 model is prefered. To establish the significance at which
we may reject the Case 2 model in favor of the Case 1 model, we
conduct Monte Carlo simulations which account for the possibility of
uncorrected systematic errors in the light curves as described in
\citet{hartman:2009}. For \hatcur{18} we reject the best-fit Case 2
model at the $\sim 4.4\sigma$ confidence level, while for \hatcur{19}
we reject the best-fit Case 2 model at the $\sim 5.1\sigma$ confidence
level. We also note that for both objects the only hierarchical triple
stellar system that could potentially fit the photometric observations
is a system where the two brightest stars have nearly equal
masses. Because both \hatcur{18} and \hatcur{19} have narrow spectral
lines ($\vsini=$\hatcurSMEiivsin{18}\,\kms\ and
\hatcurSMEiivsin{19}\,\kms\ respectively), a second component with a
luminosity ratio close to one and a RV semi-amplitude of several tens
of \kms\ would have easily been detected in the spectra of these
objects.

{\em Case 3: Hierarchical system, 2 stars, 1 planet, planet orbits
  fainter star}: In this scenario the system contains a transiting
planet, but it would have a radius that is larger than what we infer
assuming there is only one star in the system. For this case we vary
the mass of the faint planet-hosting star, the radius of the planet,
and the impact parameter of the transit. We assume the mass of the
planet is negligible relative to the mass of its faint host star. For
\hatcur{18} the best-fit Case 3 model has $\chi^2_{\rm 18,Case 3} =
1590.0$ while for \hatcur{19} the best-fit Case 3 model has
$\chi^2_{\rm 19,Case 3} = 2726.1$. For both objects the best-fit is
when the two stars in the system are of equal mass. Repeating the
Monte Carlo simulations to determine the statistical significance of
this $\chi^2$ difference, we find that for \hatcur{18} we may reject
the Case 3 model in favor of the Case 1 model at the $\sim 4.9\sigma$
confidence level, while for \hatcur{19} we may reject the Case 3 model
at the $\sim 3.6\sigma$ confidence level. As for the Case 2 model, the
only Case 3 models that could potentially fit the photometric
observations for either \hatcur{18} or \hatcur{19} are models where
both stars in the system have nearly equal mass. The narrow spectral
lines for both \hatcur{18} and \hatcur{19} means that the systemic
velocities for the putative binary star companions would need to be
very similar to those of the brighter stars not hosting the planets
(within $\sim 1$\,\kms) for the secondary stars to have gone
undetected in any of our spectroscopic observations.

{\em Case 4: Hierarchical system, 2 stars, 1 planet, planet orbits
  brighter star}: As in Case 3, in this scenario the system constains
a transiting planet, but it would have a radius that is larger than
what we infer assuming there is only one star in the system. For this
case we vary the mass of the faint contaminating star, the radius of
the planet, and the impact parameter of the transit. Again we assume
the mass of the planet is negligible relative to the mass of its host
star. For \hatcur{18} the best-fit model occurs when the mass of the
contaminating star is negligible with respect to the mass of the
planet-hosting star (which effectively corresponds to the Case 1
model), with $\chi^2_{\rm 18,Case 4}$ increasing as the mass of the
faint companion is increased. We find that a fainter companion with $M
> 0.62$\,\msun\ is rejected at the $3\sigma$ confidence level, while a
companion with $M > 0.51$\,\msun\ is rejected at the $2\sigma$
confidence level. This corresponds to $3\sigma$ and $2\sigma$ upper
limits on the $V$-band luminosity ratios of a possible contaminating
star of $0.25$ and $0.09$ respectively. At the $2\sigma$ level, the
radius of \hatcurb{18} could be $5\%$ larger than what we measure in
\refsecl{globmod} if there is an undetected faint companion star. For
\hatcur{19} we find that a fainter companion with $M >
0.81$\,\msun\ is rejected at the $3\sigma$ level, while a fainter
companion with $M > 0.69$\,\msun\ is rejected at the $2\sigma$
level. This corresponds to $3\sigma$ and $2\sigma$ upper limits on the
$V$-band luminosity ratios of a possible contaminating star of $0.84$
and $0.27$ respectively. At the $2\sigma$ level, the radius of
\hatcurb{19} could be $\sim 18\%$ larger than what we measure in
\refsecl{globmod} if there is an undetected faint companion
star. While $\chi^2$ generally increases when the mass of the faint
star increases, its minimum actually occurs when the faint star has a
mass of $\sim 0.4-0.5$\,\msun, where the value of $\chi^2$ is 4 less
than the value when the faint companion is excluded (the Case 1
model). The difference is too low to be statistically significant, but
it is nonetheless interesting that this object also exhibits a linear
drift in its radial velocity, which might indicate the presence of a
low-mass stellar companion. If there is a $0.45$\,\msun\ faint
companion, the radius of \hatcurb{19} would be $\sim 3\%$ larger than
what we measure in \refsecl{globmod}.

{\em Case 5: Chance alignment, 3 stars, background stars are
  eclipsing}: For case 5 we vary the masses of the two eclipsing
stars, the impact parameter of their eclipses, the age of the
background system, the metallicity of the background system, and the
difference in distance modulus between the foreground star and the
background binary $\Delta V$. For \hatcur{18} we find that the
best-fit model has $\chi^2_{\rm 18,Case 5} = 1561.2$ and consists of a
``background'' binary at $\Delta V = 0$ with a primary component that
has the same mass as the foreground star. This is effectively Case 2,
except that the age and metallicity of the binary are allowed to vary,
so that the result has a slightly lower $\chi^2$ value than the
best-fit Case 2 model. The value of $\chi^2_{\rm 18,Case 5}$ steadily
increases with $\Delta V$. For \hatcur{18} the best-fit Case 5 model
may be rejected in favor of the Case 1 fiducial model at the $\sim
4.1\sigma$ confidence level. We may reject models with $\Delta V >
0.2$ with greater than $5\sigma$ confidence. The $5\sigma$ lower limit
on the $V$-band luminosity ratio between the primary component of the
background binary and the foreground star is $\sim 0.75$. Such a
system would have easily been identified and rejected as a
spectroscopic double-lined object in either the Keck or TRES
spectra. For \hatcur{19} we find that the best-fit model has
$\chi^2_{\rm 19,Case 5} = 2727.7$ and consists of a background binary
at $\Delta V = 0.1$\,mag. The primary component of the binary has a
mass of 0.85\,\msun, while the secondary has a mass of
$0.14$\,\msun. The binary system has an age of 9\,Gyr and a
metallicity of [Fe/H]$=+0.4$. This model can be rejected in favor of
the fiducial model at the $3.5\sigma$ confidence level. We note that
we can reject models with $\Delta V > 0.9$\,mag at the $5\sigma$
level. The $5\sigma$ lower limit on the $V$-band luminosity ratio
between the primary component of the background binary and the
foreground star is $\sim 0.58$. As for \hatcur{18}, such a system
would have easily been identified and rejected as a spectroscopic
double-lined object in either the Keck or TRES spectra. We conclude
that for \hatcur{18} and \hatcur{19} a blend model consisting of a
single star and a background eclipsing binary is inconsistent with the
photometric observations at the $3$-$4\sigma$ level, and any models
that are not inconsistent with greater than $5\sigma$ confidence are
inconsistent with the spectroscopic observations. This reaffirms our
conclusion in the previous section about the true planetary nature of
the signals in both \hatcur{18} and \hatcur{19}.

\subsection{Global modeling of the data}
\label{sec:globmod}

This section describes the procedure we followed for each system to
model the HATNet photometry, the follow-up photometry, and the radial
velocities simultaneously.  Our model for the follow-up \lcs\ used
analytic formulae based on \citet{mandel:2002} for the eclipse of a
star by a planet, with limb darkening being prescribed by a quadratic
law.  The limb darkening coefficients for the Sloan $i$-band and Sloan
$g$-band were interpolated from the tables by \citet{claret:2004} for
the spectroscopic parameters of each star as determined from the SME
analysis (\refsecl{stelparam}).  The transit shape was parametrized by
the normalized planetary radius $p\equiv \rpl/\rstar$, the square of
the impact parameter $b^2$, and the reciprocal of the half duration of
the transit $\zrstar$.  We chose these parameters because of their
simple geometric meanings and the fact that these show negligible
correlations \citep[see][]{bakos:2010}.  The relation between
$\zrstar$ and the quantity \arstar, used in \refsecl{stelparam}, is
given by
\begin{equation}
    \arstar = P/2\pi (\zrstar) \sqrt{1-b^2} \sqrt{1-e^2}/(1+e \sin\omega)
\end{equation}
\citep[see, e.g.,][]{tingley:2005}. Our model for the HATNet data was
a simplified version of the \citet{mandel:2002} analytic
functions (an expansion in terms of Legendre polynomials), for the
reasons described in \citet{bakos:2010}.
Following the formalism presented by \citet{pal:2009}, the RVs were
fitted with an eccentric Keplerian model parametrized by the
semiamplitude $K$ and Lagrangian elements $k \equiv e \cos\omega$ and
$h \equiv e \sin\omega$, in which $\omega$ is the longitude of
periastron.

We assumed that there is a strict periodicity in the individual transit
times.  For each system we assigned the transit number $N_{tr} = 0$ to the
first complete follow-up \lc.  For
\setcounter{planetcounter}{1}
\loopand\hatcurb{18} this was the \lc\ obtained on 2008 Apr 25\loopcomma
\setcounter{planetcounter}{2}
\loopand\ for \hatcurb{19} this was the \lc\ obtained on 2009 Dec
1\loopcomma The adjustable parameters in the fit that determine the
ephemeris were chosen to be the time of the first transit center
observed with HATNet ($T_{c,-71}$, and $T_{c,-204}$ for \hatcurb{18},
and \hatcurb{19} respectively) and that of the last transit center
observed with the \flwof\ telescope ($T_{c,+69}$, and $T_{c,0}$ for
\hatcurb{18}, and \hatcurb{19} respectively).  We used these as
opposed to period and reference epoch in order to minimize
correlations between parameters \citep[see][]{pal:2008}.  Times of
mid-transit for intermediate events were interpolated using these two
epochs and the corresponding transit number of each event, $N_{tr}$.
For \hatcurb{18}, the eight main parameters describing the physical
model were thus the times of first and last transit
center, $\rpl/\rstar$, $b^2$, $\zrstar$, $K$, $k \equiv e\cos\omega$,
and $h \equiv e\sin\omega$.  For \hatcurb{19} we included as a ninth
parameter a velocity acceleration term to account for an apparent
linear drift in the velocity residuals after fitting for a Keplerian
orbit. Three additional parameters were included for \hatcurb{18} that
have to do with the instrumental configuration. For \hatcurb{19}, six
additional parameters were included.  These include the HATNet blend
factors $B_{\rm inst}$ (one for each HATNet field for \hatcurb{19}),
which accounts for possible dilution of the transit in the HATNet
\lc\ from background stars due to the broad PSF (20\arcsec\ FWHM), the
HATNet out-of-transit magnitude $M_{\rm 0,HATNet}$ (also one for each
HATNet field for \hatcurb{19}), and the relative zero-point
$\gamma_{\rm rel}$ of the Keck RVs (and the Subaru RVs for
\hatcurb{19}).

We extended our physical model with an instrumental model that
describes brightness variations caused by systematic errors in the
measurements.  This was done in a similar fashion to the analysis
presented by \citet{bakos:2010}.  The HATNet photometry has already
been EPD- and TFA-corrected before the global modeling, so we only
considered corrections for systematics in the follow-up \lcs.  We chose
the ``ELTG'' method, i.e., EPD was performed in ``local'' mode with EPD
coefficients defined for each night, and TFA was performed in
``global'' mode using the same set of stars and TFA coefficients for
all nights.  The five EPD parameters were the hour angle (representing
a monotonic trend that changes linearly over time), the square of the
hour angle (reflecting elevation), and the stellar profile parameters
(equivalent to FWHM, elongation, and position angle of the image).
The functional forms of the above parameters contained six
coefficients, including the auxiliary out-of-transit magnitude of the
individual events.  For each system the EPD parameters were
independent for all nights, implying 12, and 24 additional
coefficients in the global fit for \hatcurb{18} and \hatcurb{19}
respectively.  For the global TFA analysis we chose 20 template stars
for each system that had good quality measurements for all nights and
on all frames, implying an additional 20 parameters in the fit for
each system. In both cases the total number of fitted parameters (43,
and 49 for \hatcurb{18} and \hatcurb{19} respectively) was much
smaller than the number of data points (422, and 438, counting
only RV measurements and follow-up photometry measurements).

The joint fit was performed as described in \citet{bakos:2010}.  We
minimized \chisq\ in the space of parameters by using a hybrid
algorithm, combining the downhill simplex method \citep[AMOEBA;
  see][]{press:1992} with a classical linear least squares algorithm.
Uncertainties for the parameters were derived applying the Markov
Chain Monte-Carlo method \citep[MCMC, see][]{ford:2006} using
``Hyperplane-CLLS'' chains \citep{bakos:2010}. This provided the full
{\em a posteriori} probability distributions of all adjusted
variables. The {\em a priori} distributions of the parameters for
these chains were chosen to be Gaussian, with eigenvalues and
eigenvectors derived from the Fisher covariance matrix for the
best-fit solution. The Fisher covariance matrix was calculated
analytically using the partial derivatives given by \citet{pal:2009}.

Following this procedure we obtained the {\em a posteriori}
distributions for all fitted variables, and other quantities of
interest such as \arstar. As described in \refsecl{stelparam},
\arstar\ was used together with stellar evolution models to infer a
theoretical value for \loggstar\ that is significantly more accurate
than the spectroscopic value. The improved estimate was in turn
applied to a second iteration of the SME analysis, as explained
previously, in order to obtain better estimates of \teffstar\ and
\feh.  The global modeling was then repeated with updated
limb-darkening coefficients based on those new spectroscopic
determinations. The resulting geometric parameters pertaining to the
light curves and velocity curves for each system are listed in
Table~\ref{tab:planetparam}.

Included in each table is the RV ``jitter''. This is a component of
noise that we added in quadrature to the internal errors for the RVs
in order to achieve $\chi^{2}/{\rm dof} = 1$ from the RV data for the
global fit. It is unclear to what extent this excess noise is
intrinsic to the star, and to what extent it is due to instrumental
effects which have not been accounted for in the internal error
estimates.

The planetary parameters and their uncertainties can be derived by
combining the {\em a posteriori} distributions for the stellar, \lc,
and RV parameters.  In this way we find masses and radii for each
planet.  These and other planetary parameters are listed at the bottom
of Table~\ref{tab:planetparam}.  We find:
\begin{itemize}
\item {\em \hatcurb{18}} -- the planet has mass
  $\mpl=$\hatcurPPmlong{18}\,\mjup, radius
  $\rpl=$\hatcurPPrlong{18}\,\rjup, and mean density
  $\rho_p=$\hatcurPPrho{18}\,\gcmc.
\item {\em \hatcurb{19}} -- the planet has mass
  $\mpl=$\hatcurPPmlong{19}\,\mjup, radius
  $\rpl=$\hatcurPPrlong{19}\,\rjup, and mean density
  $\rho_p=$\hatcurPPrho{19}\,\gcmc.
\end{itemize}

Both planets have an eccentricity consistent with zero
($e=$\hatcurRVeccen{18} for \hatcurb{18}, and $e=$\hatcurRVeccen{19}
for \hatcurb{19}). As mentioned above, for \hatcur{19}, the RV
residuals from a single-Keplerian orbital fit exhibit a linear trend
in time. We therefore included an acceleration term to account for
this trend. We find $\dot{\gamma}=$\hatcurRVtrone{19}\,\msd. In
\refsecl{discussion} we consider the implication of additional
bodies (stellar or planetary) in the \hatcur{19} system.

%
%
\ifthenelse{\boolean{emulateapj}}{
    \begin{deluxetable*}{lcc}
}{
    \begin{deluxetable}{lcc}
}
\tabletypesize{\scriptsize}
\tablecaption{Orbital and planetary parameters for \hatcurb{18} and \hatcurb{19}\tablenotemark{a}\label{tab:planetparam}}
\tablehead{
    \colhead{~~~~~~~~~~~~~~~Parameter~~~~~~~~~~~~~~~} &
    \colhead{\hatcurb{18}}                            &
    \colhead{\hatcurb{19}}
}
\startdata
\noalign{\vskip -3pt}
\sidehead{\Lc{} parameters}
~~~$P$ (days)             \dotfill    & $\hatcurLCP{18}$              & $\hatcurLCP{19}$              \\
~~~$T_c$ (${\rm BJD}$)    
      \tablenotemark{b}   \dotfill    & $\hatcurLCT{18}$              & $\hatcurLCT{19}$              \\
~~~$T_{14}$ (days)
      \tablenotemark{b}   \dotfill    & $\hatcurLCdur{18}$            & $\hatcurLCdur{19}$            \\
~~~$T_{12} = T_{34}$ (days)
      \tablenotemark{b}   \dotfill    & $\hatcurLCingdur{18}$         & $\hatcurLCingdur{19}$         \\
~~~$\arstar$              \dotfill    & $\hatcurPPar{18}$             & $\hatcurPPar{19}$             \\
~~~$\zrstar$              \dotfill    & $\hatcurLCzeta{18}$       & $\hatcurLCzeta{19}$       \\
~~~$\rpl/\rstar$          \dotfill    & $\hatcurLCrprstar{18}$        & $\hatcurLCrprstar{19}$        \\
~~~$b^2$                  \dotfill    & $\hatcurLCbsq{18}$            & $\hatcurLCbsq{19}$            \\
~~~$b \equiv a \cos i/\rstar$
                          \dotfill    & $\hatcurLCimp{18}$            & $\hatcurLCimp{19}$            \\
~~~$i$ (deg)              \dotfill    & $\hatcurPPi{18}$          & $\hatcurPPi{19}$          \\

\sidehead{Limb-darkening coefficients \tablenotemark{c}}
~~~$a,i$ (linear term, $i$ filter) \dotfill & $\hatcurLBii{18}$  & $\hatcurLBii{19}$  \\ 
~~~$b,i$ (quadratic term) \dotfill & $\hatcurLBiii{18}$ & $\hatcurLBiii{19}$ \\
~~~$a,g$               \dotfill    & $\hatcurLBig{18}$             & $\hatcurLBig{19}$             \\
~~~$b,g$               \dotfill    & $\hatcurLBiig{18}$            & $\hatcurLBiig{19}$            \\

\sidehead{RV parameters}
~~~$K$ (\ms)              \dotfill    & $\hatcurRVK{18}$      & $\hatcurRVK{19}$      \\
~~~$\dot{\gamma}$ (\msd)  \dotfill    & $\ldots$                      & $\hatcurRVtrone{19}$          \\
~~~$k_{\rm RV}$\tablenotemark{d} 
                          \dotfill    & $\hatcurRVk{18}$          & $\hatcurRVk{19}$          \\
~~~$h_{\rm RV}$\tablenotemark{d}
                          \dotfill    & $\hatcurRVh{18}$              & $\hatcurRVh{19}$              \\
~~~$e$                    \dotfill    & $\hatcurRVeccen{18}$          & $\hatcurRVeccen{19}$          \\
~~~$\omega$ (deg)         \dotfill    & $\hatcurRVomega{18}$      & $\hatcurRVomega{19}$      \\
~~~RV jitter (\ms)        \dotfill    & \hatcurRVjitter{18}           & \hatcurRVjitter{19}           \\

\sidehead{Secondary eclipse parameters}
~~~$T_s$ (BJD)            \dotfill    & $\hatcurXsecondary{18}$       & $\hatcurXsecondary{19}$       \\
~~~$T_{s,14}$              \dotfill   & $\hatcurXsecdur{18}$          & $\hatcurXsecdur{19}$          \\
~~~$T_{s,12}$              \dotfill   & $\hatcurXsecingdur{18}$       & $\hatcurXsecingdur{19}$       \\

\sidehead{Planetary parameters}
~~~$\mpl$ ($\mjup$)       \dotfill    & $\hatcurPPmlong{18}$          & $\hatcurPPmlong{19}$          \\
~~~$\rpl$ ($\rjup$)       \dotfill    & $\hatcurPPrlong{18}$          & $\hatcurPPrlong{19}$          \\
~~~$C(\mpl,\rpl)$
    \tablenotemark{e}     \dotfill    & $\hatcurPPmrcorr{18}$         & $\hatcurPPmrcorr{19}$         \\
~~~$\rhopl$ (\gcmc)       \dotfill    & $\hatcurPPrho{18}$            & $\hatcurPPrho{19}$            \\
~~~$\log g_p$ (cgs)       \dotfill    & $\hatcurPPlogg{18}$           & $\hatcurPPlogg{19}$           \\
~~~$a$ (AU)               \dotfill    & $\hatcurPParel{18}$           & $\hatcurPParel{19}$           \\
~~~$T_{\rm eq}$ (K)        \dotfill   & $\hatcurPPteff{18}$           & $\hatcurPPteff{19}$           \\
~~~$\Theta$\tablenotemark{f} \dotfill & $\hatcurPPtheta{18}$          & $\hatcurPPtheta{19}$          \\
~~~$\langle F \rangle$ ($10^{\hatcurPPfluxavgdim{18}}$\ergscmsq) \tablenotemark{g}
                          \dotfill    & $\hatcurPPfluxavg{18}$        & $\hatcurPPfluxavg{19}$        \\ [-1.5ex]
\enddata
\tablenotetext{a}{
    We list the median value of each parameter from its MCMC {\em a
      posteriori} distribution. We also provide the upper and lower
    $1\sigma$ error-bars about the median for each parameter.
}
\tablenotetext{b}{
    \ensuremath{T_c}: Reference epoch of mid transit that
    minimizes the correlation with the orbital period. It
    corresponds to $N_{tr} = 24$.
    \ensuremath{T_{14}}: total transit duration, time
    between first to last contact;
    \ensuremath{T_{12}=T_{34}}: ingress/egress time, time between
    first and second, or third and fourth contact. BJD is calculated
    from UTC.
}
\tablenotetext{c}{
    Values for a quadratic law, adopted from the tabulations by
    \cite{claret:2004} according to the spectroscopic (SME) parameters
    listed in \reftabl{stellar}.
}
\tablenotetext{d}{The Lagrangian orbital parameters derived from the global modeling, and primarily determined by the RV data. 
}
\tablenotetext{e}{
    Correlation coefficient between the planetary mass \mpl\ and radius
    \rpl.
}
\tablenotetext{f}{
    The Safronov number is given by $\Theta = \frac{1}{2}(V_{\rm
    esc}/V_{\rm orb})^2 = (a/\rpl)(\mpl / \mstar )$
    \citep[see][]{hansen:2007}.
}
\tablenotetext{g}{
    Incoming flux per unit surface area, averaged over the orbit.
}
\ifthenelse{\boolean{emulateapj}}{
    \end{deluxetable*}
}{
    \end{deluxetable}
}
%



\section{Discussion}
\label{sec:discussion}

\begin{figure*}[!ht]
\ifthenelse{\boolean{emulateapj}}{
\epsscale{1.0}
}{
\epsscale{1.0}
}
\plotone{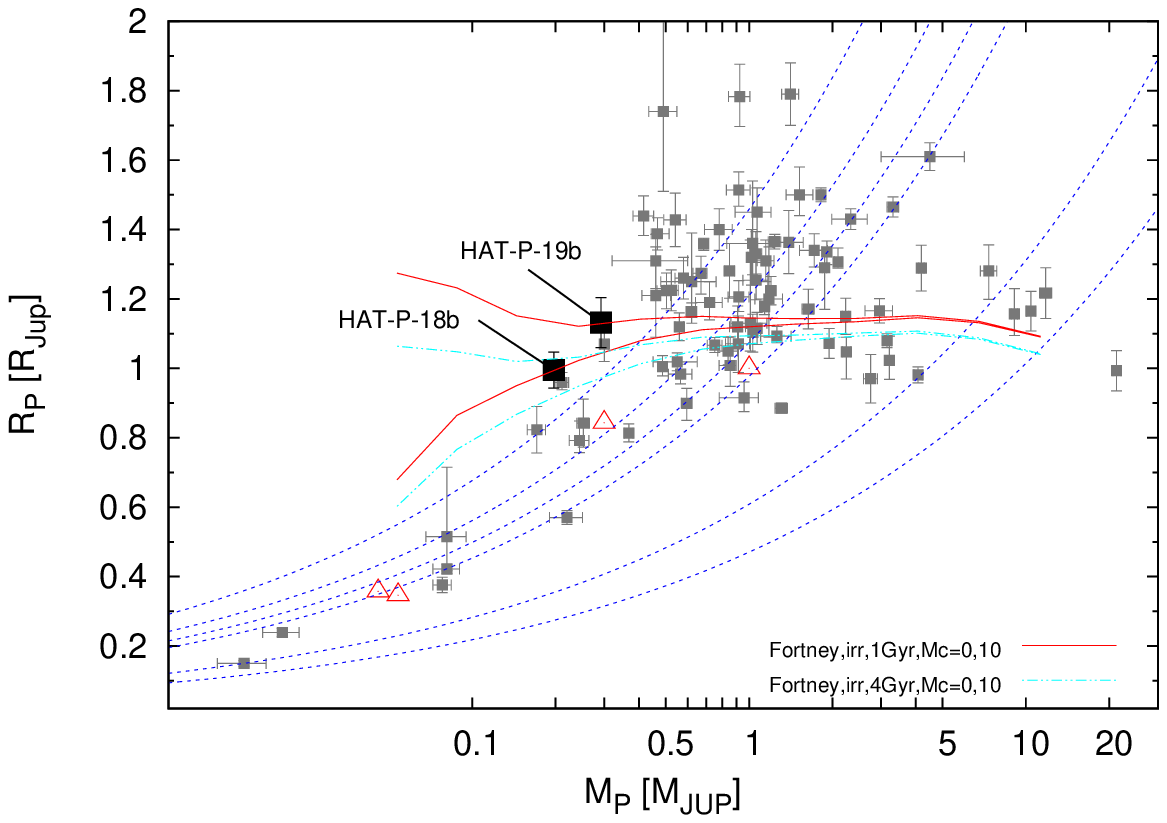}
\caption{ Mass--radius diagram of known TEPs (small filled
  squares). \hatcurb{18} and \hatcurb{19}\ are shown as large filled
  squares.  Overlaid are \citet{fortney:2007} planetary isochrones
  interpolated to the solar equivalent semi-major axis of \hatcurb{18}
  for ages of 1.0\,Gyr (upper, solid lines) and 4\,Gyr (lower
  dashed-dotted lines) and core masses of 0 and 10\,\mearth (upper and
  lower lines respectively), as well as isodensity lines for 0.4, 0.7,
  1.0, 1.33, 5.5 and 11.9\gcmc (dashed lines). Solar system planets
  are shown with open triangles.
\label{fig:exomr}}
\end{figure*}

\reffigl{exomr} compares \hatcurb{18} and \hatcurb{19} to other known
TEPs on a mass-radius diagram. We discuss the properties of each
planet in turn.
\subsection{\hatcurb{18}}
\label{sec:disc18}
From the \cite{fortney:2007} planetary models, the expected radius for
a coreless \hatcurPPm{18}\,\mjup\ planet orbiting a 4.5\,Gyr star
with a Solar-equivalent semi-major axis of
\hatcurPPaequiv{18}\,AU is $\sim 1.02$\,\rjup, which is consistent
with the measured radius for \hatcurb{18} of
\hatcurPPr{18}\,\rjup. The preferred age for \hatcur{18} from the
\hatcurisoshort{18} isochrones (\hatcurISOage{18}\,Gyr), is somewhat
older than 4.5\,Gyr, in which case the expected planetary radius would
be even smaller. If a slight core of $10$\,\mearth\ is assumed, the
expected radius of $0.91$\,\rjup\ is below the measured radius. We
conclude therefore that \hatcurb{18} is a predominately
hydrogen-helium gas giant planet, and does not possess a significant
heavy element core.

\hatcurb{18} is perhaps most similar in properties to the slightly
higher density planet HAT-P-12b ($M = 0.211\pm0.012$\,\mjup, $R =
0.959_{-0.021}^{+0.029}$\,\rjup; \citealp{hartman:2009}). Both
planets orbit K dwarfs (\hatcur{18} has mass $M_{\star} =
\hatcurISOm{18}$\,\msun, and HAT-P-12 has mass $M_{\star} =
0.73\pm0.02$\,\msun). However \hatcur{18} appears to be older than
HAT-P-12, having an isochrone age of \hatcurISOage{18}\,Gyr compared with $2.5\pm2.0$\,Gyr for HAT-P-12. \hatcur{18} is also more metal rich ([Fe/H]=$\hatcurSMEzfeh{18}$) than HAT-P-12 ([Fe/H]=$-0.36\pm0.04$).

\subsection{\hatcurb{19}}
\label{sec:disc19}
Like \hatcurb{18}, \hatcurb{19} also does not appear to possess a
signficant heavy element core. From the \cite{fortney:2007} planetary
models, the expected radius for a coreless
\hatcurPPm{19}\,\mjup\ planet orbiting a 4.5\,Gyr star with a
Solar-equivalent semi-major axis of \hatcurPPaequiv{19}\,AU is $\sim
1.02$\,\rjup, which is lower than the measured radius for \hatcurb{19}
of \hatcurPPr{19}\,\rjup. If an age of 1.0\,Gyr is assumed, the
expected planet radius increases to 1.06\,\rjup, but is still slightly
lower than, though consistent with, the measured radius.

Like \hatcurb{18}, \hatcurb{19} is also very similar in mass/radius to
another TEP, in this case WASP-21b ($M = 0.30 \pm 0.01$\,\mjup, $R =
1.07 \pm 0.05$\,\rjup; \citealp{bouchy:2010}). WASP-21b orbits a
somewhat hotter star than \hatcurb{19} (WASP-21 has $M =
1.01_{-0.025}^{+0.024}$\,\msun, while \hatcur{19} has $M =
\hatcurISOm{19}$\,\msun). \hatcur{19} is also more metal rich than
WASP-21 ([Fe/H]=$\hatcurSMEzfeh{19}$ for \hatcur{19}, while
[Fe/H]=$-0.4 \pm 0.1$ for WASP-21).

As noted in \refsecl{globmod}, the RV residuals of \hatcur{19} show a
linear trend in time, which is evidence for a third body in the
system. The evidence for this trend comes entirely from the Keck/HIRES
observations which span 144 days. Although the Subaru/HDS observations
predate the Keck/HIRES observations, the uncertain RV zero-point
difference between these two datasets prevents us from comparing the
Subaru/HDS observations with the Keck/HIRES observations to extend the
baseline for measuring the linear variation in the RV residuals. The
Subaru/HDS observations span only 3 days, so the trend is not evident
in this dataset. Following \cite{winn:2010}, we set $\dot{\gamma} \sim
G M_{c} \sin i_{c} / a_{c}^2$ to give an order-of-magnitude constraint
on the third body, assuming the orbit is circular. This gives
\begin{equation}
\left( \frac{M_{c}\sin i_{c}}{\mjup} \right) \left( \frac{a_{c}}{1 AU} \right)^{-2} \sim 0.9
\end{equation}
The time span of the RV measurements, and the lack of evidence for
jerk ($\ddot{\gamma}$) in the RV residuals, lets us put a rough
limit on the third body's orbital period of $P_{c} \ga 2\times144 =
288$\,d, or $a_{c} \ga 0.8$\,AU. This gives a rough lower limit on the
mass of the third body of $M_{c} \ga 0.6$\,\mjup, though this depends
on the eccentricity, argument of periastron, and time of
conjunction. The object could also be a low-mass star with $M >
90$\,\mjup\ if it has $a_{c} \ga 10$\,AU.

\subsection{Core Mass--Metallicity Correlation}

\begin{figure*}[!ht]
\plotone{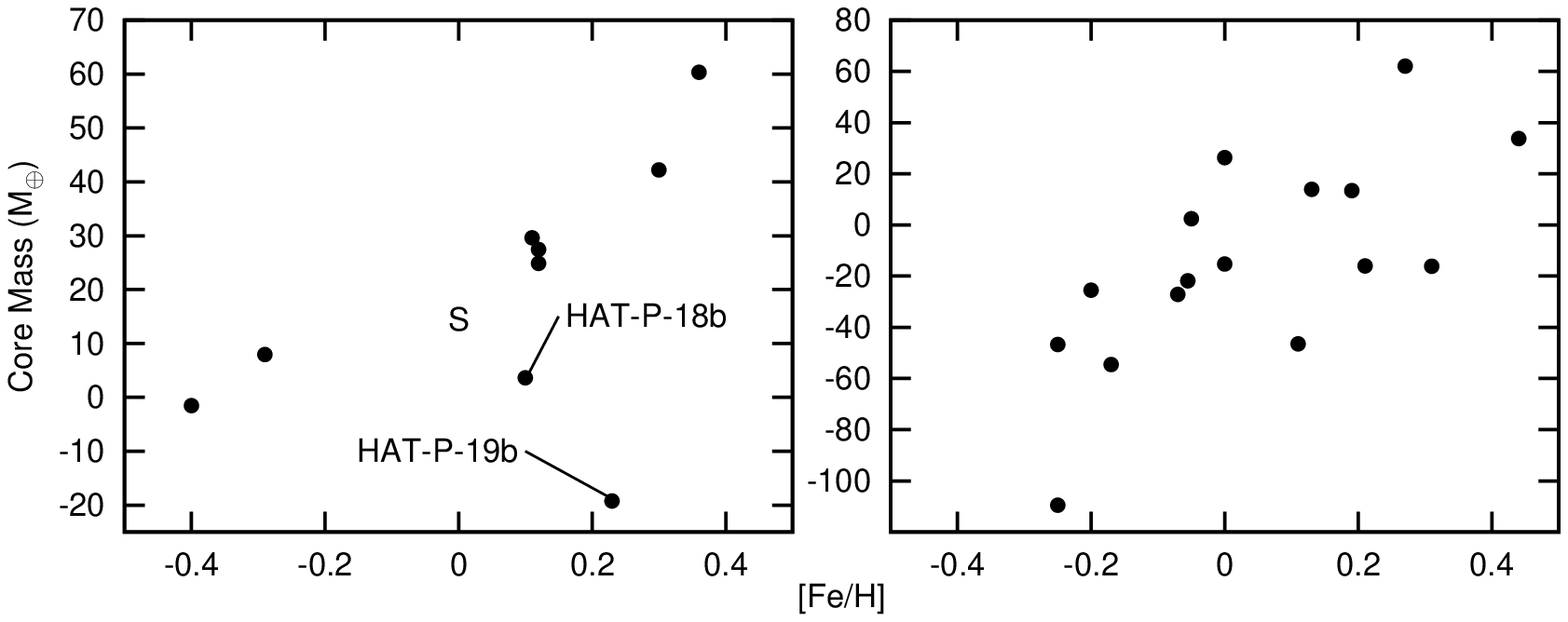}
\caption{ Planet core mass vs. host star metallicity for planets with
  $0.15 < M < 0.4$\,\mjup\ (left) and $0.4 < M <
  0.6$\,\mjup\ (right). The core mass for each planet is determined by
  linear interpolation within the \citet{fortney:2007} planet model
  tables for the estimated age, mass, and solar-equivalent semimajor
  axis of the planet. We adopt an age of $4.0$\,Gyr or $0.3$\,Gyr for
  systems with an estimated age greater or less than these limits, and
  we adopt a solar-equivalent semimajor axis of $9.5$\,AU for
  Saturn. Planets with a negative inferred core mass have radii that
  are too large to be accommodated by the \citet{fortney:2007}
  models. In this case the core mass is linearly extrapolated from the
  models, and provides a measure for the degree to which the observed
  radius disagrees with the models. The location of Saturn is
  indicated by the ``S'' in the left plot. \hatcurb{18} and
  \hatcurb{19} do not follow the previously suggestive correlation
  between core mass and host star metallicity for Saturn-mass
  planets. No correlation is apparent for planets with $M >
  0.6$\,\mjup.
\label{fig:Zcoremass}}
\end{figure*}

As noted in the introduction, the previously known Saturn-mass planets
exhibited a suggestive correlation between core mass (or density) and
host star metallicity. The two low density planets HAT-P-12b and
WASP-21b are consistent with having no core, and orbit sub-solar
metallicity stars. While the higher density planets Kepler-9b,
Kepler-9c, CoRoT-8b, WASP-29b, and HD~149026b are consistent with
having substantial cores, and orbit super-solar metallicity stars. The
apparent correlation between planet core mass and host star
metallicity was previously noted by \citet{guillot:2006} and
\citet{burrows:2007} for all TEPs known at the time (nine and fourteen
respectively). Many of the planets with $M \ga 0.4$\,\mjup{} have
radii that are larger than can be accommodated by theoretical models,
so it is unclear whether the inferred core masses are physically
meaningful for these planets. Nonetheless, for planets in the
mass-range $0.4$--$0.7$\,\mjup\ \citet{enoch:2010} find that planet
radius is inversely proportional to host star metallicity, which is
what would be expected if the heavy element content of these planets
(or core mass) is proportional to host star
metallicity. \reffigl{Zcoremass} shows the relation between core mass
inferred from the \citet{fortney:2007} models and stellar metallicity
for planets with $0.15 < M < 0.4$\,\mjup, and $0.4 < M <
0.6$\,\mjup. \hatcurb{18} and \hatcurb{19} do not follow the
correlation that was previously seen for the other Saturn-mass
planets. However, since the sample size of known Saturn-mass TEPs is
still quite small, further discoveries are needed to illuminate the
properties of planets in this mass-range.

\clearpage

\acknowledgements 

HATNet operations have been funded by NASA grants NNG04GN74G,
NNX08AF23G and SAO IR\&D grants. Work of G.\'A.B.~and J.~Johnson were
supported by the Postdoctoral Fellowship of the NSF Astronomy and
Astrophysics Program (AST-0702843 and AST-0702821, respectively). GT
acknowledges partial support from NASA grant NNX09AF59G. We
acknowledge partial support also from the Kepler Mission under NASA
Cooperative Agreement NCC2-1390 (D.W.L., PI). G.K.~thanks the
Hungarian Scientific Research Foundation (OTKA) for support through
grant K-81373. This research has made use of Keck telescope time
granted through NOAO (programs A146Hr, A201Hr, and A264Hr), NASA
(programs N018Hr, N049Hr, N128Hr, and N167Hr), and through the NOAO
Keck-Gemini time exchange program (program G329Hr).



\end{document}